\documentclass[twocolumn]{aastex63}

\usepackage{graphicx}  
\usepackage{subfigure} 
\usepackage{changepage} 
\usepackage{graphicx}   
\usepackage{subfigure}
\usepackage{multirow}
\usepackage{amsmath}
\usepackage{lineno}

\usepackage{fancyhdr}
\usepackage{xcolor} 
\usepackage[normalem]{ulem} 
\graphicspath{{./}{figures/}}
\shorttitle{outflows in M82}
\shortauthors{Li, Zhu, Wang, \& Feng.}

\begin{document}
\title{Revisiting the Galactic Winds in M82 II: Development of Multiphase Outflows in Simulations}

\correspondingauthor{Weishan Zhu}
\email{zhuwshan5@mail.sysu.edu.cn}

\author{Xue-Fu, Li}
\affil{School of Physics and Astronomy, Sun Yat-Sen University, Zhuhai campus, No. 2, Daxue Road \\
Zhuhai, Guangdong, 519082, China}

\author[0000-0002-1189-2855]{Weishan, Zhu}
\affil{School of Physics and Astronomy, Sun Yat-Sen University, Zhuhai campus, No. 2, Daxue Road \\
Zhuhai, Guangdong, 519082, China}

\author{Tian-Rui, Wang}
\affil{School of Physics and Astronomy, Sun Yat-Sen University, Zhuhai campus, No. 2, Daxue Road \\
Zhuhai, Guangdong, 519082, China}

\author{Long-Long, Feng}
\affil{School of Physics and Astronomy, Sun Yat-Sen University, Zhuhai campus, No. 2, Daxue Road \\
Zhuhai, Guangdong, 519082, China}



\begin{abstract}
We performed a suit of three-dimensional hydrodynamical simulations with a resolution of $\sim10$ parsecs to investigate the development of multiphase galactic wind in M82. The star formation and related feedback processes are solved self-consistently using a sink particle method, rather than relying on various assumptions that were used in previous studies. Our simulations produce a starburst event lasting around 25 Myr, which has a total stellar mass of 1.62 - 3.34 $\times 10^8\, \rm{M_{\odot}}$, consistent with observational estimates. The total injected supernova energy is between $1.14\times 10^{57}$ and $2.4\times 10^{57} \rm{erg}$. Supernova (SN) feedback heats portions of the cool gas in the central disc to warm and hot phases, and then drives the gas in all three phases out, eventually forming multiphase outflows. These outflows can replicate key properties of the winds observed in M82, such as morphology, mass outflow rates of cool and hot phases, and X-ray emission flux, provided the gas return from star-forming clumps to the diffuse interstellar medium is implemented appropriately. The maximum mass outflow rate of all gas (hot) is about 6-12 (2-3)$\rm{M_{\odot}/yr}$ at $r\sim4.0\,$ kpc, corresponding to a mass loading factor of 2-4. However, the outflow velocities in our simulations are slower than observational estimates by $\sim 20\%-60\%$. The gas return process significantly influences the outflow properties, while the initial gas distribution in the nuclear region has a moderate effect. Yet, our results face some challenges in achieving convergence as the resolution increases. We discuss potential improvements to address these issues in future work. 
\end{abstract}
\keywords{Galactic winds (572); Galaxy evolution (594); Starburst galaxies (1570); Hydrodynamical
simulations (767);Stellar feedback (1602); Circumgalactic medium (1879)}



\section{Introduction} \label{sec:intro}
Galactic-scale outflows, commonly referred to as galactic winds, are believed to play a critical role in modern galaxy formation and evolution theories by suppressing star formation efficiency, redistributing baryons on galaxy-halo scales, and enriching the circumgalactic and intergalactic medium (e.g., \citealt{1986ApJ...303...39D,1994MNRAS.271..781C,1999ApJ...519L.109C,2003ApJ...599...38B,2005ARA&A..43..769V,2017ARA&A..55...59N}). Galactic winds have been observed in numerous local and high redshift galaxies (e.g., \citealt{1987AJ.....93..264M,1990ApJS...74..833H,1999ApJ...513..156M,2000ApJS..129..493H,2005ApJS..160..115R,2009ApJ...692..187W,2012ApJ...760..127M,2014A&A...568A..14A,2015ApJ...809..147H,2019ApJ...873..122D,2019ApJ...886...74M,2021MNRAS.505.5753F,2023ApJ...949....9P,2023arXiv231006614X,2024A&A...685A..99C,2024MNRAS.531.4560W}). These winds are primarily powered by feedback from starbursts and accretion onto supermassive black holes, with the former channel dominating in less massive galaxies, such as M82, where galactic winds were first detected (\citealt{1963ApJ...137.1005L}). This work focuses on galactic winds driven by starbursts in relatively low-mass galaxies.

Thanks to efforts from both the observation and the theoretical sides in the past over six decades, we now have an overall picture of the launching and development of galactic winds driven by starburst (\citealt{1985Natur.317...44C,1988ApJ...330..695T,1992ApJ...401..596L,Tomisaka1993,1994ApJ...430..511S,Strickland2000,2005ApJ...618..569M, 2005ARA&A..43..769V,2011ApJ...735...66M,2017arXiv170109062H,2018Galax...6..114Z, 2020A&ARv..28....2V,2024ARA&A..62..529T}). In a relatively short period, massive stars and supernovae release large amounts of momentum and energy into the interstellar medium (ISM) in the central region of a galaxy. A portion of this feedback energy is transferred to the surrounding ISM as thermal energy, heating some of the ISM to warm ($2 \times 10^4 <\rm{T}<5 \times 10^5\,$K) and hot (T$>5 \times 10^5$ K) phases. The exact temperature ranges for the different phases vary moderately in the literature. In this study, we adopt the ranges used in several previous works. (e.g. \citealt{schneider2020physical,schneider2024cgols}). Stellar feedback drives the warm and hot ISM outward at velocities ranging from a few hundred to a few thousand km/s, and eventually generates multiphase galactic outflows that can extend up to tens of kiloparsecs. 

Despite significant progress, several key aspects of starburst-driven galactic winds remain uncertain (e.g., \citealt{2017arXiv170109062H, 2018Galax...6..138R, 2020A&ARv..28....2V,2024ARA&A..62..529T}). For instance, there are notable uncertainties in the measured properties of galactic winds in individual galaxies, including the velocity, mass, momentum, and energy outflow rates of different gas phases. The origin and acceleration mechanisms of the outflowing cool gas also remain unresolved. Furthermore, the role of radiation pressure in shaping galactic winds requires further investigation. In addition, while cosmic rays have the potential to play an important role in driving galactic winds (e.g. \citealt{2008ApJ...687..202S,2014MNRAS.437.3312S} ), the specifics of cosmic-ray transport remains an open question. This uncertainty leads to varying predictions about their impact on the properties of multiphase galactic winds. (\citealt{2018ApJ...868..108B,2022MNRAS.510..920Q,2022MNRAS.510.1184Q}).  

A more comprehensive and in-depth understanding of starburst-driven galactic outflows is urgently needed, both to explain the properties of individual galaxies experiencing winds and to strengthen current galaxy formation and evolution theories. Ongoing and future multiwavelength observations with advanced facilities such as Keck, ALMA, and JWST will enhance the accuracy and reliability of measured galactic wind properties in both the local and distant universe. Meanwhile, numerical simulations, particularly those modelling well-observed outflows, are crucial for incorporating relevant physics with greater fidelity, comparing directly with observations and theoretical models. These efforts will help uncover the detailed physical mechanisms driving galactic winds. To contribute to this, we have conducted a series of three-dimensional hydrodynamical simulations to investigate the winds in M82, triggered by a recent starburst in its nuclear region. 

In fact, the galactic winds in M82 or M82-like galaxies have been extensively studied using one- and two-dimensional axisymmetric simulations (e.g.\citealt{1988ApJ...330..695T, 1994ApJ...430..511S,1996ApJ...463..528S, 1998MNRAS.293..299T, 1999ApJ...513..142M, Strickland2000, 2003ApJ...597..279T, 2009ApJ...697.2030S}), and three-dimensional simulations (e.g.\citealt{2008ApJ...674..157C,2008ApJ...689..153R,  melioli2013evolution,schneider2020physical,schneider2024cgols}). Each of these studies successfully replicated certain properties of M82’s winds, demonstrating how the characteristics of galactic winds depend on factors like ISM distribution, star formation history, and the distribution and clustering of super star clusters. Notably, the simulations in \cite{schneider2020physical}; \cite{schneider2024cgols} were able to drive a multiphase outflow on a 10 kpc scale, with velocities comparable to those observed in M82, though the mass outflow rates for the cool and hot phases were somewhat lower and the injected energy was somewhat higher than the usually assumed value. These simulations have also contributed to validating and improving theoretical models (e.g.\citealt{1985Natur.317...44C,2005ApJ...618..569M,2022ApJ...924...82F}). However, earlier simulations of M82’s winds have some limitations. For instance, in all these simulations, the recent starburst is not resolved in real-time according to the state of ISM but imposed by hand based on various assumptions. Furthermore, key physical processes, such as radiative cooling and stellar feedback, could be more accurately modeled. Many simulations have restricted volumes, typically around 1 kpc in size, and some use outdated mass models for M82.

Based on advancements from previous studies of the M82 wind—spanning simulations, observations, and theoretical work—our study aims to deepen the understanding of outflows in M82 using higher-resolution three-dimensional hydrodynamic simulations. These simulations feature a more realistic treatment of the recent starburst process in the nuclear region, a refined implementation of key physical processes, and updated initial conditions. The first paper in this series (\citealt{Wang2024}) focuses on the recent starburst and the launch of outflows. As the second paper, this work primarily explores the development of multiphase outflows. The paper is organized as follows: Section 2 details the methods and simulations. Section 3 provides an overview of the outflow evolution in our simulations. Section 4 presents quantitative properties of the outflows, such as filling factors, density and metallicity profiles, and outflow rates for the cool, warm, and hot gas. In Section 5, we discuss the effects of resolution and gas metallicity, compare our findings with previous studies, and desribe potential improvements for future work. Finally, Section 6 summarizes the key findings.

\section{Methodology} \label{sec:method}
\subsection{Overall Framework}
The framework of our method is detailed in \cite{Wang2024}, to which we refer the readers for further details. Here, we provide a brief overview. We begin by constructing the mass model of M82, accounting for contributions to the mass and gravitational potential from the gas disk, the pre-existing stellar disk, and the dark matter halo. Our set up of the mass model and gas disc largely follows methods outlined in previous studies (e.g., \citealt{Strickland2000,2008ApJ...674..157C,melioli2013evolution,schneider2020physical}). While early works such as \cite{Strickland2000,2008ApJ...674..157C,melioli2013evolution} often neglected the contribution of the dark matter halo, incorporating it is necessary to reproduce the latest rotation curve reported by \cite{Greco2012}. The initial total gas mass in our simulations, $\rm{M_{gas,t}}$, ranges from $2.4 \times 10^{9}$ to $4.7 \times 10^{9}\,\rm{M_{\odot}}$. Approximately $95\%$ of the gas mass is initially concentrated in the central plane of the disk, within 400 pc of height. The uncertainty in the molecular gas mass, which is estimated to be between $1.3 \times 10^{9}$ to $2.4 \times 10^{9} \rm{M_{\odot}}$ (\citealt{2002ApJ...580L..21W,2013PASJ...65...66S,Adam2015,2021ApJ...915L...3K}), is taken into account by setting the initial total gas mass between $2.4 \times 10^{9}$ to $3.2 \times 10^{9}$ $ \rm{M_{\odot}}$. Additionally, given the likelihood that a significant amount of gas was funneled into M82's core prior to its most recent starburst—likely due to tidal interactions with M81 or the presence of a bar—we ran simulations with an initial total gas mass exceeding $3.2 \times 10^{9}$ $\rm{M_\odot}$, to increase the gas density in the nuclear region. The gravitational effects of the stellar disk and dark matter halo are modeled as a static background in our simulations. While there are uncertainties regarding the estimated masses of M82's stellar disk ,$\rm{M_{sd}}$, and dark matter halo, $\rm{M_{dmh}}$, we adopt values of $\rm{M_{sd}}\sim 2-4 \times 10^9\rm{M_\odot}$ and $\rm{M_{dmh}}\sim 5-8 \times 10^{10}\rm{M_\odot}$. The stellar disk is assumed to follow a Miyamoto–Nagai profile, while the dark matter halo is assumed to follow a core-Navarro–Frenk–White profile.

The density distributions of the gas disk and the hot gas halo are given by
\begin{equation}
\begin{split}
    &\rho_{d}(r_{ax},z) =  \rho_{d,0}  \times \\
    & \rm{exp}\left[ -\frac{\Phi_{tot}(r_{ax},z)-e_{d}^{2}\Phi_{tot}(r_{ax},0)-(1-e_{d}^{2})\Phi_{tot}(0,0)}{c_{s,d}^{2}} \right]
\end{split}
\end{equation}

\begin{equation}
\begin{split}
    &\rho_{h}(r_{ax},z) =  \rho_{h,0}   \times \\
    & \rm{exp}\left[ -\frac{\Phi_{tot}(r_{ax},z)-e_{h}^{2}\Phi_{tot}(r_{ax},0)-(1-e_{h}^{2})\Phi_{tot}(0,0)}{c_{s,h}^{2}} \right] 
\end{split}
\end{equation}

where $\rho_{d,0}$ represents the characteristic density of the gas disk, which is determined by $\rm{M_{gas,t}}$. The values of $\rho_{d,0}$ used in different simulations are listed in Table \ref{tab:simulations}. $\rho_{h,0}$ is the characteristic density of hot gas halo, and equals to $2\times 10^{-3}$ $\rm g/cm^{-3}$. $\Phi_{tot}$ is the total gravitational potential of stellar disk, gas disk and dark matter halo. The radial distance from the disk's central axis is denoted as $r_{ax}=\sqrt{x^2+y^2}$,  and z refers to the coordinate along the axis perpendicular to the disk. The effective sound speeds for the gas disk and the hot gas halo are $c_{s,d}$ = $3.3 \times 10^{6}$ cm/s and $c_{s,h}$ = $3.0 \times 10^{7}$ cm/s, respectively.
The rotational factors $e_{d}$ and $e_{h}$ are 

\begin{equation}
    e_{d} = e_{rot,d} \  \rm{exp}(-z/z_{rot}),
\end{equation}

\begin{equation}
    e_{h} = e_{rot,h} \  \rm{exp}(-z/z_{rot}),
\end{equation}
where $e_{rot,d}$ = 0.92, $e_{rot,h}$ = 0.2 and $z_{rot}$ = 5 kpc.  

In each simulation, we first generate the initial gas distribution based on the gas disk and halo models. We then track the evolution of the gas, star formation, and feedback processes using the Athena++ code \footnote{https://github.com/PrincetonUniversity/athena} (\citealt{Stone2020}). The gravitational potential from the pre-existing stellar disk and dark matter halo is treated as a fixed background, while the gas gravity is calculated using Athena++'s built-in module. During the simulation, the gravity of newly formed stars and their interaction with the gas are computed at each time step. Radiative cooling and heating are handled by the Grackle library \footnote{https://grackle.readthedocs.io} (\citealt{Smith2016}), and we impose a temperature floor of 300 K throughout the simulation. 

The star formation process is modeled using the sink particle method (e.g.,\citealt{1995MNRAS.277..362B,2004ApJ...611..399K,Federrath2010,Gong2012,2014MNRAS.445.4015B,Howard2016}). We implemented a sink particle module in the Athena++ code, primarily following the approach outlined in \cite{Federrath2010} and \cite{Howard2016}. A sink particle represents a molecular gas cloud that continuously accretes gas from nearby grid cells within its control volume, converting the accumulated gas into stellar mass at a rate of 20\% per free-fall time. Once the stellar mass in a sink particle exceeds a certain threshold, the sink particle is converted into a star particle (or star cluster). Various criteria, such as Jeans instability, gas flow convergence, gravitational potential minima, and volume control, are applied to ensure that sink particles form at the centers of dense, collapsing gas clumps. Minor adjustments to the sink particle model were made compared to \cite{Wang2024} due to the lower resolution in this work. Specifically, the mass threshold for converting a sink particle into a star particle has been changed from a resolution dependent value in \cite{Wang2024} to $4\times 10^{4}$ $\rm M_{\odot}$ in this work. 

Our simulations incorporate feedback mechanisms such as radiation pressure, thermal heating, stellar wind, and core-collapse supernovae. However, the implementation of supernova feedback and gas return from star particles in this work differs slightly from that in \cite{Wang2024}, primarily due to the lower resolution. 
In \cite{Wang2024}, the focus is on the recent starburst and the initial launching of outflows, which adopt a smaller simulation box of 4 kpc with a high resolution of 4 pc in the central region. In contrast, this study is primarily concerned with the development of multiphase outflows, so we expanded the simulation volume to $(8\, \text{kpc})^{3}$, using a lower resolution of 8 to 16 pc to save computational resources. Additionally, to further enhance computational efficiency and reduce memory usage, we merge star particles that are separated by a distance less than three times the grid size $\Delta X$, and have an age difference of less than 0.1 Myr.

\subsection{Feedback from Supernovae and Stellar Winds} \label{sec:sn}

Feedback from core-collapse supernovae (CCSN) is widely considered the primary driver of outflows in starburst galaxies. In our simulations, we determine the total mass, $M_{ej}$, the energy, $\Delta \rm{E_{sn}}$, and the metal mass, $M_{metal, ej}$, ejected from CCSN ,  for each star particle (star cluster) at each time step as follows. For each star particle, we generate a set of virtual massive stars with masses ranging from $M \sim 10-80$ $\rm M_{\odot}$, according to the \cite{Kroupa01} initial mass function (IMF). The fraction of total mass contributed by these massive stars is also determined by the \cite{Kroupa01} IMF. Rather than explicitly creating smaller refined particles, we store the mass and expected lifetime of these virtual massive stars in lists. These stars serve as candidates for core-collapse supernovae. At each time step, we update the age of each star particle and search the lists to identify any virtual massive stars that have exceeded their expected lifetimes. For each of these stars, a supernova is triggered. The injected energy, ejected mass, and metals from each supernova are calculated using the \cite{Sukhbold2015} model. For each star particle, we sum all supernova events caused by its virtual massive stars within the last time step to determine its $M_{ej}$, $\Delta \rm{E_{sn}}$ and $M_{metal, ej}$. Virtual massive stars that have already exploded are removed from the list in the following time step.

The method used to couple supernova feedback with the surrounding interstellar medium (ISM) in simulations significantly affects the evolution of galactic winds, especially when the resolution is limited. The expansion of supernova remnants (SNRs) in a uniform medium has been extensively studied using spherically symmetric models, which outline several well-known phases of evolution. However, a simulation’s resolution may not always be sufficient to resolve all these stages. If all the supernova energy is injected into the ambient ISM purely as thermal energy, it can lead to an over-cooling issue in low-resolution simulations, where excessive energy is artificially radiated away instead of driving hot bubbles and outflows. Additionally, the simulation's time step can shrink dramatically due to the sudden energy injection. A resolution of just a few parsecs is typically necessary, if most of the energy from supernovae is injected into the surrounding ISM as thermal energy. In \cite{Wang2024}, we split the supernova energy equally between thermal and kinetic modes, since the central region’s resolution reached 4 parsecs. In this work, given our coarser resolution, we adopt the method proposed by \cite{kim2017three} to more effectively handle supernova feedback. 

The method that deals with supernovae feedback in \cite{kim2017three} assigns thermal energy and momentum to neighboring gas cells based on the local gas properties and the simulation's resolution. In regions where resolution is high and the ambient gas density is low, supernova energy is injected as thermal energy. Otherwise, it is treated as kinetic energy. The supernovae feedback algorithm in \cite{kim2017three} has benefit from the results of \cite{Kim_2015}, which used high-resolution 3D simulations to study the evolution of internal and kinetic energy during the Sedov-Taylor phase of supernova explosions. In our simulations, we adopt the method of \cite{kim2017three} with several modifications. First, we ensure both momentum and energy conservation, as suggested by \cite{Hopkins18}. Additionally, we account for the enhanced feedback effects from clustered supernovae, which have been shown to significantly impact outflows (e.g., \citealt{Gentry17}; \citealt{2018MNRAS.481.3325F}).

In galaxy-scale simulations with a resolution of around 10 parsecs, we cannot resolve the physics at the scale of individual stars. In our simulations, each star particle, with a mass between $\sim 10^4$ to $\sim 10^6$ $\rm{M_{\odot}}$, actually represents a star cluster, where all member stars are assumed to form simultaneously. \cite{Gentry17} conducted hydrodynamic simulations to study the effects of clustered supernova feedback from such star clusters. Their results show that the cumulative momentum per supernova from a cluster containing 10 to 1,000 supernovae can be up to ten times greater than that from a single supernova. This highlights how clustered supernovae significantly amplify the momentum injected into the interstellar medium (ISM). In our simulations, a single star particle is expected to release a few hundred supernovae over approximately 25 Myr. Thus, the total final radial momentum for a single star particle in our simulation is set to $p_{snr} = 2.8\times 10^{5} \rm{M_{\odot}}\,$km s$^{-1}$ ($n_{H}$/cm$^{-3})$$^{-0.17} \cdot f_{boost}$, where $f_{boost}$ is the boost factor accounting for the enhanced feedback effects from clustered supernovae founded in \citealt{Gentry17}. We apply a default boost factor of $f_{boost}=10$, while a lower value of $f_{boost}=3$ is used in a simulation for comparison. More details on our treatment of supernova feedback can be found in Section \ref{sec:sn_feedback}.

Before massive stars explode as supernovae (SN) or collapse into black holes, they can eject mass, energy, and metals into the surrounding ISM via stellar winds. This process is particularly significant for stars with masses exceeding $20\, \rm{M_{\odot}}$. In our simulations, we implement a metallicity-dependent stellar wind model based on \citealt{Jorick2021}. At each time step, we calculate the total mass loss $m_{wind}$ and the average terminal velocity $v_{wind}$ of the stellar winds produced by virtual massive stars within each star particle. Stars that have already undergone supernova explosions are excluded from these calculations. The total metal mass ejected by the stellar wind for a given star particle at each time step is expressed as $m_{wind,metal}=Z_{mean,p} \times m_{wind}$, where $Z_{mean,p}=M_{metal,p}/M_{total,p}$ represents the star particle's average metallicity, with $M_{metal,p}$ and $M_{total,p}$ being the metal mass and total mass of the star particle, respectively. We distribute the mass, momentum, and metal mass ejected by stellar winds to the surrounding grid cells within a radius of twice the grid size $[-2\Delta X, 2\Delta X]$, using mass-weighted allocation. Overall, the mass and metal ejected by stellar wind is about a few percent of that by SN in our simulations. This is because most stars have metallicities below 0.3 $Z_{\odot}$ and mass loss from stellar winds is less efficient for low-metallicity stars (e.g. \citealt{2022ARA&A..60..203V}).

\subsection{Gas Return from Star Particle}
Observations indicate that giant molecular clouds (GMCs) convert only a fraction of their mass into stars over their lifetimes. Estimates of the net star formation efficiency (SFE) in GMCs vary significantly, with observed values ranging from a few percent to as much as ten percent, although with considerable uncertainty (\citealt{2020MNRAS.493.2872C}). In contrast, theoretical and simulation studies predict a broader range of SFEs, from a few percent up to $80\%$, depending on the specific feedback mechanisms and the physical properties of the GMCs, such as surface density  (e.g. \citealt{Dale2014,Kim_2018,li2019disruption,2021MNRAS.506.5512F}). The remaining gas in these clouds is either partially or entirely dispersed by the energetic feedback from newly formed massive stars, i.e. the destruction of GMCs.  

Despite a significant process, many key aspects of the destruction of giant molecular clouds (GMCs) remain poorly understood and constrained. Gas expulsion occurs when the feedback energy from a star cluster exceeds the gravitational binding energy of the gas within the cloud (\citealt{krause2016}). Furthermore, the fraction of unbound gas mass within a molecular cloud shows a linear relationship with the mass of the molecular cloud itself (\citealt{Dale2014}).


Recent high-resolution simulations, down to 0.1 parsec, have provided detailed insights into the evolution of GMCs (e.g. \citealt{Kim_2018,li2019disruption,2021MNRAS.506.5512F}). These studies reveal that star formation within GMCs occurs continuously during the first free-fall timescale $\tau_{\text f \text f}$, with the star formation rate (SFR) peaking around t = $\tau_{\text f \text f}$. As young stars form, their feedback—through stellar winds, radiation, and supernovae—begins to drive the remaining gas out of the cloud. Over time, the fraction of unbound gas increases steadily. In GMCs with shallow potential wells, nearly all of the remaining gas is eventually dispersed. However, in more massive and compact clouds, some ionized gas can remain gravitationally bound due to the deeper potential well (\citealt{2021MNRAS.506.5512F}).

In previous simulation studies using the sink particle method, gas return from sink/star particles to the ISM is often neglected. Some works assume that all the gas within sink particles will convert into stars (e.g. \citealt{2010ApJ...709...27W,2011ApJ...730...40P,Gong2012,kim2017three}), which is reasonable when the resolution is sufficiently high—at sub-parsec scales—where a sink particle represents the dense cores within GMCs. However, at lower resolutions, this assumption can overestimate the star formation efficiency. On the other hand, some studies have suggested that gas retained in star particles remains gravitationally bound to the newly formed cluster within the GMC over long timescales, and therefore did not account for gas return to the ISM (e.g. \citealt{Howard2016}). Given the limited resolution in our work, we cannot fully resolve the physical processes within individual molecular clouds. Additionally, the fraction of gas in GMCs that ultimately becomes unbound due to stellar feedback remains uncertain, with significant variation across different studies (e.g. \citealt{Dale2014,Kim_2018,li2019disruption,2021MNRAS.506.5512F}). To address this, we implemented a simple model for the fraction of gas mass that can escape from sink/star particles, based on their initial gas mass $M_{gas,0}$
\begin{equation}
   f_{return} = \left\{\begin{matrix} (\frac{M_{gas,0}}{10^{4}})^{\beta }
  & M_{gas,0} > 10^{4} \text M_{\odot}\\ 1
  &M_{gas,0} \le 10^{4} \text M_{\odot}
\end{matrix}\right .
\label{eq:freturn}
\end{equation}

\begin{deluxetable*}{cccccccc}[htbp]
\tablenum{1}
\tablecaption{The name and parameters of different simulations. $\rm{M_{gas,t}}$ is the initial total mass of gas, of which the disc contributes $\sim 90\%$. $\rho_{d,0}$ is the characteristic density of the gas disk. $f_{return}$ and $\tau_{return}$ indicate various gas return fraction and time scale (see the text in Section 2.3). $f_{boost}$ is the momentum boost factor in the feedback module. $\rm{Z_{gas,i}}$ is initial metallicity of gas. }
\label{tab:simulations}
\tablewidth{\linewidth}
\tablehead{
\colhead{Name} & \colhead{$\rm{M_{gas,t}}$ ($\rm{M_{\odot}}$)} & \colhead{$\rho_{d,0}$ ($\rm{g/cm^{-3}}$)} & \colhead{resolution (pc)} & \colhead{$f_{return}$} &\colhead{$\tau_{return}$ (Myr)} & \colhead{$f_{boost}$} &\colhead{ $\rm{Z_{gas,i}}$ ($Z_{\odot}$)} } 
\startdata
M2R16NGX& $2.4\times 10^{9} $ & 300&16 & none & none & 10 & 0.02\\
M3R16NGX& $3.2\times 10^{9} $ & 400&16 & none & none & 10 & 0.02\\
M3R16DGS& $3.2\times 10^{9} $ & 400&16 & default & 10 & 10 & 0.02\\
M3R16DGF& $3.2\times 10^{9} $ & 400&16 & default & 5 & 10 & 0.02\\
M3R16HGS& $3.2\times 10^{9} $ & 400&16 & high & 10 & 10 & 0.02 \\
M4R16DGS& $3.8\times 10^{9} $ & 485&16 & default & 10 & 10 & 0.02 \\
M5R16NGX& $4.7\times 10^{9} $ & 600&16 & none & none & 10 & 0.02\\
M5R16DGS& $4.7\times 10^{9} $ & 600&16 & default & 10 & 10 & 0.02\\
M3BOOST3& $3.2\times 10^{9} $ & 300&16 & default & 10 & 3  & 0.02\\
M3R8NGX& $3.2\times 10^{9} $ & 300&8 & default & none &10 & 0.02 \\
M3R8DGS& $3.2\times 10^{9} $ & 300&8 & default & 10 & 10 & 0.02 \\
M3HZ& $3.2\times 10^{9} $ & 300&16& default & 10 & 10 & 0.1 \\
\enddata
\end{deluxetable*}

In most of our simulations, we set the parameter $\beta=-0.798$, corresponding to a gas return fraction of $16.0\%$ for a GMC with a mass of $10^5 \rm{M_{\odot}}$. We refer to this as the default gas (DG) return scenario. For comparison, in one simulation, we use $\beta=-0.496$ , representing a higher gas (HG) return scenario, with a return fraction of $32.0\%$ for a GMC of the same mass. Once a star particle is formed, its gas component is gradually returned to the surrounding grid cells at a specified rate, as described below.
\begin{equation}
    \dot{M}_{gas}=-f_{return}\frac{M_{gas}}{\tau_{return}} .
\label{eq:dM}
\end{equation}
This formula is inspired by simulation studies on the evolution of GMCs (e.g., \citealt{Kim_2018,li2019disruption,2021MNRAS.506.5512F}). Observational evidence suggests that the destruction of GMCs by stellar feedback occurs over a timescale of $5-10$ Myr (\citealt{2020MNRAS.493.2872C}).
Along with the returning of gas mass, metals in a sink/star particles is also returned to nearby ISM at a rate of $Z_{mean,p} \times \Delta t \times \dot{M}_{gas}$ during each time step $\Delta t$. For conservation purposes, we adopt a gas return timescale of $\tau_{return}=10$ Myr (denoted as Slow or `S') in our simulations. Additionally, we use $\tau_{return}=5$ Myr (denoted as Fast or `F') to explore the effect of a shorter return timescale. To provide further comparison, we also run simulations without any gas return (labelled as `NG'). The gas returned from a sink/star particle is injected into neighboring grid cells within a radius of two cells, with its temperature set to 8000 K and its velocity matching that of the host sink/star particle.

\subsection{Simulation Suits} 

We conducted a suite of simulations to explore how various factors, such as the initial total gas mass, gas return model, resolution, boost factor, and initial metallicity, affect the development of M82-like multiphase outflows driven by starburst. The parameters for each simulation are listed in Table \ref{tab:simulations}. Each simulation name encodes key information about the initial conditions.  For example, in the name `M2R16NGX', `M2' indicates the initial total gas mass in units of $10^{9}$ $\rm{M_{\odot}}$, `R16' signifies fora resolution of 16 parsecs in the central region, `NG'(`DG'/`HG') denotes no gas return (default/Higher gas return fraction), `N'(`S'/`F') representing a gas return timescale of $\tau_{return}=\infty(5/10)$ Myr. Additionally, the simulation `M3BOOST3' follows the same setup as `M3R16DGS', but with a smaller boost factor $f_{boost}=3$, while `M3HZ' also mirrors `M3R16DGS', except for a higher initial gas metallicity of $\rm{Z_{gas,i}}=0.1$ $\rm{Z_{\odot}}$. 

The simulations are carried out within a cubic box with a side length of 8 kpc. To optimize computational resources, we implement a Static Mesh Refinement (SMR) grid structure. In the central region, defined by a radial distance from the disc's central axis, $r_{ax}=\sqrt{x^2+y^2}$, of less than 1000 pc and a vertical height $|z|$ below 120 pc, the resolution of each grid cell is 16 pc. Beyond this central region, the resolution decreases sequentially to 32, 64, and 128 pc, allowing for greater computational efficiency in the outer areas.

\section{Overview of Outflow Development}

\begin{figure*}[htbp]
    \centering
    \includegraphics[width=0.45\textwidth]{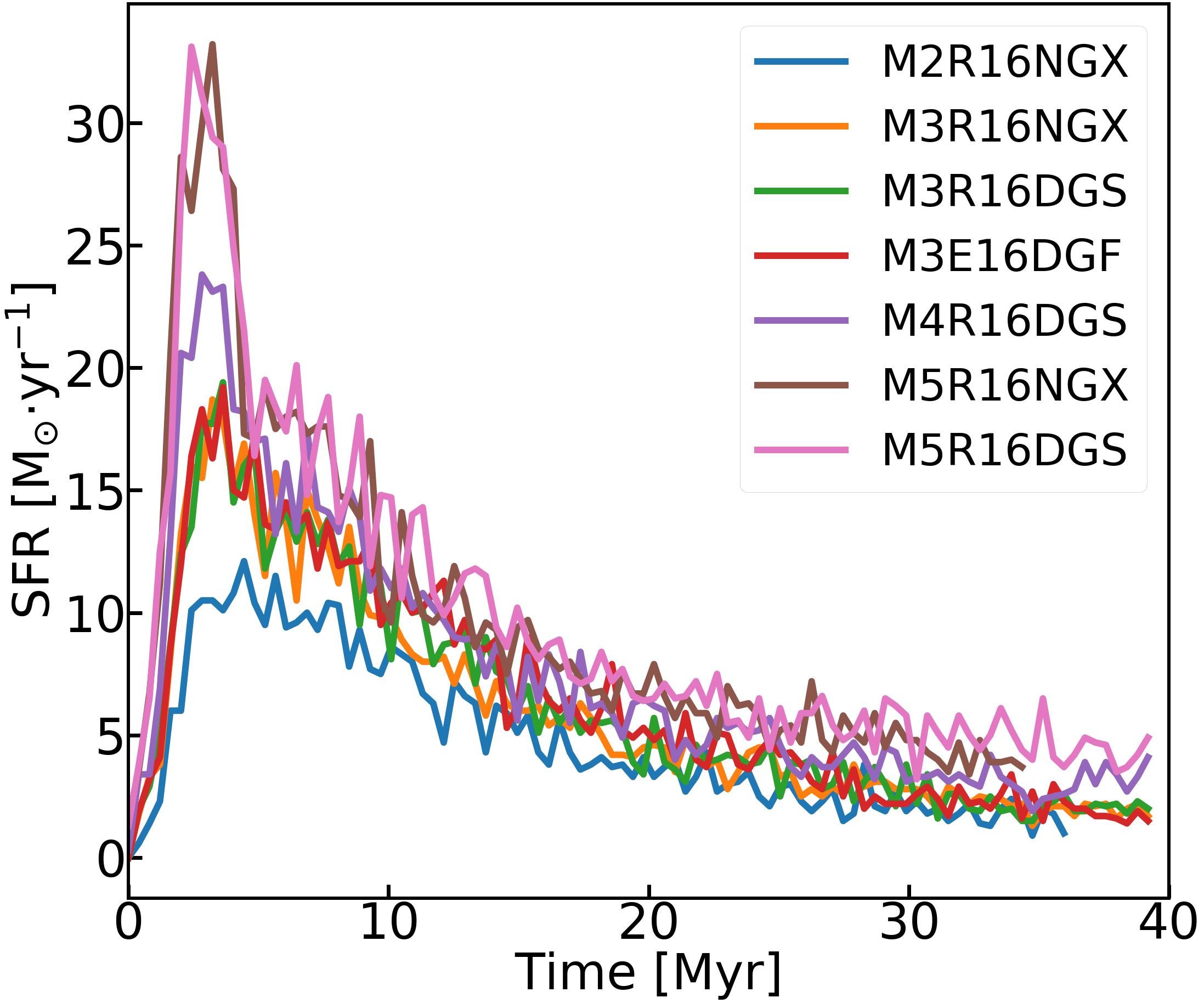}
    \includegraphics[width=0.45\textwidth]{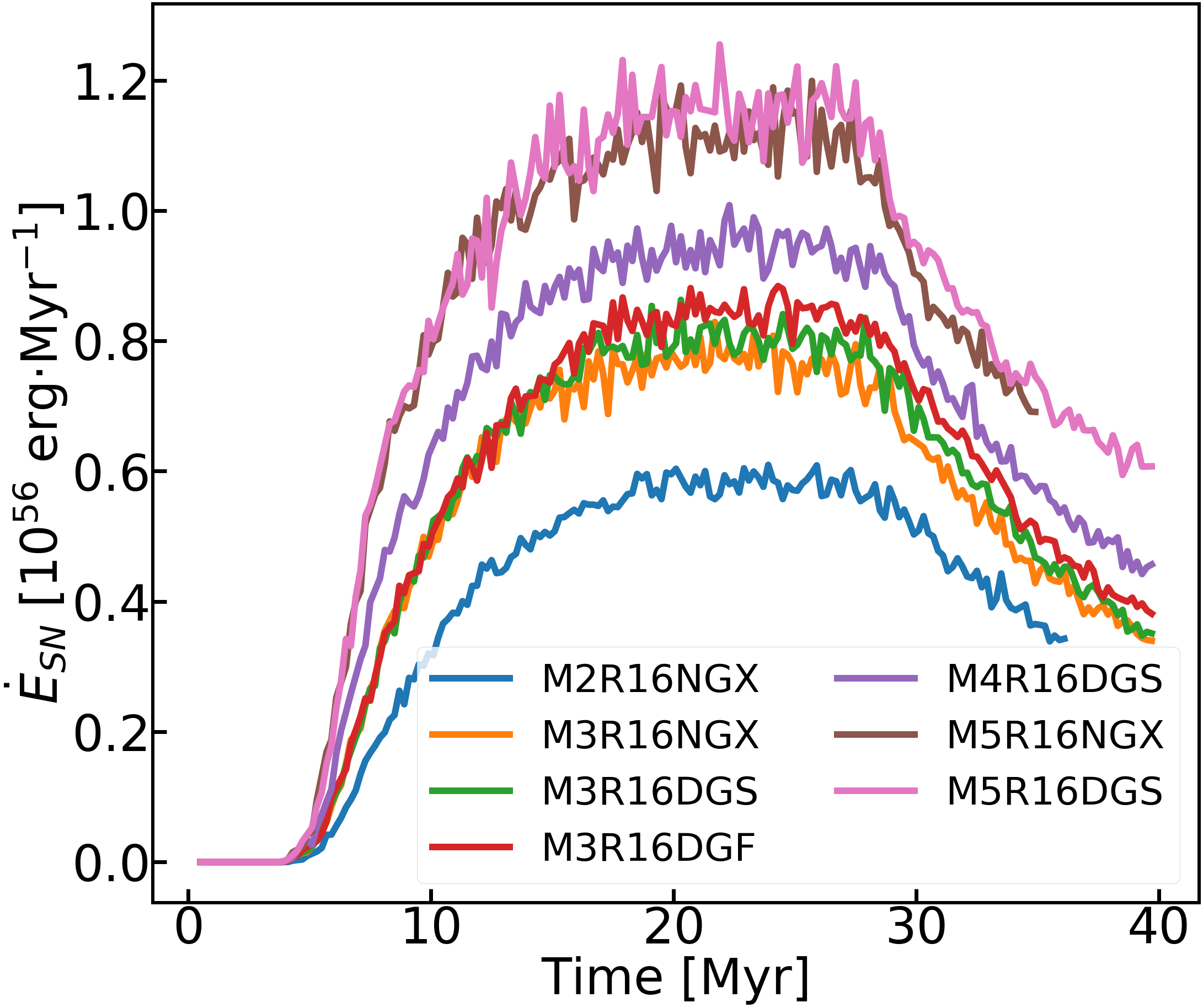}\\
    \includegraphics[width=0.45\textwidth]{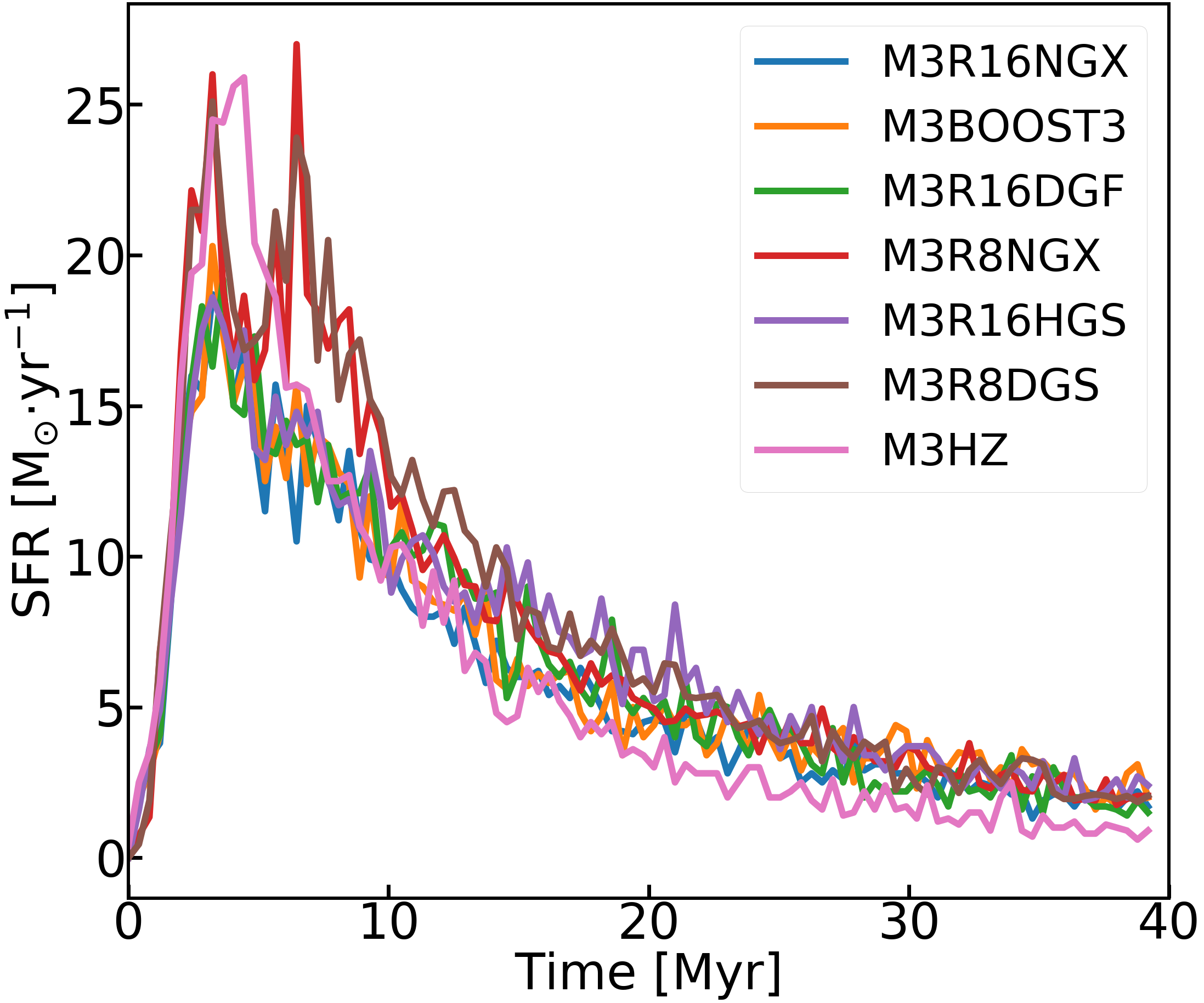}
    \includegraphics[width=0.45\textwidth]{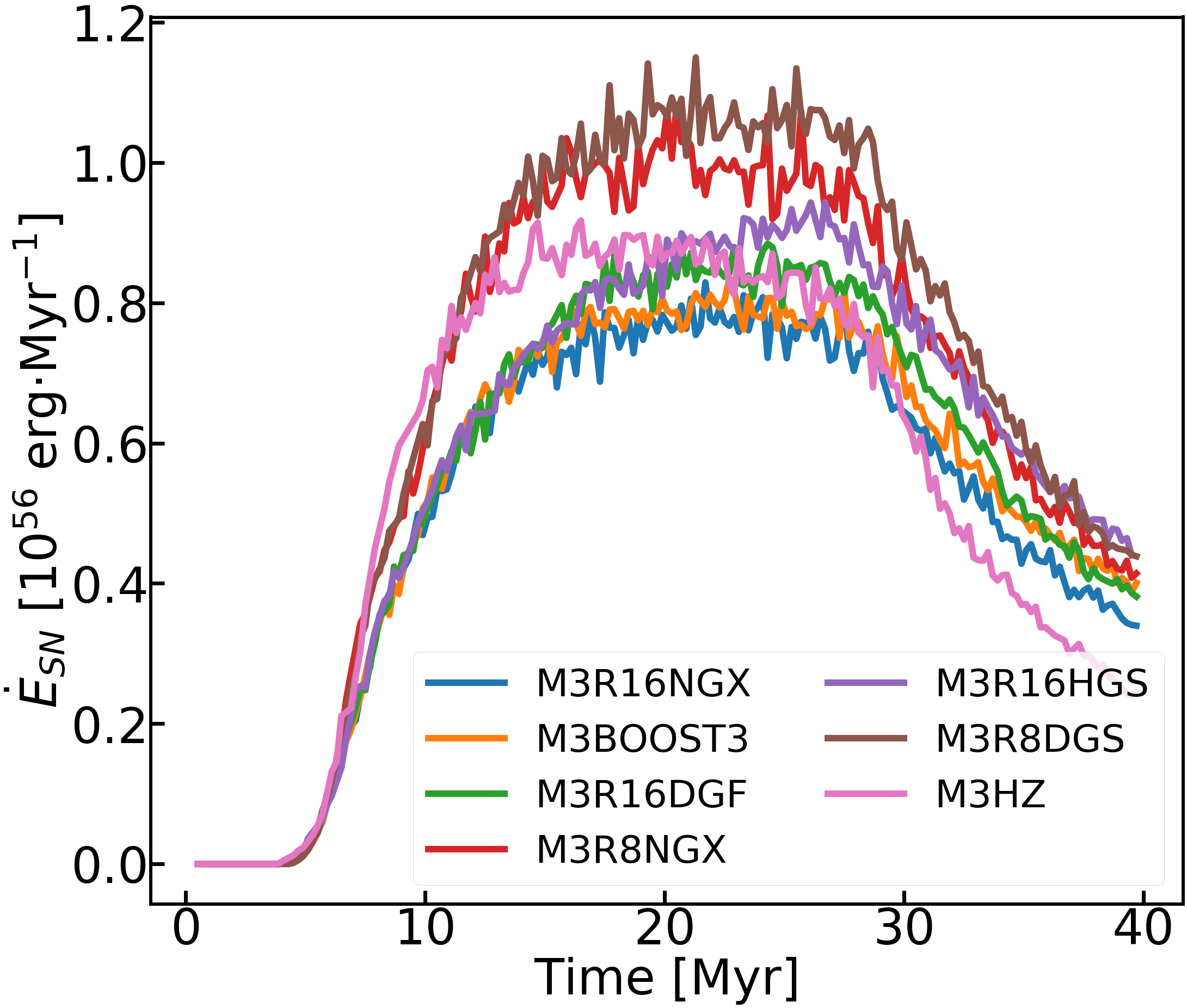}
    \caption{Star formation history (left) and energy injection rate (right) from core-collapse supernovae in different simulations.}
    \label{fig:SFR_SNII}
\end{figure*}

\subsection{Starburst and Supernovae Energy}
To begin, we briefly describe the energy source driving the outflows in our simulations, namely, the starburst and the corresponding SN energy feedback in our simulations. Figure \ref{fig:SFR_SNII} illustrates the star formation rate (SFR) and core-collapse supernova (CCSN) energy injection rate as functions of time across different simulations. The left column of Figure \ref{fig:SFR_SNII} indicates that star formation begins shortly after the start of simulations, rising rapidly to a peak of 10 - 35 $\rm{M_{\odot}\ yr^{-1}}$ at time 3 - 5 Myr, followed by a sharp decline over the next $\sim$10 Myr, eventually tapering off to around 1.5 - 5.0 $\rm{M_{\odot}\ yr^{-1}}$ at $t\sim30$ Myr. This decrease in SFR after the peak is primarily due to the declining gas accretion rate onto sink particles after $t\sim 5$ Myr. The total stellar mass formed by $t=30$ Myr in different simulations ranges from $1.62 \times 10^{8}$ to $3.34 \times 10^{8}$ $\rm M_{\odot}$, which observational estimates for the recent starburst in the nucleus of M82, ranging from 2.0 - 3.5 $\times 10^8$ $\rm{M_\odot}$ (e.g. \citealt{1993ApJ...412...99R,2003ApJ...599..193F,2009ApJ...697.2030S}). Additionally, the late-stage SFR in our simulations aligns with the estimated range of $\sim$2.0 - 10.0 $ \rm{M_\odot yr^{-1}}$ reported in the literature (e.g. \citealt{1998ApJ...498..541K,2006MNRAS.369.1221B,2010MNRAS.408..607F,Adam2015,2023MNRAS.518.4084Y}), where the significant uncertainty stems in part from variations in the assumed IMF. The energy injection rate from CCSN, $\rm{\dot{E}_{SN}}$, is shown in the right column of Figure \ref{fig:SFR_SNII}. Across different simulations, the total energy released by CCSN by $t=30$ Myr ranges from $1.14 \times 10^{57} \rm$ to $2.4 \times 10^{57}$ erg, which is comparable to values used in previous studies (e.g., \citealt{Strickland2000,2009ApJ...697.2030S,2008ApJ...674..157C,melioli2013evolution}). This energy is approximately $20\%-40\%$ of the value used in \cite{schneider2020physical} and \cite{schneider2024cgols}.

However, our simulations show a moderate discrepancy compared to some studies in the literature regarding the starburst history of M82. In our work, only one burst occurs over the past $\sim25-30$ Myr, while other studies suggest a preference for two bursts (\citealt{1993ApJ...412...99R,2003ApJ...599..193F}). Additionally, several studies propose that M82’s nuclear starburst has been active for approximately 10 Myr(e.g., \citealt{2003ApJ...599..193F,mayya2009m82}), although some features point to a longer duration (\citealt{2009ApJ...697.2030S}). More reliable observations are needed to resolve the disagreement. Moreover, the characteristics of the starburst in our simulations are influenced by the initial parameters. We find that the initial total gas mass, $\rm{M_{gas,t}}$, can strongly affect the SFR at fixed resolution. A more massive gas disc leads to a more dense gas distribution in the central region, triggering an earlier and more intense burst. In simulations with higher $\rm{M_{gas,t}}$,the corresponding supernova energy injection rates increase proportionally with the SFR. In contrast, the gas return process has only a minor effect on the SFR, though it significantly impacts outflow rates and X-ray emissions, as we will demonstrate in Section 4. 

\begin{figure}[htbp]
    \centering
    \includegraphics[width=0.85\columnwidth,trim=20 10 5 5, clip]{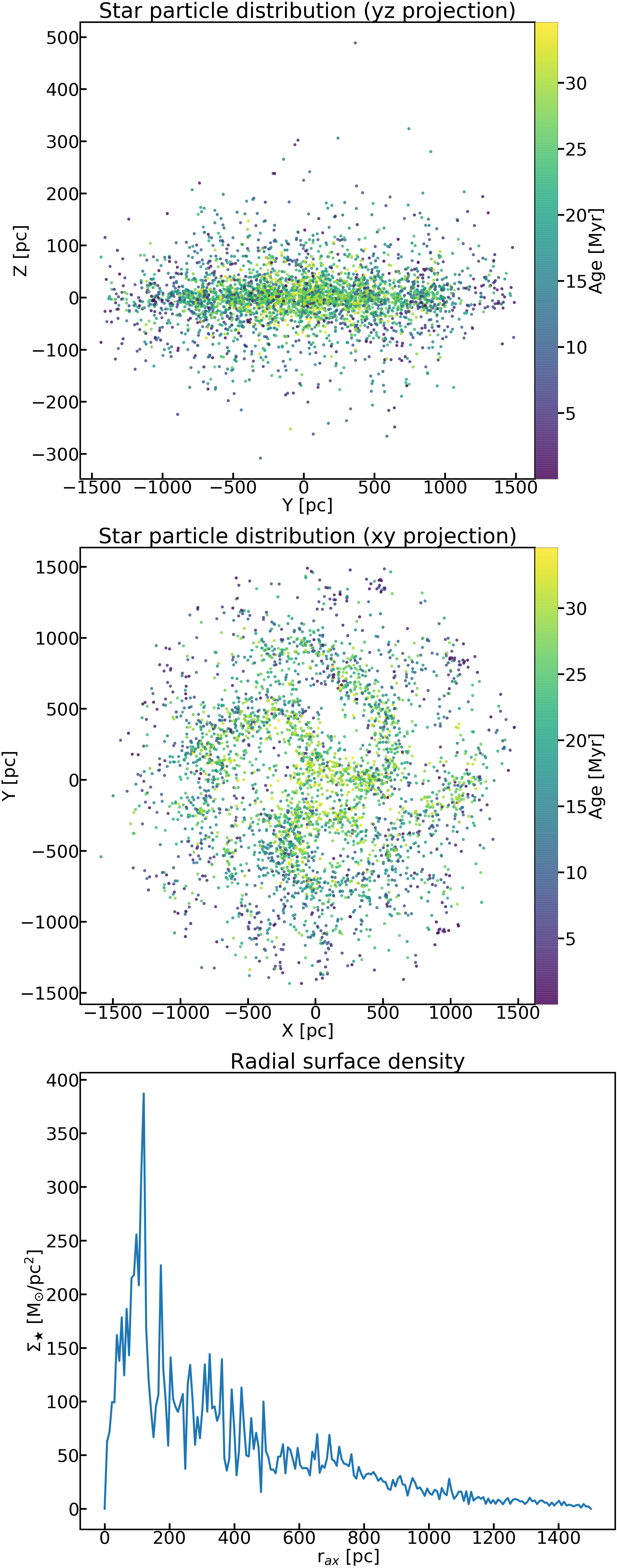}
    \caption{Radial (top) and axial (middle) distribution of star particles at 35 Myr in simulation M3R16NGX. The bottom panel shows the radial surface density of stars born in the burst.}
    \label{fig:particle}
\end{figure}

The bottom panel of Figure \ref{fig:SFR_SNII} shows the simulations with the same $\rm{M_{gas,t}}=3.2\times10^9\, \rm{M_{\odot}}$. The SFR changes shows only mild variations when either a higher gas return fraction or a shorter gas return timescale is implemented. Meanwhile, the SFR in the early stage of simulation will increases noticeably when a higher initial gas metallicity of $\rm{Z_{gas,i}}=0.1$ $\text{Z}_{\odot}$ is adopted,as the elevated metallicity enhances the cooling rate. Additionally, increasing the resolution moderately boosts the SFR, as our star formation prescription depends on the gas density, which improves with higher resolution. Nonetheless, a resolution of 8 pc is still insufficient to achieve full convergence.

\begin{figure}[htbp]
    \centering
    \includegraphics[width=0.85\columnwidth]{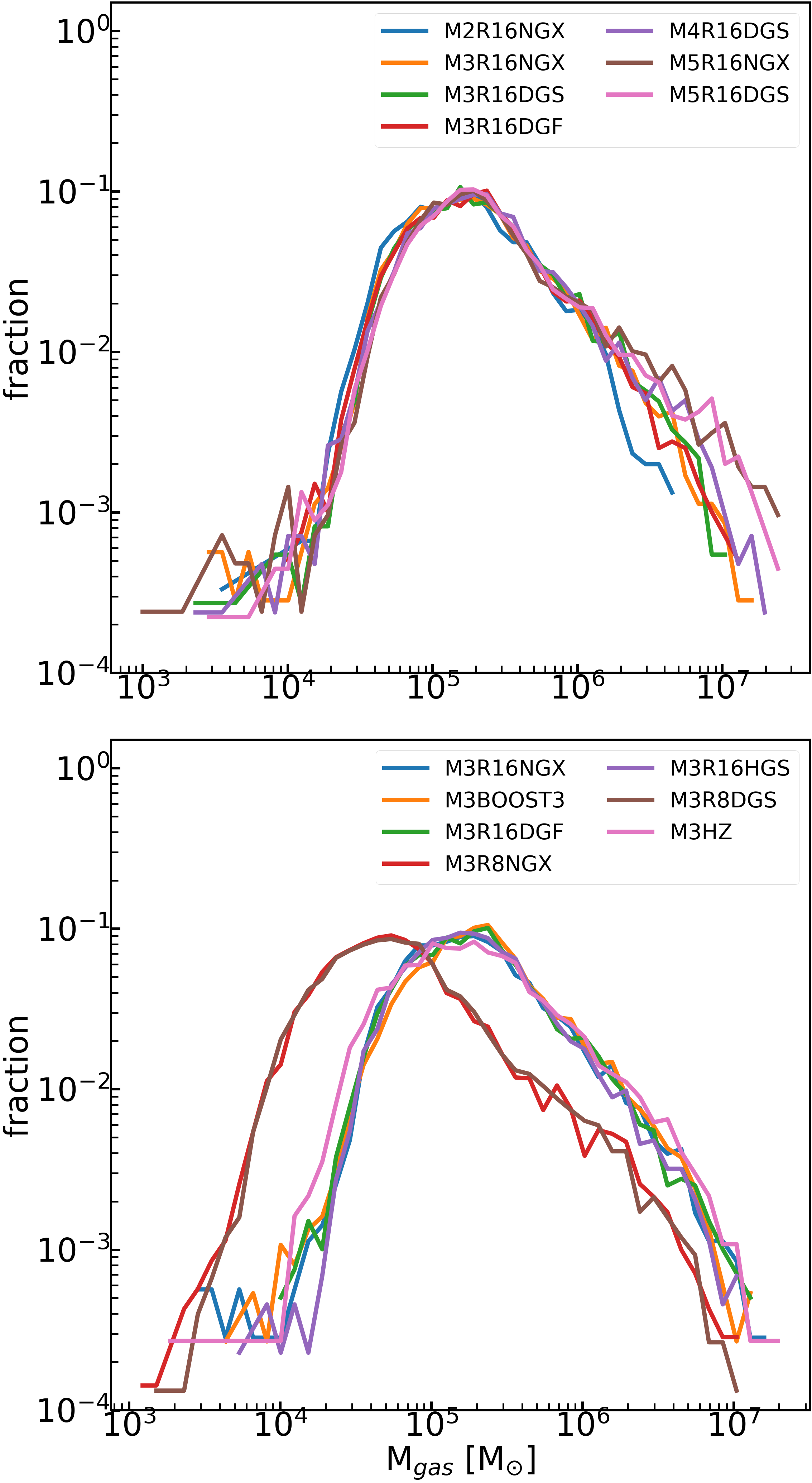}
    \caption{The distribution of initial gas mass of star particles at 30 Myr in different simulations.}
    \label{fig:massF}
\end{figure}

In addition to the total stellar mass formed and the corresponding supernova energy, the spatial distribution of newly formed stars can also influence the development of outflows (e.g. \citealt{schneider2024cgols}). Figure \ref{fig:particle} presents the distribution of star particles and their radial surface density at 35 Myr in the 
 M3R16NGX simulation, which is similar to our other simulations. The majority of star particles are concentrated in the inner region, within $r_{ax}<1000$ pc, while a smaller number are located in the outer region, $1000<r_{ax}<1500$ pc. This distribution aligns broadly with observations, though the starburst in our simulations exhibits a somewhat wider spatial extent. Observations suggest that most of the star clusters younger than 30 Myr are found in the nuclear region ($r_{ax}<625$ pc, with some younger clusters distributed on the disk (\citealt{lim2013star};\citealt{rodriguez2010spatially}). 

\begin{figure*}[!htbp]
    \centering
    \includegraphics[width=0.75\textwidth]{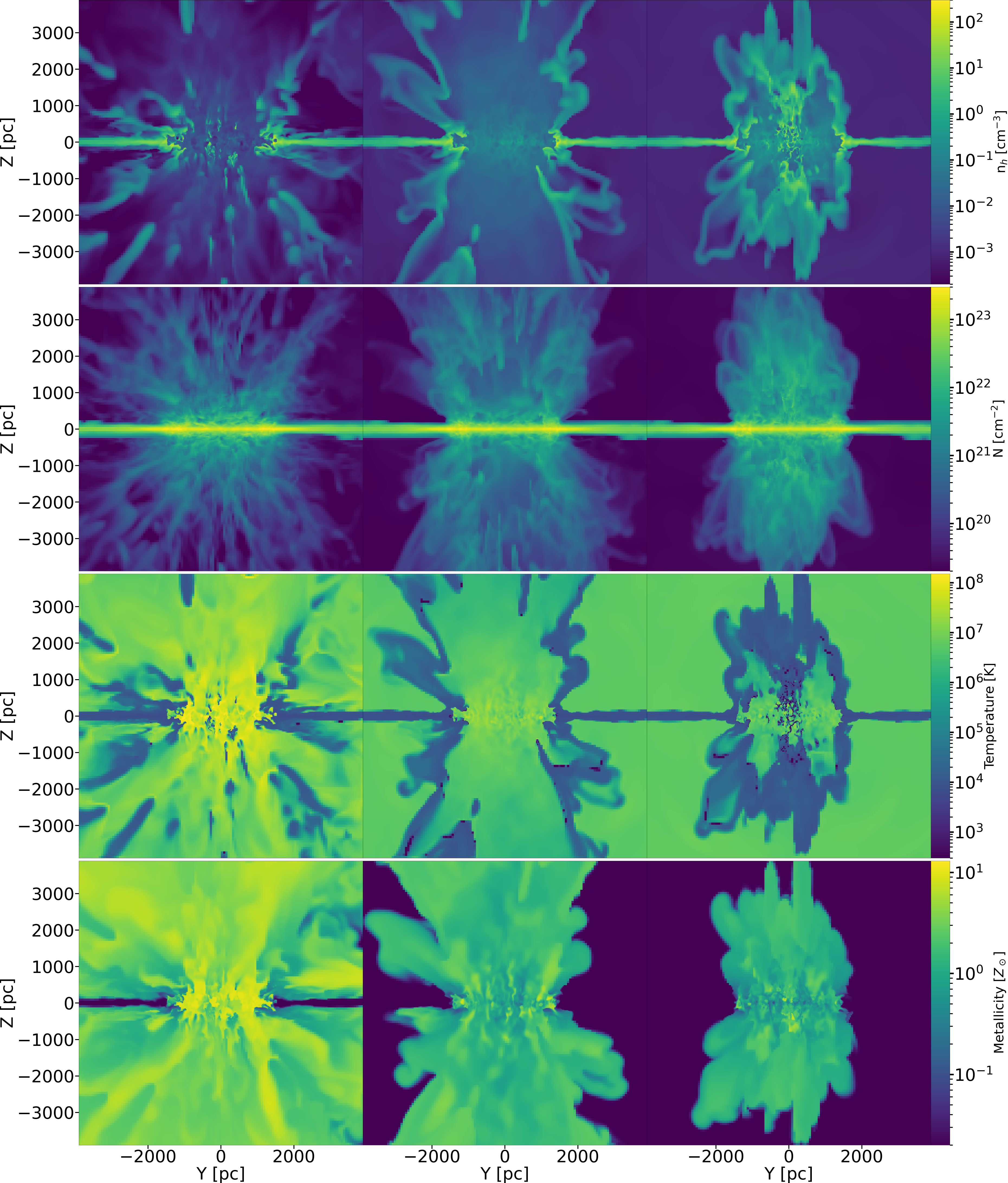}
    \caption{Edge on view of the number density (top), temperature (third row) and metallicity (bottom) in a slice of depth 16 pc, and the column density (second row) at 30 Myr in simulations M3R16NGX (left), M3R16DGS(middle) and M3R16HGS (right). }
    \label{fig:outflowimageall}
\end{figure*}

As shown in the bottom panel of Figure \ref{fig:particle}, the surface density of star particles peaks at $r_{ax}\sim 200$ pc, then declines exponentially outward, which is similar to the distribution of super star clusters in M82 (\citealt{mayya2009m82}). In our simulations, some sink particles can form at heights of $z\sim100$ pc, where they are likely influenced by strong supernova feedback. In certain regions, the gas outflow velocity can reach several hundreds km/s, propelling sink particles outward at high speeds. Some of these particles may later accrete surrounding gas and transform into star particles. Interestingly, observational evidence also supports ongoing star formation in the halo of M82 (\citealt{mayya2009m82}), and young stars have been detected in its halo region (\citealt{lim2013star}). Therefore, the spatial distribution of recently formed stars in our simulations aligns with these observations.

We further examine the distribution of the initial gas mass of the star particles in simulations at 30 Myr, as shown in Figure \ref{fig:massF}. The upper panel suggests that a larger $\rm{M_{gas,t}}$ results in a broader mass range, since higher gas densities can support the formation of more massive star particles. In contrast, the bottom panel shows that increasing the resolution causes the initial gas mass of star particles to shift towards the lower mass end. In simulations with a resolution of 16 (8) pc, the peak mass falls between $8 \times 10^{4} - 5 \times 10^{5}$ $\rm M_{\odot}$ ($2 \times 10^{4} - 10^{5}$ $\rm M_{\odot}$ ).  This shift occurs because higher resolution allows for the formation of smaller gas particles. As a result, the gas return fraction increases with improved resolution, given that the $\beta$ parameter in Eqn. \ref{eq:freturn} is the same. However, similar to the effect on SFR, variations in the gas return scenario and time scale have little impact on the mass distribution of star particles.

\begin{figure*}[!htbp]
    \centering
    \includegraphics[width=0.75\textwidth]{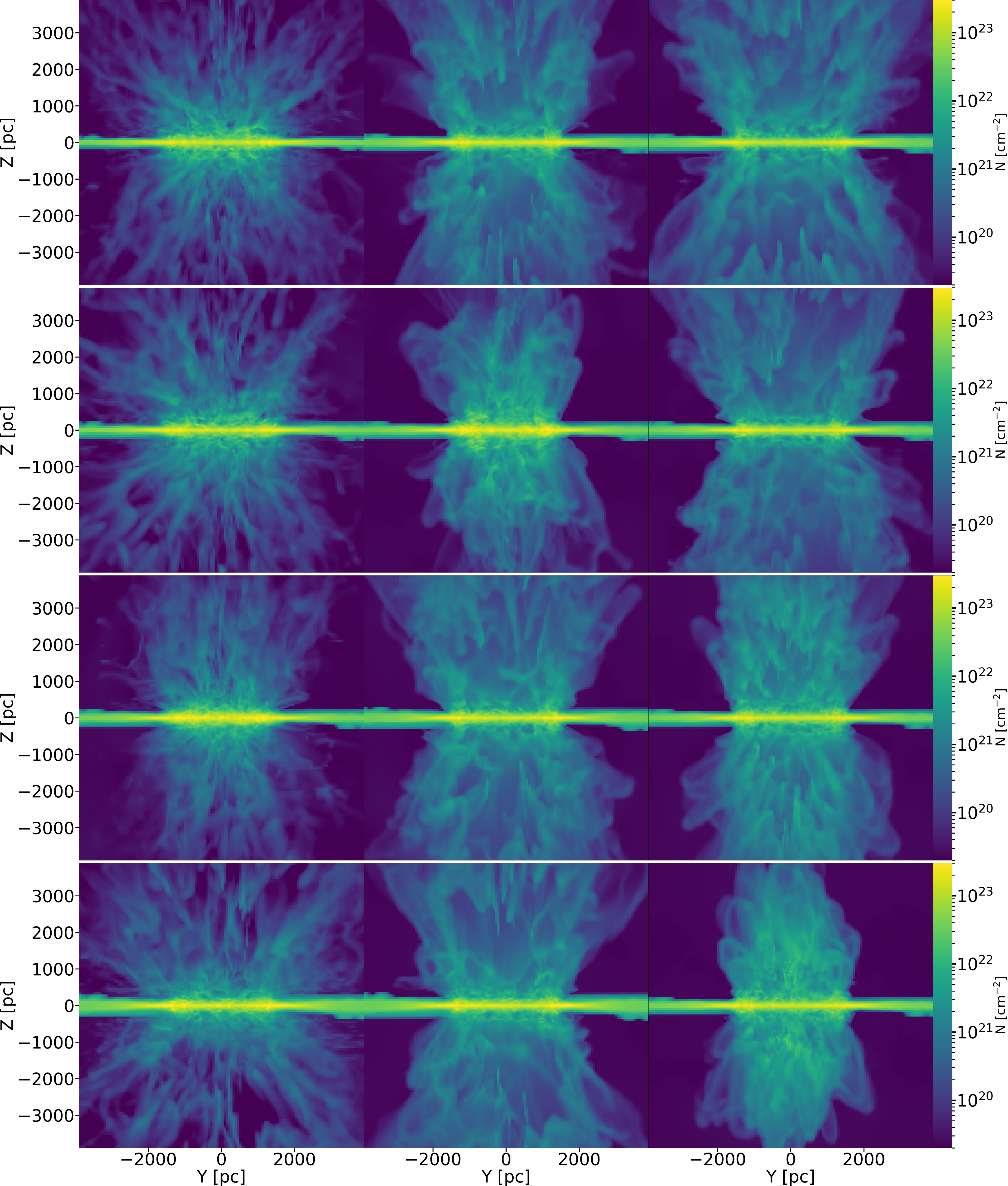}
    \caption{Edge on view of the column density for our simulations at 30 Myr. From top to bottom, the left column are M2R16NGX, M3R16NGX, M3R8NGX, M5R16NGX; the middle column are M3R16DGS, M3R8DGS, M4R16DGS, M5R16DGS; the right column are M3HZ, M3BOOST3, M3R16DGF, M3R16HGS, respectively.}
    \label{fig:outflowimageall2}
\end{figure*}

\subsection{Overview of Outflows}
In all of our simulations, the starburst successfully generates large-scale galactic outflows. However, differences in star formation histories (SFHs) and total feedback energy have led to significant variations in the outflow properties. Figure \ref{fig:outflowimageall} provides an edge-on view of the outflows at t = 30 Myr in three simulations, M3R16NGX, M3R16DGS and M3R16HGS. The first, third, and fourth rows display the number density, temperature, and metallicity of gas in the central slice with a depth of 16 pc. The second row shows the column density of gas, highlighting the substantial amount of gas within the biconical wind extending above the galactic disk. The outflows exhibit a highly inhomogeneous and filamentary structure. The temperature distribution reveals that the outflows are multiphase, with temperatures ranging from 300 to around $10^{8}$ Kelvin. The metallicity of the gas also varies significantly, from around 0.1 to several times $\rm{Z_{\odot}}$. 

Figure \ref{fig:outflowimageall} further illustrates how gas return from star particles affects the characteristics of outflows. When gas return is neglected, the outflow exhibits a wide opening angle, and much of the outflow volume is filled by hot gas at temperatures exceeding $T>10^7$ K. The cool and cold gas tend to form clouds and short filaments, embedded within this hot medium. The metallicity of the hot gas can reach up to $5-10$ $\text{Z}_{\odot}$. In simulation M3R16DGS, where gas return is incorporated at the default rate, the outflow's opening angle and occupied volume are notably reduced. The interior of the outflow is dominated by gas with temperatures ranging between $10^{5.5}$ and $10^{7.0}$ K, and is covered by a cool and cold outer layer. encased in a cooler and colder outer layer. Although gas return has a minimal effect on SFH, it slows the expansion of the outflow, particularly along the major axis. The right column of Figure \ref{fig:outflowimageall} demonstrates that a higher gas return fraction ($\beta=-0.496$) can further narrows the opening angle and reduce the volume of the outflow at the same epoch. 

Figure \ref{fig:outflowimageall2} presents the column density at t=30 Myr for all simulations. The left column shows that
in simulations with a 16 pc resolution and without gas return, the wind has occupied nearly the entire volume. As $M_{gas,t}$ increases, the outflow expands to fill a larger volume due to the increased feedback energy (see Figure \ref{fig:SFR_SNII}). In the simulations with a higher resolution, the outflow's opening angle is significantly reduced. As discussed in the next subsection, this reduction is due to less gas being accreted onto star particles, leaving more gas within the disc. The middle column of Figure \ref{fig:outflowimageall2} indicates that simulations with default or slow gas return have a smaller outflow opening angle compared to those without gas return. Additionally, comparing the upper middle and upper right panels of Figure \ref{fig:outflowimageall2}, we observe that a higher initial gas metallicity leads to a wider outflow opening angle. This is because higher metallicity results in a greater total stellar mass formed during the burst and consequently higher supernova feedback energy (see Figure \ref{fig:SFR_SNII}). 

\begin{figure*}[htbp]
    \centering
    \includegraphics[width=0.75\textwidth]{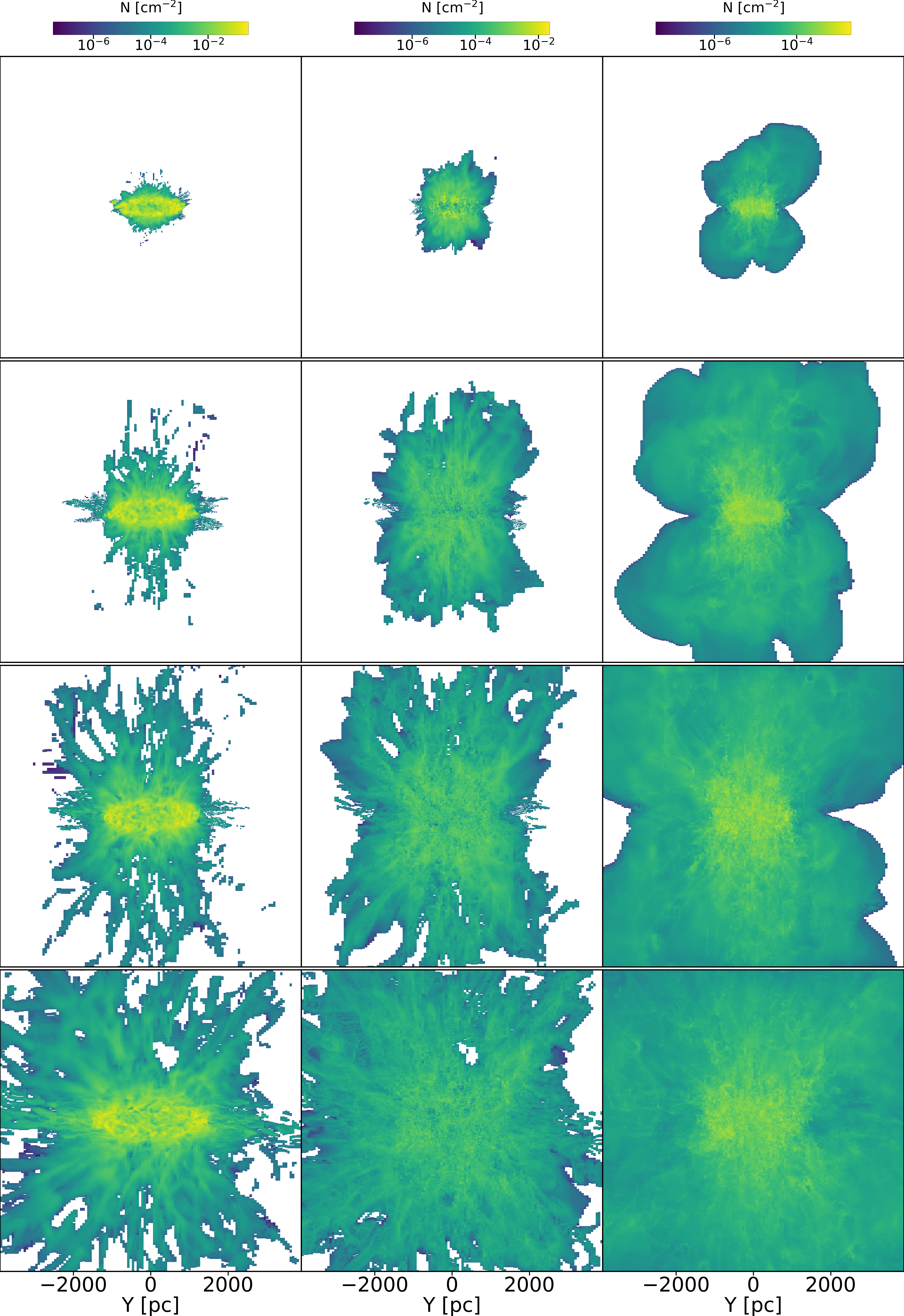} 
    \caption{Projected gas density in M3R16NGX at 10 (top), 15 (second row), 20 (third row) and 30 (bottom) Myr. The left, middle and right columns are the cool ($\rm{T} < 2\times 10^{4}$ K), warm ($2\times 10^{4}$ K $< \rm{T} < 5\times 10^{5}$ K) and hot ($\rm{T} > 5\times 10^{5}$ K) phases, respectively.} 
    \label{fig:outflowimage_bianhua}   
\end{figure*}

\begin{figure*}[htbp]
    \centering
    \includegraphics[width=0.75\textwidth]{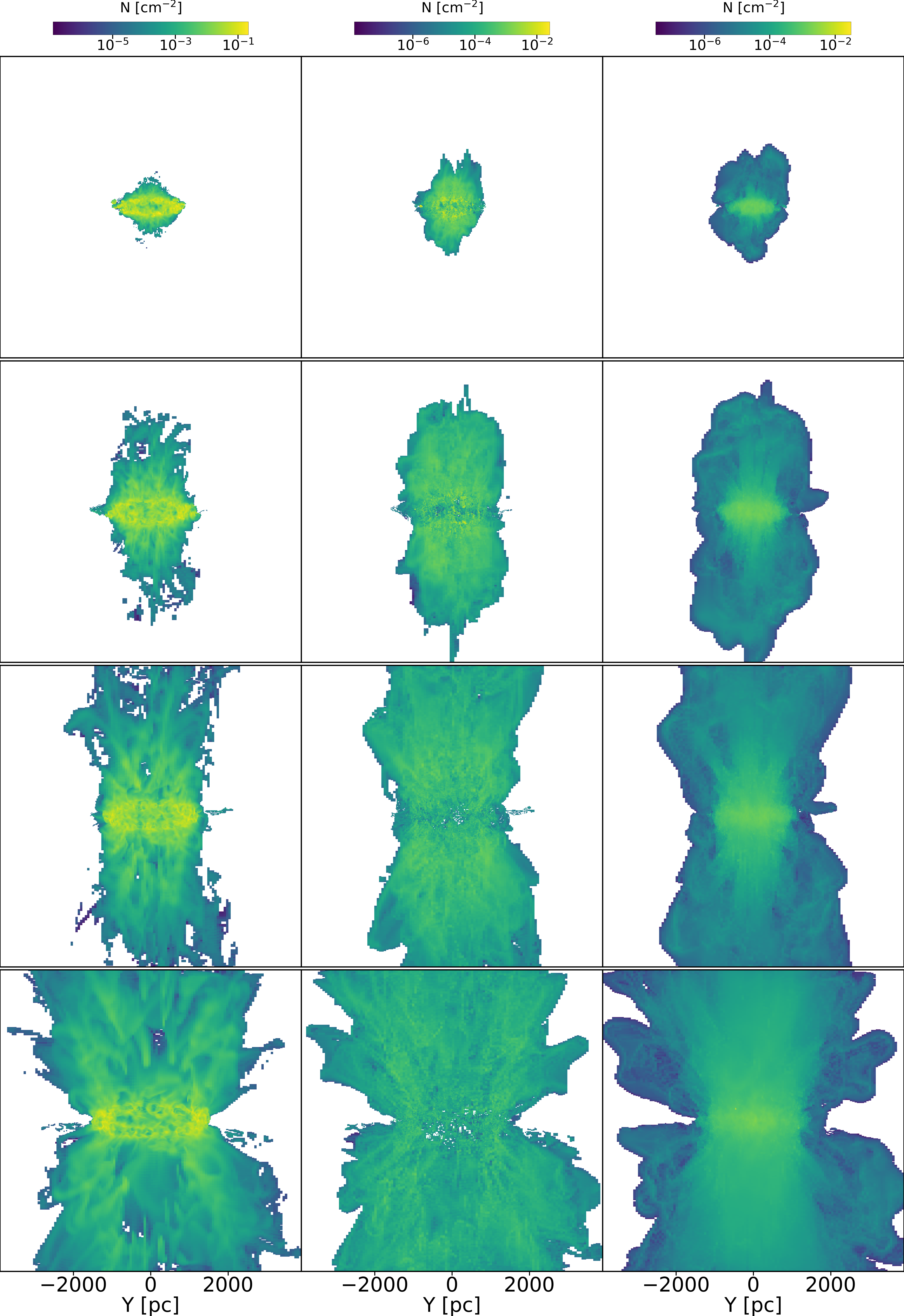} 
    \caption{The same as Figure \ref{fig:outflowimage_bianhua}, but in simulation M3R16DGS. } 
    \label{fig:outflowimage_bianhua2}   
\end{figure*}

\subsection{Development of the Multi-phase Outflows}
To illustrate the evolution of the outflow, Figure \ref{fig:outflowimage_bianhua} shows the development of gas in three phases, that is, cool ($\rm{T} < 2\times 10^{4}$ $\rm{K}$), warm ($2\times 10^{4}$ $ \rm{K} < \rm{T} < 5\times 10^{5}$ $\rm{K}$), hot ($\rm{T} > 5\times 10^{5}$ $\rm{K}$), in the simulation M3R16NGX at four different epochs (10, 15, 20 and 30 Myr). The figure depicts the projected column density of enriched gas (with metallicity higher than the initial value) viewed at an inclination of 77 degrees. The outflow in all three phases originates from the nuclear region, with the hot gas expanding outward at the fastest rate. By t = 10 Myr, a bubble of enriched hot gas has already extended beyond 2 kpc in height. At the same time, the warm gas occupies a slightly smaller volume, reaching a height of about 1.5 kpc, while only a small fraction of the volume above $|z|=500$ pc is filled with cool gas. Between t = 10 and t = 20 Myr, the supernova energy injection rate gradually increases (see Figure \ref{fig:SFR_SNII}), causing the hot bubble to expand rapidly and reach the upper boundary of the simulation box by t = 15 Myr. The warm phase gas moves outward at a slightly slower pace, with some escaping the simulation box before t = 20 Myr. In contrast, the cool gas outflow progresses at the slowest rate. After t = 20 Myr, continued supernova feedback and radiative cooling maintain the steady development of multiphase outflows. Most of the gas in the outflow, including the cool phase, originates from the nuclear region of the gas disc, as evidenced by the growing low-density void there. This indicates that the majority of cool gas in the outflow comes directly from the disc, rather than cooling from the warm and hot phases. 

For comparison, Figure \ref{fig:outflowimage_bianhua2} illustrates the evolution of outflow in the M3R16DGS simulation. Unlike the case without gas return, the column density of all three phases in the outflow is noticeably enhanced. However, the opening angle, the outflow velocity of the hot phase, and the volume occupied by the outflow are reduced. This is expected, as the default gas return scenario only slightly boosts the star formation rate and supernova feedback, while a significantly larger amount of gas is entrained into the wind. 

\begin{figure*}[htbp]
    \centering
    \includegraphics[width=0.75\textwidth]{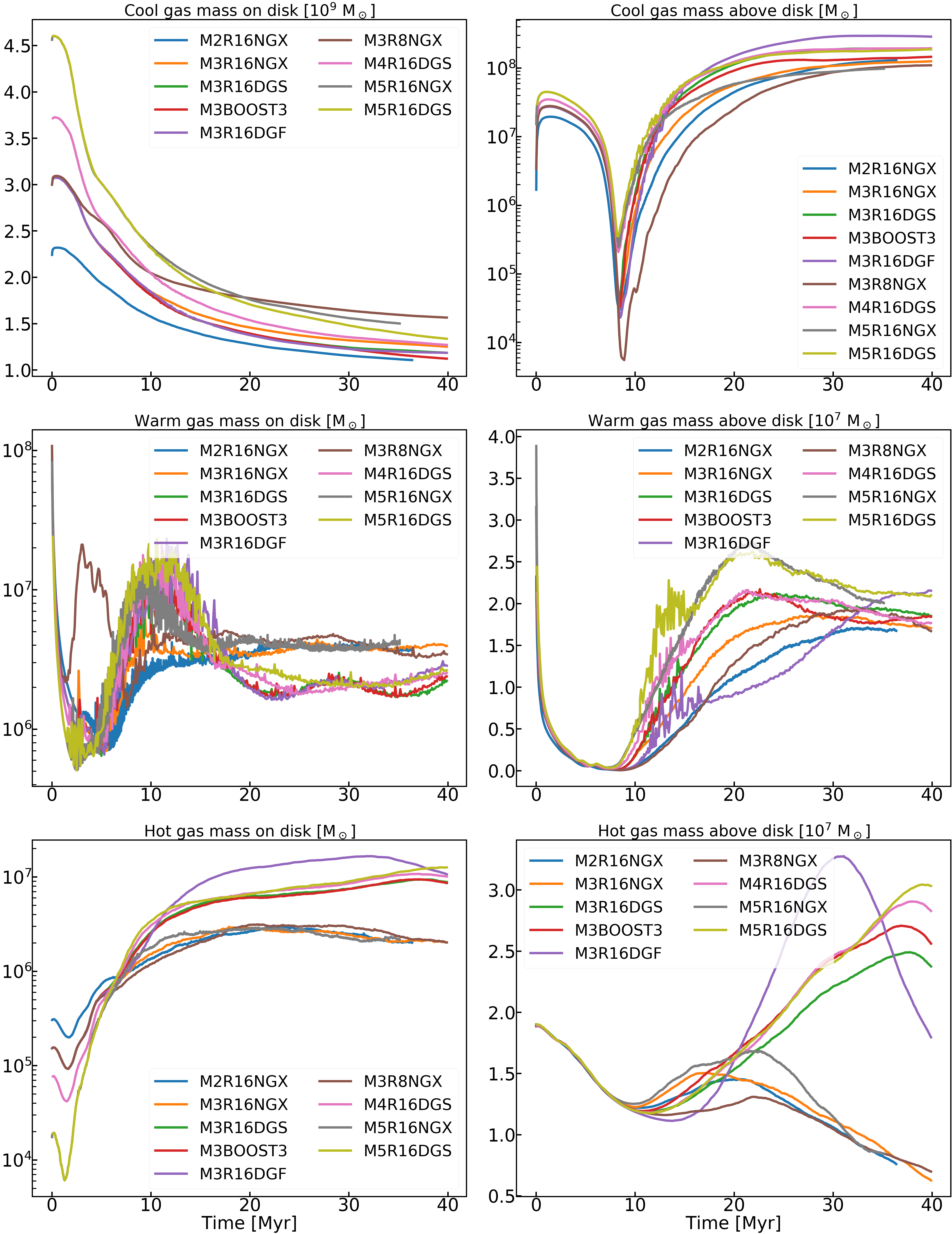}
    \caption{Total mass of cool (top), warm (middle), and hot (bottom) gas in the disk (left) and above the disk (right) as functions of time.}
    \label{fig:gaswithtime}
\end{figure*}

The evolution of multiphase outflows can be quantitatively assessed by tracking the gas mass in different phases within and above the disc, as depicted in Figure \ref{fig:gaswithtime}. To maintain clarity, we present results from representative simulations, while others are omitted for brevity. For convenience, a height of $z=400$ pc is used to delineate the division between the disc and the gas halo. In the upper left panel of Figure \ref{fig:gaswithtime}, we observe a rapid decline in cool gas on the disc before $\text{t}=10$ Myr, followed by a gradual decrease until the simulation's end. Most of this cool gas has been accreted into sink particles. Table \ref{tab:Gasreturnmass} lists the mass of gas retained in sink/star particles at t = 30 Myr,amounting to approximately $30\%-60\%$ of the initial gas disc mass. This fraction appears somewhat higher than the unbound gas fraction typically observed after star formation in individual GMCs in previous studies. However, we argue that this result may not be entirely unexpected, given that M82 still harbors approximately $1.3-2.4$ $\times10^9$ $\rm{M_{\odot}}$ of molecular gas (\citealt{2002ApJ...580L..21W,2013PASJ...65...66S,Adam2015,2021ApJ...915L...3K}). Between $5\%$ to $10\%$ of the initial gas disc mass has converted into stars, with an additional few percent expelled from the disc due to supernova feedback.

\begin{table}[htbp]
        \tablenum{2}
	\centering
	\caption{The total gas mass stored in sink/star particles at 30 Myr in different simulations.}
	\begin{tabular}{lccr}
		\hline
		Name & Gas mass stored in particles ($10^{9} \rm M_{\odot}$)  \\
		\hline
		M2R16NGX& $0.962 $  \\
        M3R16NGX& $1.544 $   \\
        M3R16DGS& $1.492 $   \\
        M3R16DGF& $1.406 $   \\
        M3R16HGS & $1.355 $  \\
        M4R16DGS& $2.027  $   \\
        M5R16NGX& $2.879  $   \\
        M5R16DGS& $2.806  $   \\
        M3R8NGX& $1.219 $   \\
        M3R8DGS & $1.077 $  \\
        M3BOOST3& $1.545 $ \\
        M3HZ & $1.611 $  \\      
		\hline
	\end{tabular}
	\label{tab:Gasreturnmass} 
\end{table}

We observe that a higher initial total gas mass, $\rm{M_{gas,t}}$, leads to a more rapid decline in the cool gas within the disc. This is because a larger $\rm{M_{gas,t}}$ indicates greater gas density in the central region, accelerating the collapse of cool gas into molecular clouds (represented as sink particles in the simulation) and subsequently speeding up star formation. The inclusion of gas return slightly reduces the amount of cool gas remaining on the disc and in sink particles, primarily due to a modest increase in SFR and enhanced stellar feedback. While gas return somewhat slows the expansion of the outflow (see Figures \ref{fig:outflowimage_bianhua} and \ref{fig:outflowimage_bianhua2}), the returned gas eventually contributes to the outflow after approximately 10 Myr. Figure \ref{fig:gaswithtime} shows that simulations with gas return result in a higher mass of hot and cool gas above the disc. 

The top right panel of Figure \ref{fig:gaswithtime} displays the evolution of cool gas above $z=400$ pc. At $t\sim0$, increase in cool gas above the disc, triggered by the rapid radiative cooling of the warm gas (as seen in the middle panels). Following this, the amount of cool gas above $z=400$ pc decreases steadily until $t\sim 7-8$ Myr, likely due to gravitational collapse pulling the gas toward the disc. Between $t\sim 8$ and $t\sim 15$ Myr, the amount of cool gas above the disc rises sharply, driven primarily by outflows from the disc. After $t\sim 15$ Myr, the accumulation of cool gas slows down, even though the supernova (SN) energy injection rate remains near its peak. This deceleration is likely due to the depletion of cool gas in the nuclear region as time progresses beyond $t\gtrsim 15$ Myr. 

Significant differences in the mass of cool gas above the disc emerge across various simulations after 10 Myr, with gas return being the primary factor. In the absence of gas return, the mass of cool gas above the disc at $t=30$ Myr remains around $\sim1.0\times 10^8$ $\rm{M_\odot}$, and shows little variation between simulations with different $\rm{M_{gas,t}}$. However, when gas return is included, this mass increases to approximately $1.8\times 10^8$ $\rm{M_\odot}$ for the default case of $\tau_{return}=10 $ Myr, and further rises to around $\sim 3.0\times 10^8$ $\rm{M_\odot}$ for $\tau_{return}=5 $ Myr. have a moderate effect on the growth rate of cool gas above the disc during the period between 10 and 20 Myr. 

Figure \ref{fig:gaswithtime} shows that the amount of warm gas on and above the disc increases rapidly starting at $t\sim 3$ Myr and $t \sim 8$ Myr, respectively. For hot gas, the corresponding times are $t\sim 2$Myr and $t \sim 8$ Myr. The development of warm outflows is slightly slower than that of the hot phase, which aligns with the visual impression seen in Figures \ref{fig:outflowimage_bianhua} and \ref{fig:outflowimage_bianhua2}. However, the growth in the mass of warm and hot gas above the disc lags behind what is visually depicted in those figures. This delay occurs because a substantial region above the disc remains unaffected by the outflow until around $t=5$ Myr, and much of the gas in this area undergoes radiative cooling. The gas return process increases the amount of warm gas on the disc between $t=5-15$ Myr, but after $t=15$ Myr, the effect reverses. The impact of gas return on the warm gas above the disc is more complex, while higher $\rm{M_{gas,t}}$ results in more warm gas above the disc. As shown in the bottom row of Figure \ref{fig:gaswithtime}, gas return can increase the mass of hot gas on and above the disc, though $\rm{M_{gas,t}}$ has a relatively minor effect. 

The results from M3R8NGX and M3R16NGX in Figure \ref{fig:gaswithtime} suggest that higher resolution leads to more cool, warm, and hot gas remaining on the disc (with $|z|<400$ pc), while less gas is accreted into sink particles or entrained in the wind. This is likely because the physical volume represented by a sink/star particle in simulations with 16 pc resolution is eight times larger than that in 8 pc resolution. Higher resolution better resolves gas clumps, resulting in moderately higher peak densities, though unlikely by a factor of eight. Consequently, sink particles in the 8 pc resolution simulations accrete and store less gas than those in the 16 pc resolution simulations. Meanwhile, the integrated star formation efficiency for each sink particle increases with resolution. As a result, the amount of gas remaining on the disc in M3R8NGX is approximately $30\%$ higher than in M3R16NGX after $t=20$ Myr, while the mass of stars formed in M3R8NGX is around $\sim 20\%$ greater (see Figure \ref{fig:SFR_SNII}). Although the higher star formation rate (SFR) in M3R8NGX releases about $\sim20\%$ more energy released from CCSN disc, this is outpaced by the increase in gas mass left on the disc. As shown in Figure \ref{fig:outflowimageall2}, the development of outflow in M3R8NGX is indeed slower than in M3R16NGX.

\section{profile, outflow rates and X-ray emission}
This section presents a more detailed quantitative analysis of the outflow properties. For clarity, we will focus primarily on simulations with a resolution of 16 pc in the central region and examine the impact of two key factors: the initial total gas mass,  $\rm{M_{gas,t}}$, which represents different models of gas distribution in the nuclear region, and the gas return process. The effects of other factors, such as resolution and convergence, initial gas metallicity, and the boost factor, will be discussed in the next section. We begin by analyzing the outflow profiles and rates at a fixed time, t = 30 Myr, when the mass outflow rates are close to their peak in most simulations. Following this, we examine the time evolution of profiles and outflow rates in three representative simulations: M3R16NGX, M3R16DGS, and M3R16HGS. It is important to note that the geometric shape and volume occupied by the outflows vary significantly between simulations. To standardize our analysis, we measure the properties of gas cells with a metallicity greater than 0.025 $\rm{Z_\odot}$ (the initial gas metallicity is 0.02 $\rm{Z_\odot}$, except for M3HZ), as these cells represent most of the regions affected by outflows. In addition to outflow rates, we will also explore the X-ray emission from the hot wind in our simulations and compare these results with observations.
\begin{figure*}[htbp]
    \centering
    \includegraphics[width=0.90\textwidth]{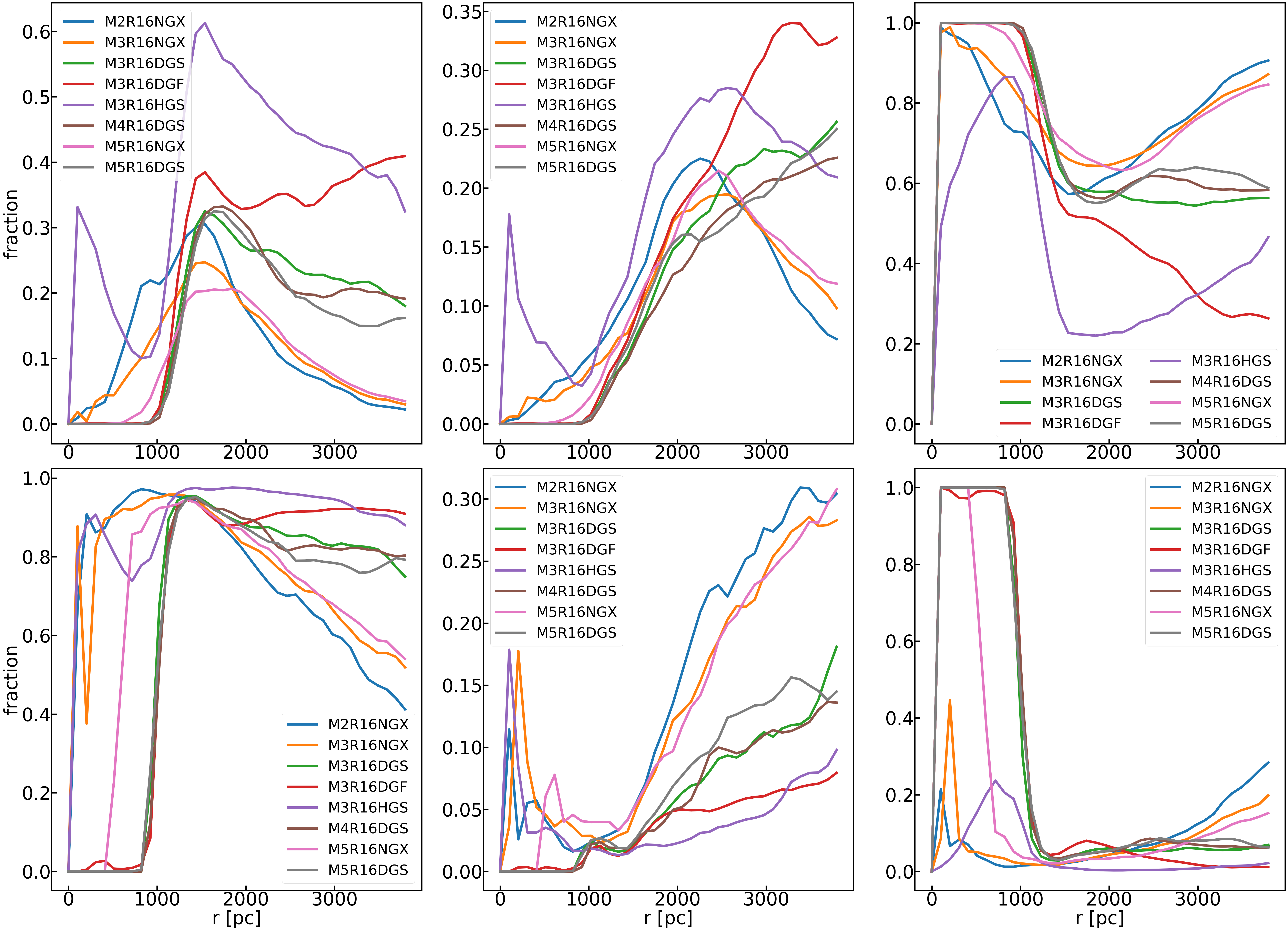}
    \caption{Volume (top) and mass (bottom) fraction of cool (left), warm (middle), hot (right) gas as a function of three dimensional radial distance at 30 Myr in simulations with a resolution of 16 pc. }
    \label{fig:outflowdata_percent}
\end{figure*}

\begin{figure*}[htbp]
    \centering
    \includegraphics[width=0.90\textwidth]{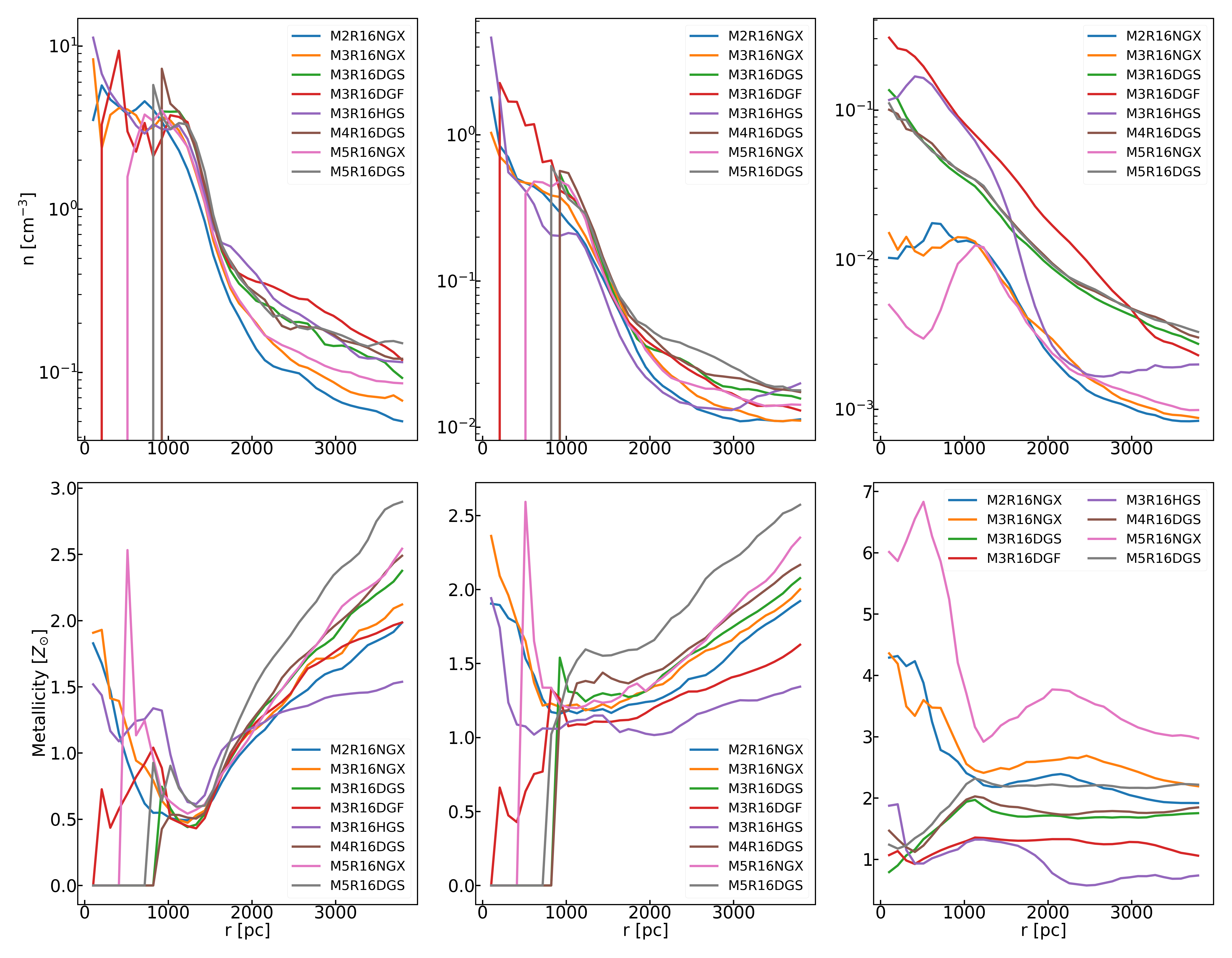}
    \caption{Radial distribution of the volume weighted mean number density and mass-weighted metallicity for gases of three phase at 30 Myr. The left, middle and right columns are the cool, warm and hot gas, respectively.}
    \label{fig:outflowdata_nz}
\end{figure*}

\subsection{Profile at t=30 Myr}
Figure \ref{fig:outflowdata_percent} illustrates the volume filling factor (FF) and mass fraction (MF) of three gas phases as functions of the radial distance from the galaxy center, $r=\sqrt{x^2+y^2+z^2}$, at 30 Myr. In all simulations, cool gas dominates the mass fraction, while hot gas dominates the volume fraction, consistent with previous studies. The filling factor of cool gas, $\rm{FF_{cool}(r)}$, typically increases from nearly 0 at $r=0$ to a maximum value around 0.3-0.4 at $r\sim 1.2$ kpc, before decreasing further out. In contrast, \cite{xu2023radial} reports that the FF of gas with $T\sim 10^4$ K drops from $\sim10^{-3}$ to $10^{-4}$ from r = 0.5 to 2.2 kpc. $\rm{FF_{cool}(r)}$ in all our simulations is much higher than \citealt{xu2023radial} at $r\gtrsim 1.2$ kpc. In simulations with the default gas return process, $\rm{FF_{cool}(r)}$ is very low within $r\lesssim 1.0$ kpc at 30 Myr, consistent with \cite{xu2023radial}. Without gas return, however, $\rm{FF_{cool}}$ will be enhanced (reduced) at $r\lesssim 1.2$ ($r\gtrsim$ 1.2) kpc. As we will  see later, $\rm{FF_{cool}}$ within $r\lesssim 1.0$ kpc was around 0.3 at $t\sim 10$ Myr in  simulations both with and without gas return. Differences in the outflow's opening angle and the fraction of gas returned can lead to a notable discrepancy on $\rm{FF_{cool}}$ within $r\lesssim 1.0$ kpc during the later stages of our simulation. 

The mass fraction of the cool gas, $\rm{MF_{cool}}$ fluctuates around $80\%-90\%$ within $r\lesssim 1.2$ kpc, gradually decreasing to about $\sim 60\%$ beyond $r\sim 3.5$ kpc when gas return from sink particles is not included. While the inclusion of gas return does not alter the peak value of $\rm{MF_{cool}}$, it significantly reduces $\rm{MF_{cool}}$ to near zero within $r\lesssim 1.0$ kpc and causes a more rapid decline beyond $r\gtrsim 1.2$ kpc. A faster gas return process increases both $\rm{FF_{cool}}$ and $\rm{MF_{cool}}$ at $r\gtrsim1.2$ kpc, while a higher gas return fraction enhances these values at all radial distances. 

The filling factor of warm gas, $\rm{FF_{warm}(r)}$, follows a similar pattern to that of cool gas in simulations without gas return, though with a lower peak of around $20\%$. The corresponding $\rm{MF_{warm}(r)}$ shows spikes within the inner 1.2 kpc region, then increases sharply to $\sim 30\%$ at $r\sim 3.5 $ kpc. In simulations with gas return, $\rm{FF_{warm}(r)}$ shows an S-shaped (sigmoid) growth curve. The corresponding $\rm{MF_{warm}(r)}$ gradually grows from 0 to $\sim15\%$ at $r\sim 3.5 $ kpc. A faster or higher gas return has a similar impact on $\rm{FF_{warm}(r)}$ as it does on $\rm{FF_{cool}(r)}$, but it leads to a reduction in $\rm{MF_{warm}(r)}$. 

In simulations without gas return, $\rm{FF_{hot}(r)}$ gradually decreases from nearly $100\%$ at $r=0$ to $60\%$ at $r\sim 2 $ kpc , then rises moderately at greater distances. In these simulations, $\rm{MF_{hot}(r)}$ also exhibits spikes at $r\lesssim 1.2 $ kpc and  slowly increases to $10\%-20\%$. When the default gas return process is included, $\rm{FF_{hot}(r)}$ remains near $100\%$ within $r\lesssim 1 $ kpc, drops sharply to $60\%$ at $r\sim 1.2 $ kpc, and then fluctuates around this level in the outer region. Consequently, $\rm{MF_{hot}(r)}$ is close to $100\%$ at $r\sim 1 $ kpc, and decreases significantly to around $5\%$ at $r\gtrsim1.2$ kpc. The gas return process, whether through a higher return fraction or faster return rate, tends to reduce both the filling factor and mass fraction of the hot gas, in contrast to its effect on cool gas. The initial total gas mass $\rm{M_{gas,t}}$ has a minor effect on the filling factor and the mass fraction across all three phases.

The upper and lower rows of Figure \ref{fig:outflowdata_nz} show the radial profiles of the number density and the mass-weighted metallicity for three gas phases at $t=30$ Myr, respectively. In some simulations, the profiles exhibit significant fluctuations within the inner 1 kpc due to the limited or absent amounts of cool or warm gas. The density profiles for the cool and warm gases remain similar across different simulations. The number density of cool gas, $n_{cool}(r)$, decreases from 5-10 $\rm{cm^{-3}}$ at $r\sim0.2$ kpc to around 0.07 $\rm{cm^{-3}}$ at $r\sim4.0$ kpc in simulations without gas return, which is lower than the results of \cite{schneider2020physical} at $r\lesssim0.2$ kpc, but comparable to their results at $0.2 \lesssim r \lesssim 4$ kpc. It’s important to note that \cite{schneider2020physical} measured the properties within a 30 degree half-opening angle cone, which differ from our procedure. The inclusion of gas return increases the density of cool gas by a factor of two at $r\gtrsim 1.5 $ kpc. Additionally, a higher initial gas mass, $\rm{M_{gas,t}}$, can moderately raise $n_{cool}$ in the outer region. 

At $r\lesssim 1.2$ kpc, $n_{cool}(r)$ decreases approximately as $r^{-0.4}$. This decline sharpens to approximately $r^{-4.0}$ in the range of $1.2\lesssim r \lesssim 2.0$ kpc, followed by a further decrease of about $r^{-1.6}$ at $r\gtrsim 2.0 $ kpc. These patterns contrast with the profile expected from a constant-velocity wind, which would exhibit a decline of $r^{-2.0}$. The subsequent subsection will demonstrate that cool gases continue to accelerate as they move outward in our simulations. In comparison, \cite{schneider2020physical} reports a decrease in $n_{cool}(r)$ that follows approximately $r^{-2.0}$ at $r\gtrsim 1.0$ kpc. The number density of the warm gas, $n_{warm}(r)$, is roughly $10\%$ of that of the cool gas. Within $r\lesssim1.0$ kpc, $n_{warm}(r)$ decreases approximately as $r^{-1.5}$, as $r^{-3.3}$ between $r\sim 1.0$ and $2.0$ kpc, and as $r^{-1.0}$ at $r\gtrsim 1.0$. 

The number density of hot gas, $n_{hot}(r)$,  in simulations that incorporate gas return decreases from approximately 0.1-0.3 $\rm{cm^{-3}}$ at $r\sim0.2$ kpc to around $3-4 \times 10^{-3} \rm{cm^{-3}}$ at $r\sim4.0$ kpc. It exhibits a decline approximately following $r^{-1.6}$  at $r\lesssim 2.0$ kpc and $r^{-1.0}$ at $r\gtrsim 2.0$ kpc. In contrast, $n_{hot}(r)$ in simulations without gas return tends to fluctuate around $10^{-2} \rm{cm^{-3}}$ at $r\lesssim 1.0$ kpc, then decreases as $r^{-1.8}$ between $1.0-2.0$ kpc, and follows a slope of $r^{-1.0}$ at $r\gtrsim 2.0$ kpc, reaching around $10^{-3} \rm{cm^{-3}}$ at $r\sim4.0$ kpc. Overall, the magnitude and trend of $n_{hot}(r)$ in our simulations without gas return are generally consistent with the findings of \cite{schneider2020physical}. In most cases, $n_{hot}(r)$ decreases with a slope shallower than $-2$, i.e., which deviates from the expectations of a purely adiabatic expansion model (\citealt{1985Natur.317...44C}). The presence of gravity and cooling, along with the mixing of cool, warm, and hot phases, has contributed to the observed deviations from the $r^{-2.0}$ profile (e.g., \citealt{2022ApJ...924...82F}). 

In our simulations with gas return, $n_{hot}(r)$ aligns well with the values inferred from X-ray observations by \citealt{lopez2020temperature} at $r < 1.5$ kpc, though it falls below their results in the $1.5 < r < 2.5$ kpc range. It is important to note that in \cite{lopez2020temperature}, the distance 
is measured from the central plane of the M82 disc, not the galaxy center, which could partly explain this discrepancy. A faster gas return process further amplifies $n_{hot}$ across almost all radial distances, while a higher gas return fraction increases $n_{hot}$ at $r\lesssim 1.5$ kpc, but reduces it at $r\gtrsim1.5$ kpc. Variations in $\rm{M_{gas,t}}$ lead to only minor differences in the density profiles of the three phases in the outflow. 


\begin{figure*}[htbp]
    \centering
    \vspace{-0.3cm}
    \includegraphics[width=0.42\textwidth,trim=20 10 50 50, clip]{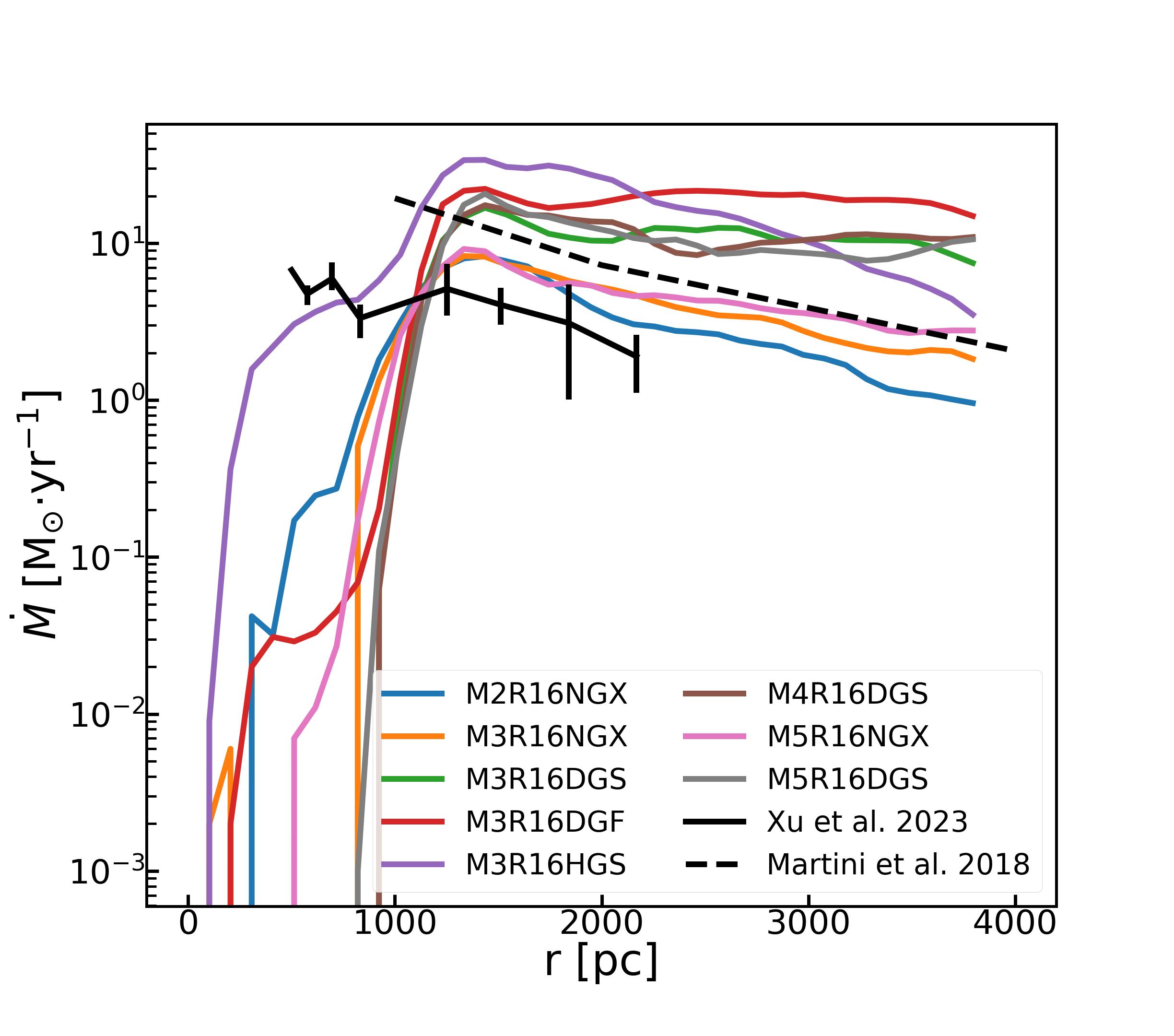}
    \includegraphics[width=0.42\textwidth,trim=20 10 50 50, clip]{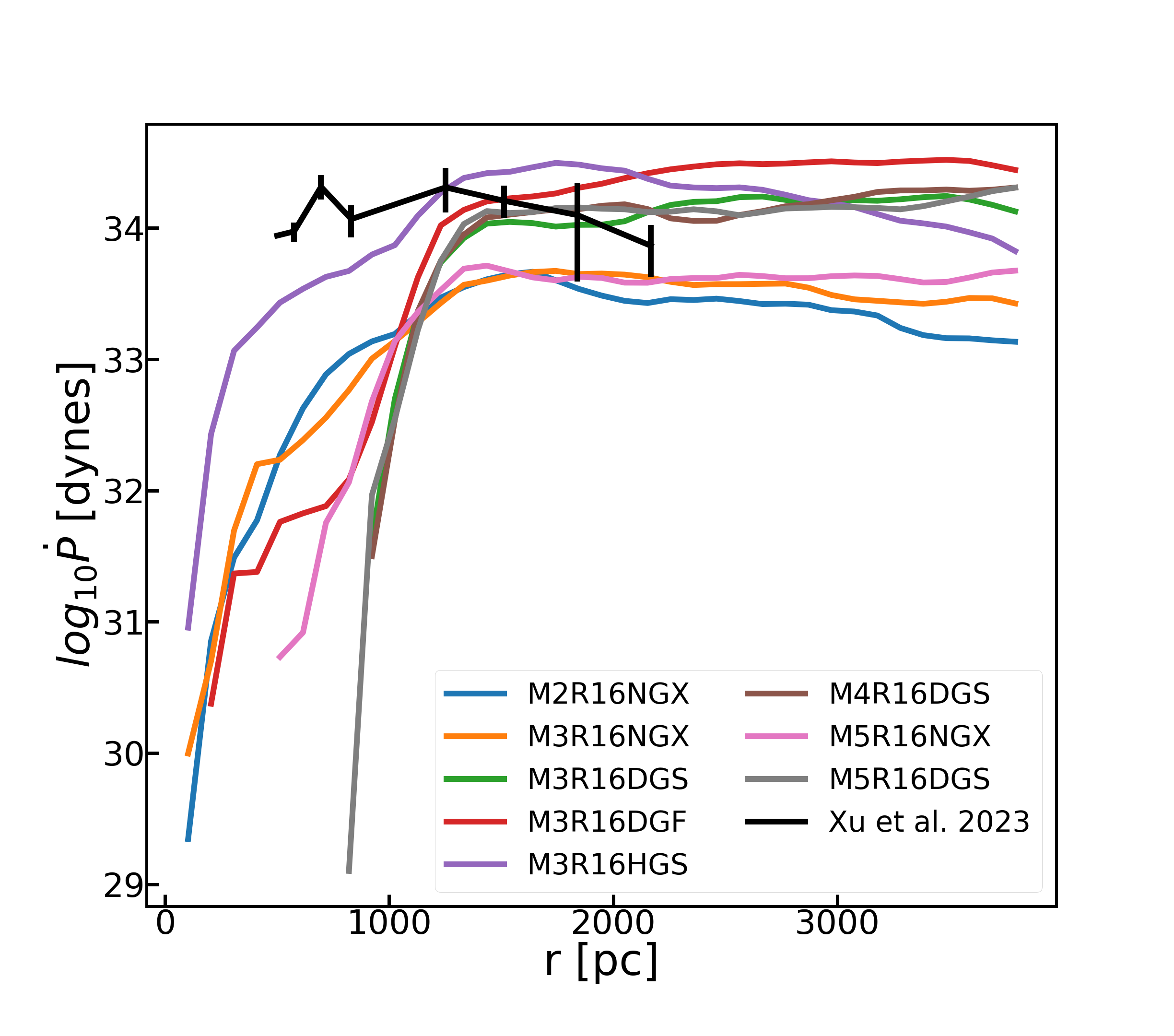}\\
    \vspace{-0.4cm}
    \includegraphics[width=0.42\textwidth,trim=20 10 50 50, clip]{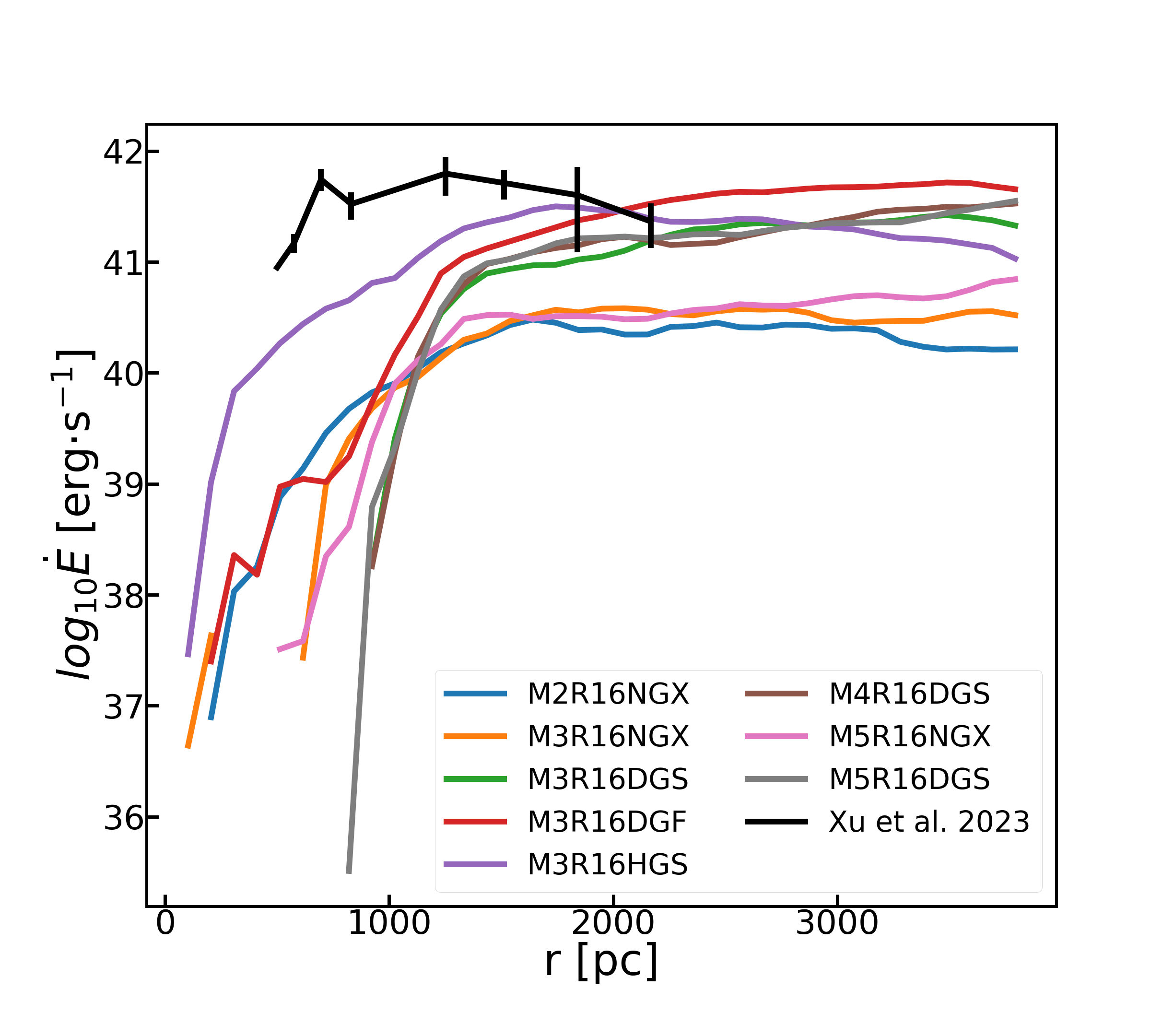}
    \includegraphics[width=0.42\textwidth,trim=20 10 50 50, clip]{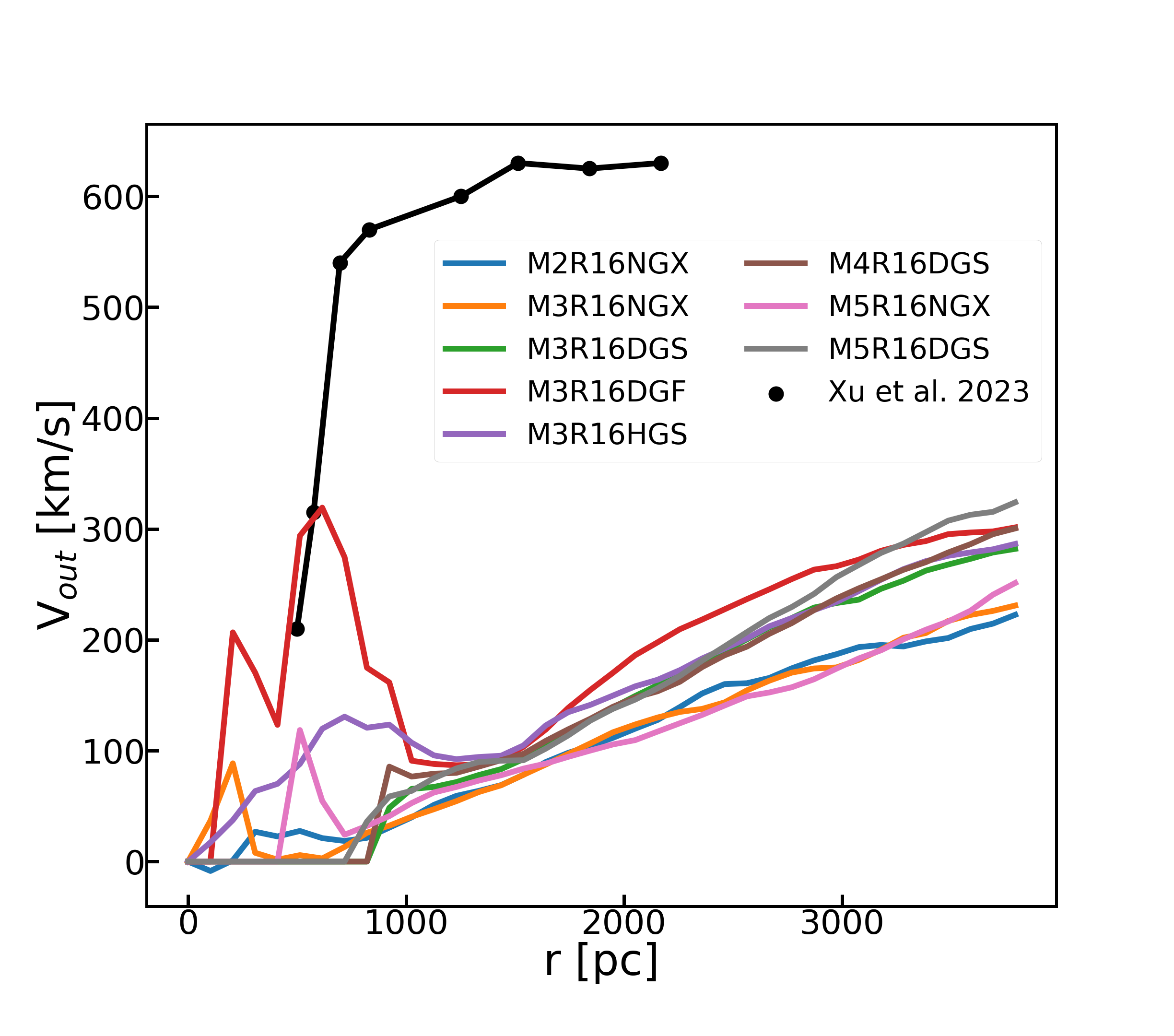}
    \caption{The mass, momentum, energy outflow rates and outflow velocity for the cool phase ($\rm{T} < 2\times 10^{4}$ K) at $t=30$ Myr in our simulations. The black solid and dashed lines represent the results of gas with $\rm{T}\sim 10^4$ K and $\rm{T}\lesssim 5000$ K that inferred from observation in \cite{xu2023radial}(\citealt{2018ApJ...856...61M}).}
    \label{fig:outflowdata_vs1}
\end{figure*}

In our simulations, most of the cool and warm gas in the outflow has a metallicity between 0.5 and 2.5 $\rm{Z_\odot}$. Both phases exhibit similar metallicity profiles: the metallicity decreases initially from about $\sim 2.0$ $\rm{Z_\odot}$ to $\sim 0.5$ $\rm{Z_\odot}$ for cool gas, and to roughly to $\sim 1.2$ $\rm{Z_\odot}$ for warm gas at $r\sim 1.2$ kpc, before increasing at larger radii. In some simulations, metallicity for cool and warm gas are unavailable within $r\lesssim 1.0$ kpc due to the absence of these phases. The hot gas has a metallicity ranging from 1.0 to 7.0 $\rm{Z_\odot}$, consistent with previous simulations (e.g., \citealt{melioli2013evolution}) and X-ray observations, which estimate values between 1.0 - 3.5 $\rm{Z_\odot}$ (e.g., \citealt{SuzakuMetal};\citealt{lopez2020temperature}). With gas return included, the metallicity of the hot gas rises with radius to about 1.0 - 3.0 $\rm{Z_\odot}$ at $r\sim 1.2$ kpc, then fluctuates at larger distances. Without gas return, the metallicity decreases with distance. A higher initial gas mass $\rm{M_{gas,t}}$ modestly increases the metallicity across all phases. The default gas return process enhances the metallicity of cool and warm gas while reducing that of the hot gas. However, a higher gas return fraction or faster return process dilutes the metallicity across all phases.

\subsection{Outflow Rates at t=30 Myr}
\begin{figure*}[htbp]
    \centering
    \vspace{-0.3cm}
    \includegraphics[width=0.42\textwidth,trim=20 10 50 50, clip]{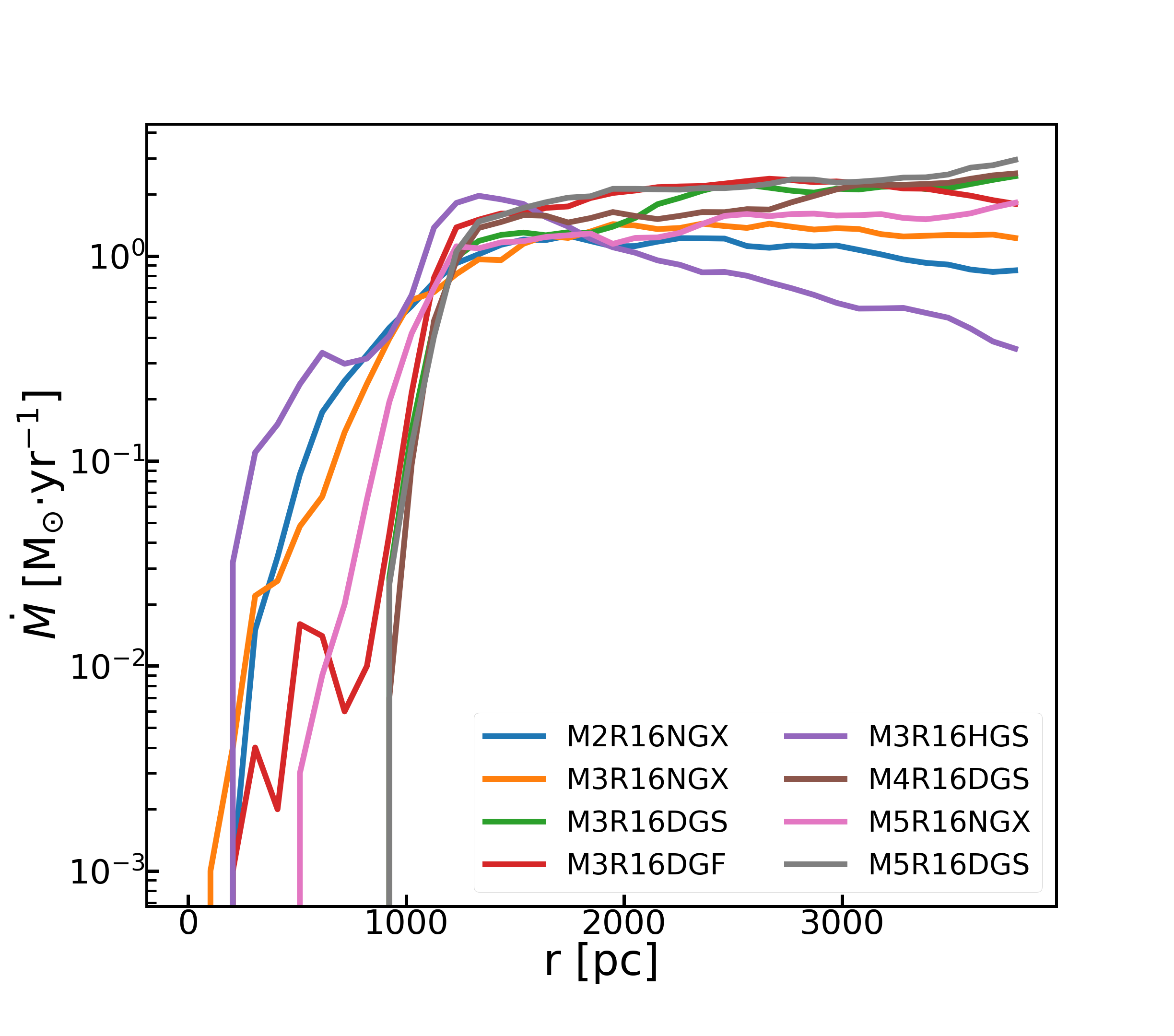}
    \includegraphics[width=0.42\textwidth,trim=20 10 50 50, clip]{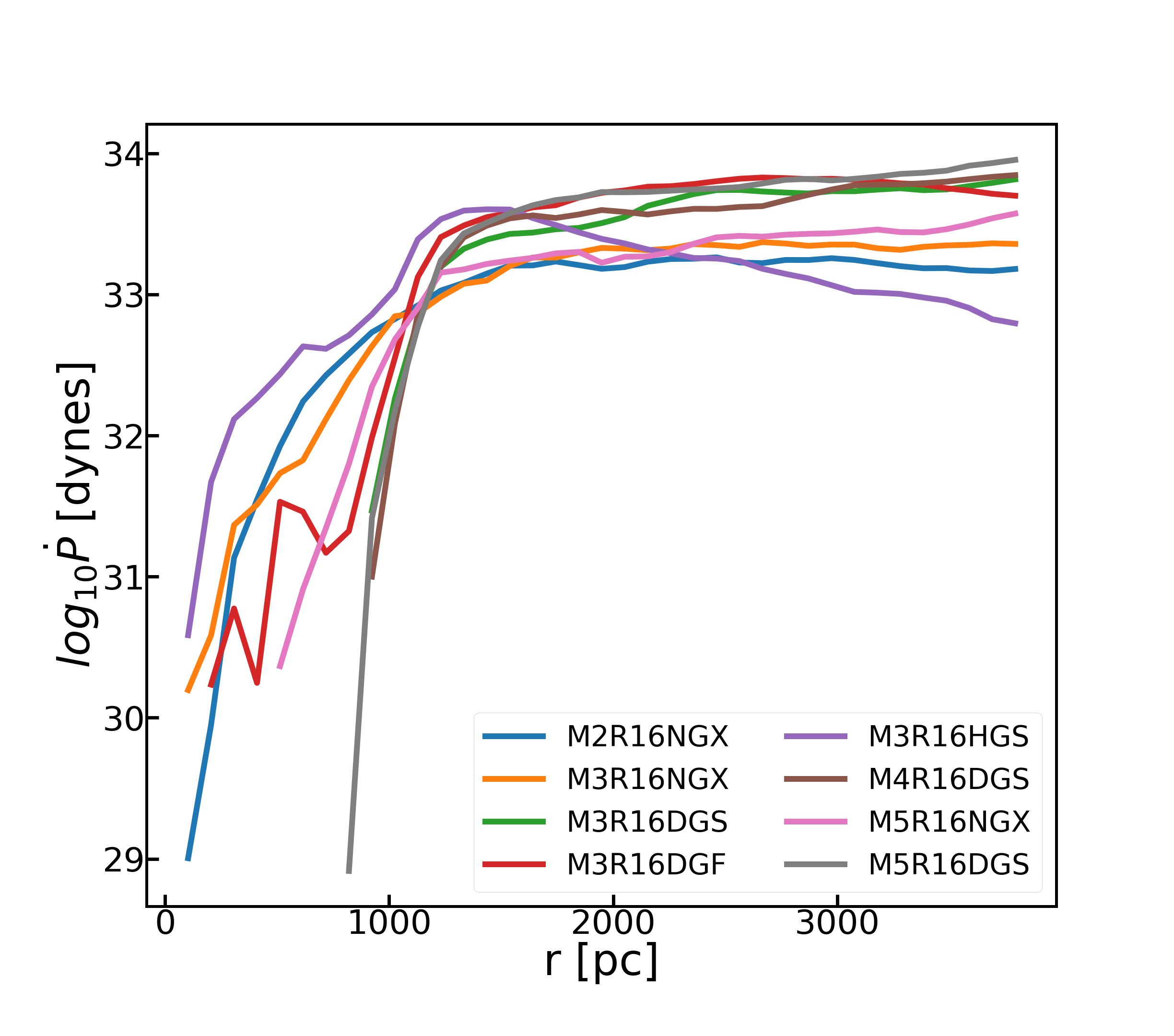}\\
     \vspace{-0.4cm}
    \includegraphics[width=0.42\textwidth,trim=20 10 50 50, clip]{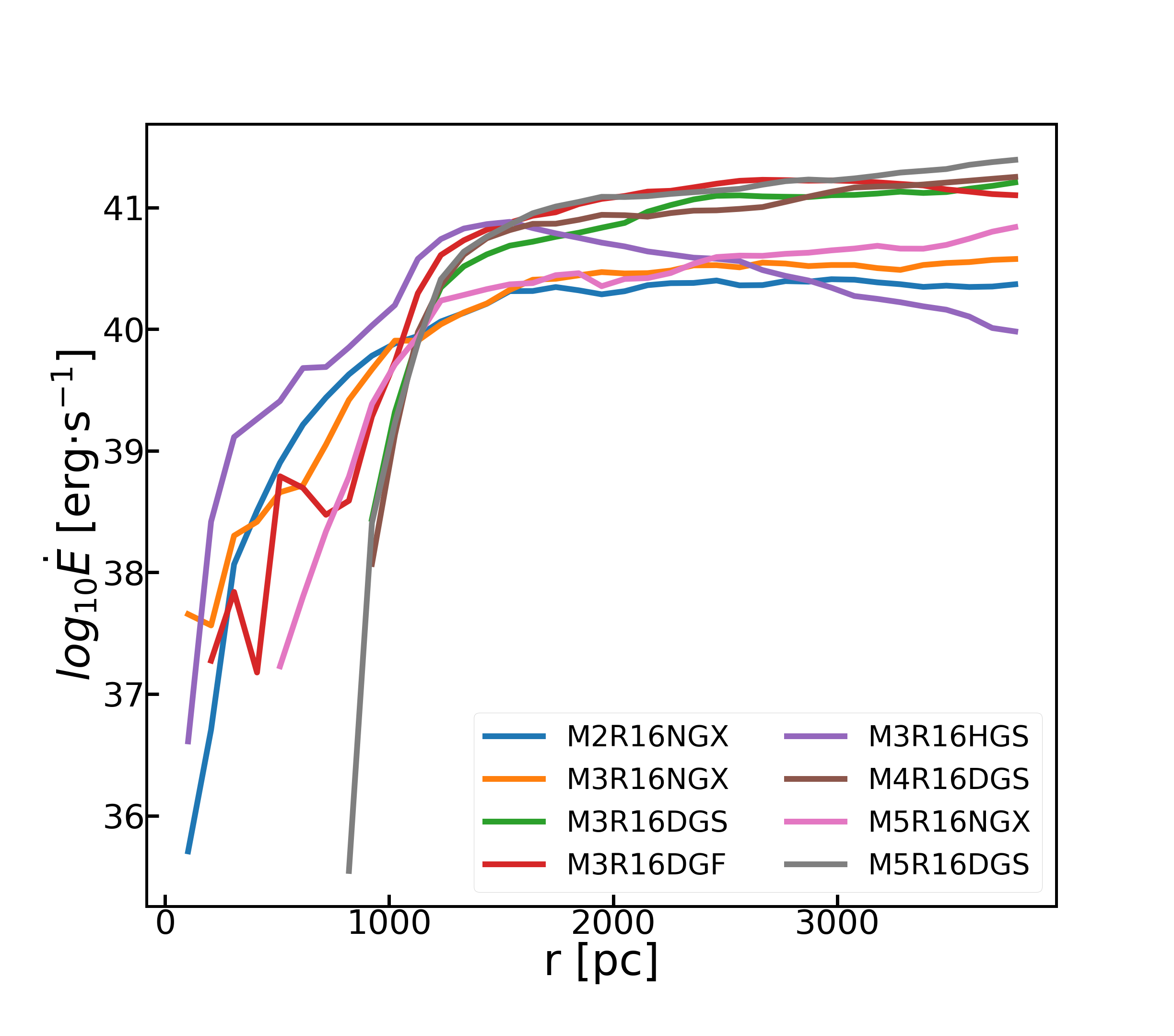}
    \includegraphics[width=0.42\textwidth,trim=20 10 50 50, clip]{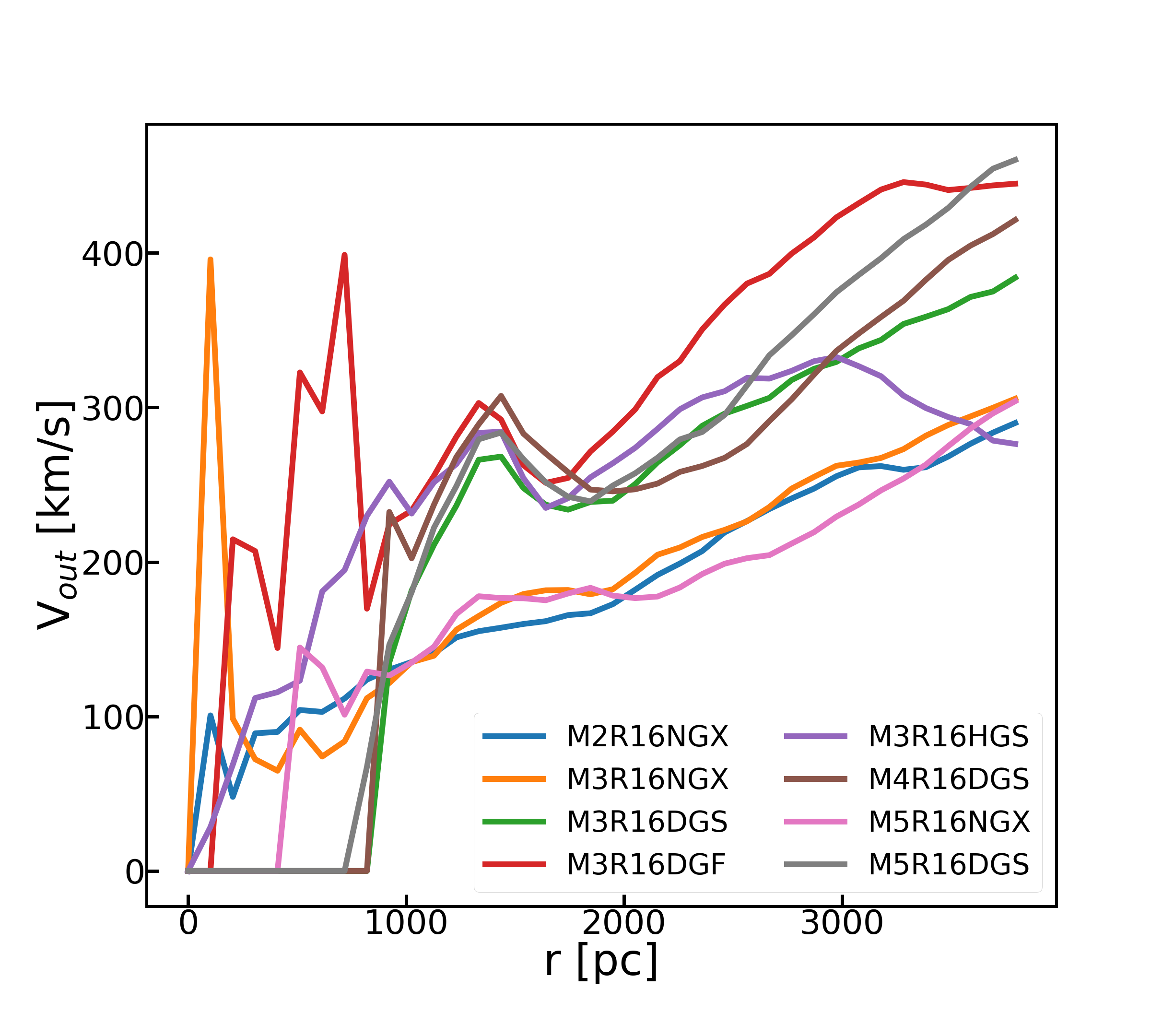}
    \caption{The same as Figure \ref{fig:outflowdata_vs1}, but for the warm phase ($2\times 10^{4} \text{K} < \rm{T} < 5\times 10^{5}$ K) at t=30 Myr in our simulations. }
    \label{fig:outflowdata_warm}
\end{figure*}

\begin{figure*}[htbp]
    \centering
    \vspace{-0.3cm}
    \includegraphics[width=0.42\textwidth,trim=20 10 50 50, clip]{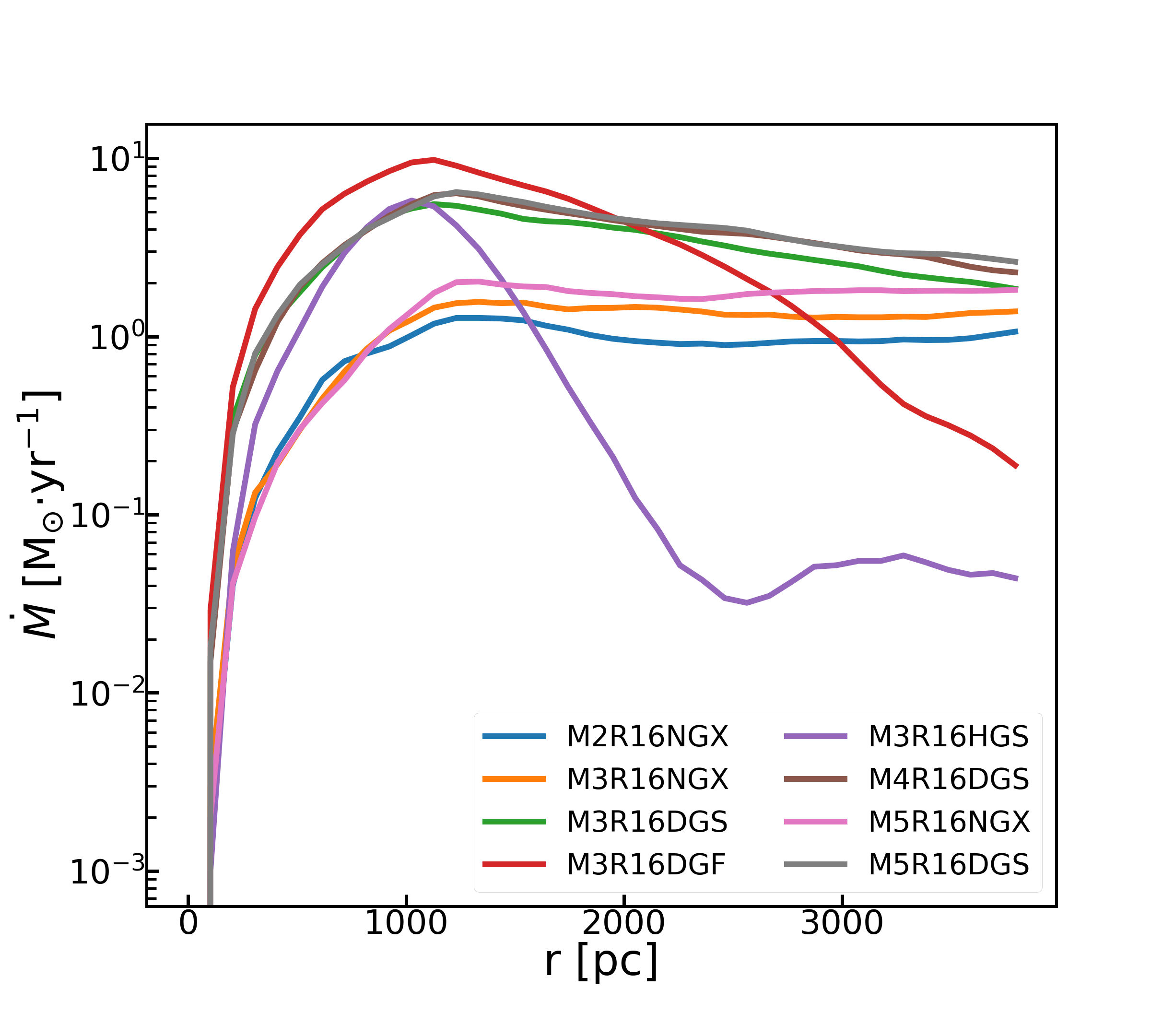}
    \includegraphics[width=0.42\textwidth,trim=20 10 50 50, clip]{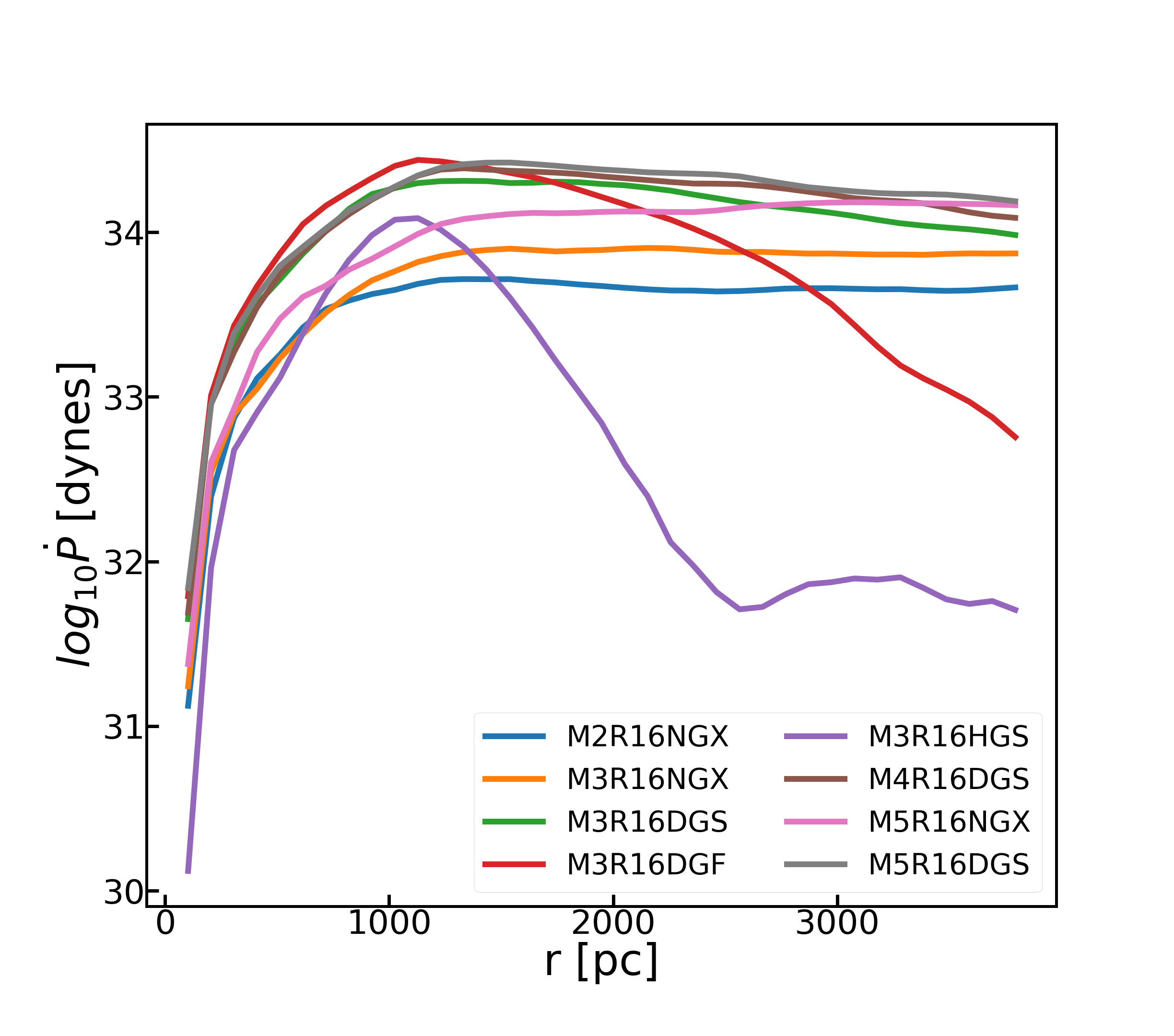}\\
     \vspace{-0.4cm}
    \includegraphics[width=0.42\textwidth,trim=20 10 50 50, clip]{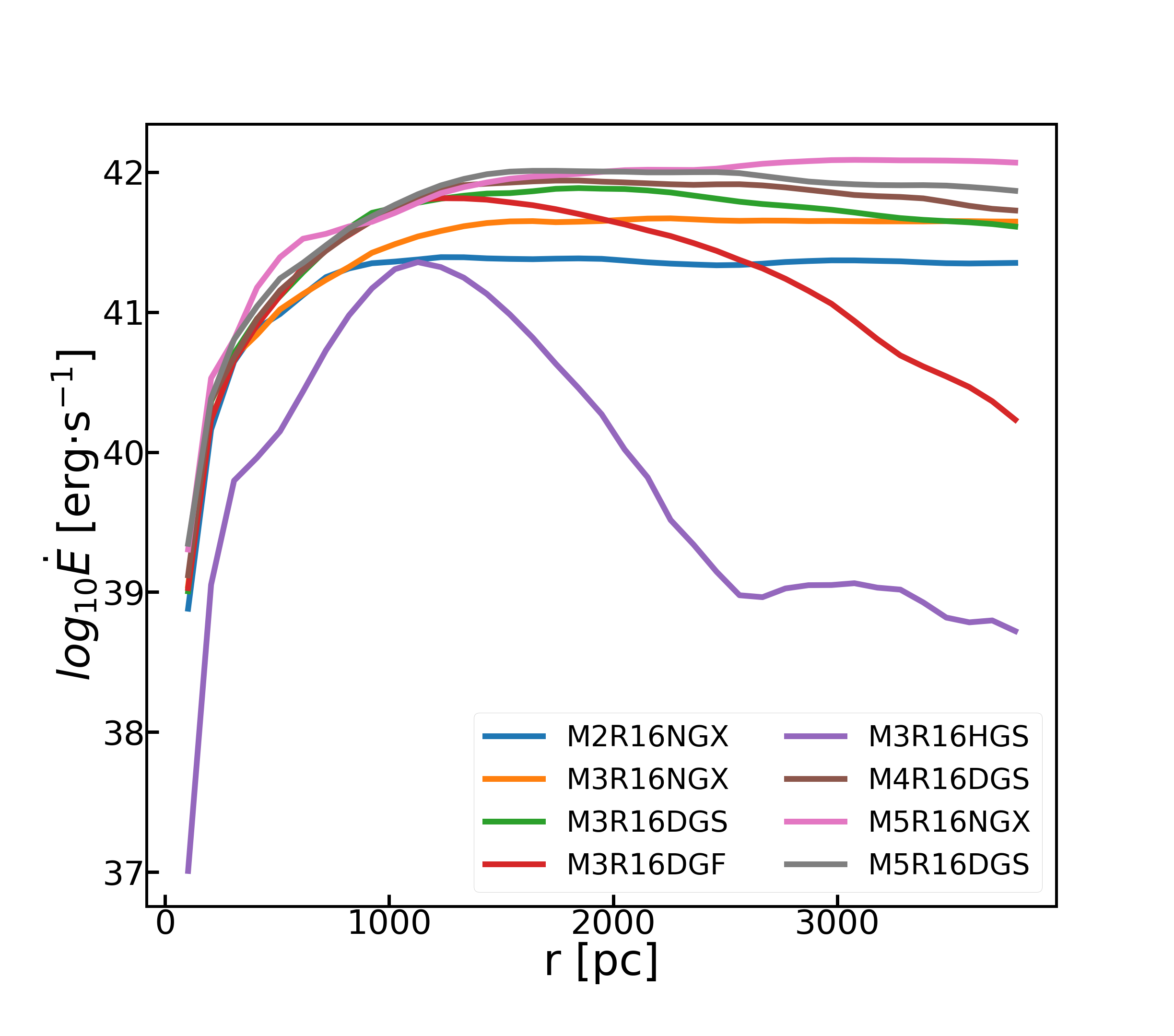}
    \includegraphics[width=0.42\textwidth,trim=20 10 50 50, clip]{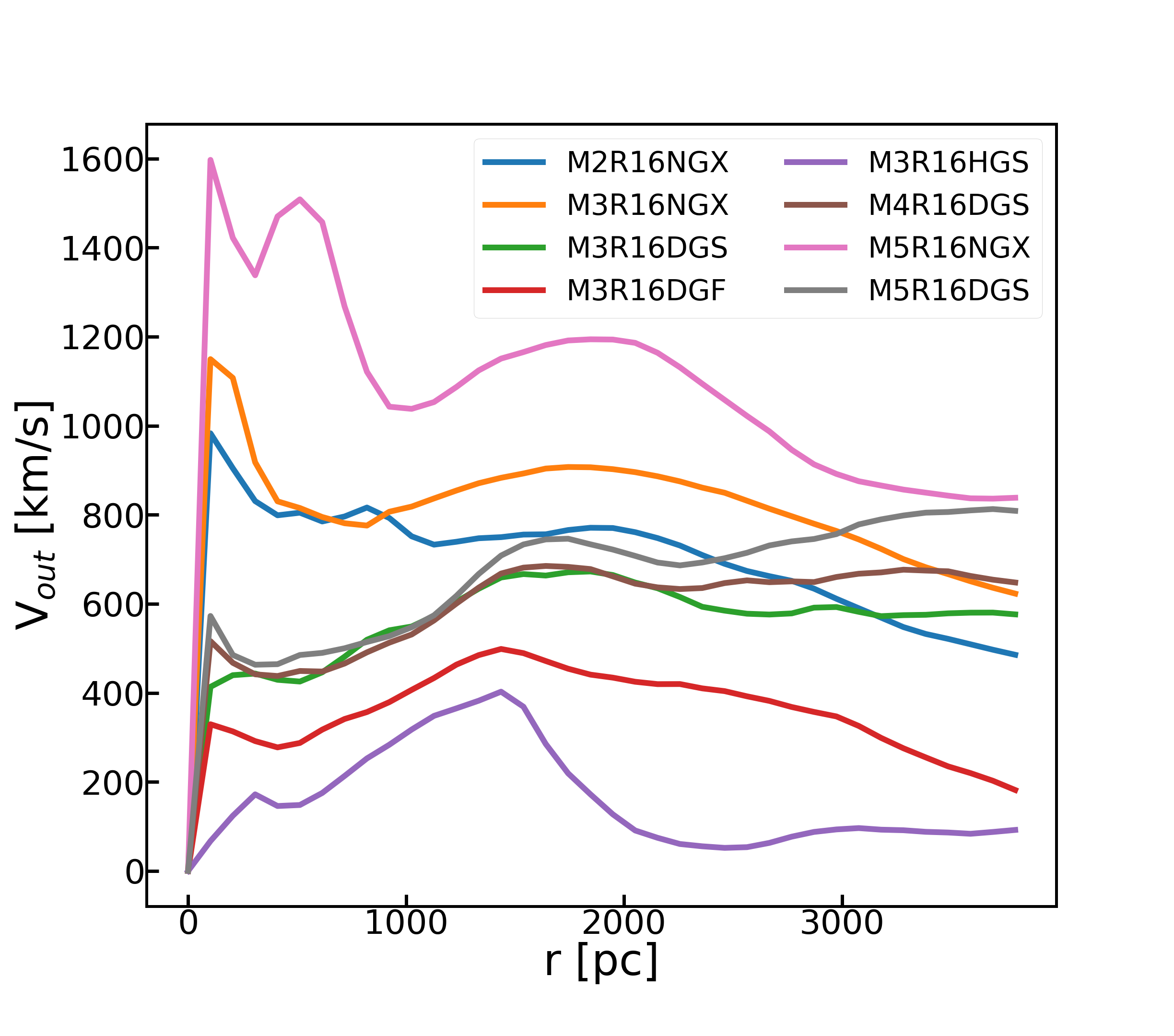}
    \caption{The same as Figure \ref{fig:outflowdata_vs1}, but for the hot phase ($\rm{T} > 5\times 10^{5}$ K) at 30 Myr. }
    \label{fig:outflowdata_hot}
\end{figure*}

The influence of galactic winds on the evolution of their host galaxies is largely determined by the outflow rate, mass loading factor, and escape fraction. In our simulations, the outflow rates generally increase from low levels around $t\sim 10$ Myr, reaching their peak between $t\sim 25-35$ Myr, with the exact timing varying moderately between different gas phases and simulations. The outflow rates and radial, volume-weighted outflow velocities are computed following the methods of \cite{pandya2021characterizing} and \cite{xu2023radial}, using a radial bin width of $\Delta r = 0.1$ kpc. For each shell between $r$ and $r+\Delta r$, we sum the products of mass and radial outflow velocity for all gas cells with a metallicity higher than the initial value, dividing by $\Delta r$ to calculate the mass outflow rate $\dot{M}$. Consistent with \cite{xu2023radial}, the radial outflow velocity of a gas cell is defined as $V_{out,i} = \vec{V_i} \cdot \vec{n}_{r,i}$, where $\vec{V_i}$ is the gas velocity of this cell and $\vec{n}_{r,i}$ is the unit vector pointing from the galaxy center to the cell. The volume-weighted average radial outflow velocity $V_{out}(r)$ is obtained by averaging $V_{out, i}$ over all grid cells in each conical shell.
Figures \ref{fig:outflowdata_vs1}, \ref{fig:outflowdata_warm}, and \ref{fig:outflowdata_hot}, illustrate the mass, momentum, and energy flow rates, along with the average outflow velocity for the cool, warm, and hot gas phases, respectively. 

We first examine the outflow rates of the cool gas in Figure \ref{fig:outflowdata_vs1}. These rates rise sharply from the innermost region to about $r \sim 1.2 \, $ kpc and then fluctuate or gradually decline with increasing radial distance. Simulations with gas return from sink particles maintain a cool gas mass outflow rate $\dot{M}_{cool}$ $\sim 10-20\rm{M_{\odot}/yr}$, while those without gas return sustain a lower rate of $2-8\rm{M_{\odot}/yr}$ at $r >$ 1.2 kpc. The overall trend and magnitude in the simulations without gas return are comparable to those in \cite{schneider2024cgols}(after multiplying their results by a factor of 3 to account for the full $4\pi$ solid angle), and align with the predictions of the high hot-phase density scenario of the multiphase wind model from \cite{2022ApJ...924...82F}, though they are moderately higher than the results of \cite{schneider2020physical}. Variations in $\rm{M_{gas,t}}$ lead to only minor discrepancies in mass outflow rates. In contrast, the simulation with $\tau_{return}=5$ shows a higher $\dot{M}_{cool}$, while an increased fraction of gas return boosts $\dot{M}_{cool}$ at $r<$ 2 kpc but decreases it beyond $r >$ 2 kpc. The overall patterns of momentum and energy outflow rates, as functions of radius, follow similar trends to those of the mass outflow rates, and are similarly affected by initial gas mass and gas return. Peak values of $\dot{P}_{cool}$, $\dot{E}_{cool}$ in simulations with gas return are $10^{34}\,\rm{dynes}$, $10^{41.5}\,\rm{erg/s}$, respectively. In simulations without gas return, these values are lower by 0.5-0.8 dex. Therefore, we conclude that the gas return process significantly influences the outflow rates of cool gas, whereas $\rm{M_{gas,t}}$ has a relatively minor impact. 

Figure \ref{fig:outflowdata_vs1} also presents observational estimates from \cite{xu2023radial} and \cite{2018ApJ...856...61M}. \cite{xu2023radial} focuses on ionized gas outflows at temperatures around $10^{4}$ K.  \citealt{2018ApJ...856...61M} investigates atomic hydrogen outflows, with their results measured as a function of projected distance along the minor axis, and the corresponding profile as a function of three dimensional radialdistance, r, should be somewhat flattened. Additionally, the mass outflow rates in \cite{2018ApJ...856...61M} might be overestimated, as suggested by (\citealt{2023MNRAS.518.4084Y}). In our simulations, $\dot{M}_{cool} $, which includes gas below $2\times 10^{4}$ K, is expected to exceed the combined outflow rates from these two observational studies. At $r\gtrsim$ 1.2 kpc, our results with gas return align well with this expectation, whereas simulations without gas return fall below the observed values. In the inner region ($r\lesssim1.2$ kpc), $\dot{M}_{cool}$ in most of our simulations lower than the estimates from \cite{xu2023radial}, which may be partly due to projection effects.

The momentum outflow rates of cool gas, $\dot{P}_{cool} $, in our simulations with gas return are comparable to or higher than \cite{xu2023radial} in the outer region, while those without gas return are lower. However, $\dot{P}_{cool}$ in all our simulations falls below the estimates of \cite{xu2023radial} at $r <$ 1.2 kpc. Similarly, the energy outflow rates, $\dot{E}_{cool} $, in most of our simulations are lower than their results within the range $0.5\,\rm{kpc}<r<2.0\,\rm{kpc}$. This discrepancy is attributed to the much slower outflow velocity of the cool gas, $V_{out,cool}$, as shown in the bottom right panel of Figure \ref{fig:outflowdata_vs1}. At $t=30$ Myr, $V_{out,cool}$ is close to zero or fluctuates significantly at $r<$ 1.0 kpc, gradually increasing from around 50 - 100 km/s at $r \sim$ 1.0 kpc to approximately 200 - 300 km/s at $r \sim$ 3.5 kpc. The increasing trend of $V_{out,cool}$with radial distance at $r>$ 1.0 kpc is consistent with \cite{schneider2020physical} and \cite{2022ApJ...924...82F}, although the velocities in our simulations are $\sim20\%-60\%$ lower. The mixing/interaction between the cool and hot phases can accelerate the cool gas in the wind. Observations of H$\alpha$ and other bands estimate outflow velocities of about 150 km/s at $r\sim 0.5$ kpc, rising to 500-600 km/s at $r\sim 1.0$ kpc, and then approximately remaining constant between $r=1.0-2.2$ kpc (\citealt{1998ApJ...493..129S,2018ApJ...856...61M,xu2023radial}). The anomalously low $V_{out,cool}$ in our simulations at $r<$ 1.0 kpc can be explained by the absence of cool gas in this region, as indicated by the number density profile in Figure \ref{fig:outflowdata_nz}. The reasons behind the much slower $V_{out,cool}$ at $r>$ 1.0 kpc in our simulations, along with potential improvements, will be discussed in the next section.

Figure \ref{fig:outflowdata_warm} shows that the overall trend of the outflow rates and velocity of the warm gas as functions of r resemble that of the cool gas. In simulations with and without gas return, $\dot{M}_{warm}$ fluctuates around 2 $\rm{M_{\odot}/yr}$ and 1 $\rm{M_{\odot}/yr}$, respectively, at $r >$ 1.2 kpc. The peak values of $\dot{P}_{cool}$, $\dot{E}_{cool}$ and $V_{out,cool}$ in simulations with gas return are $10^{33.5}\,\rm{dynes}$, $10^{41.0}\,\rm{erg/s}$ and 400 km/s, respectively. Compared to the cool gas, the velocity discrepancy between simulations with and without gas return is larger, while the opposite trend is observed for the other outflow rates. Overall, the influence of gas return remains dominant compared to variations in $\rm{M_{gas,t}}$. Additionally, a higher gas return fraction tends to enhance outflow rates in the inner region while reducing them in the outer region. Significant fluctuations in the warm gas outflow rates within $r<$ 1 kpc are due to the low filling factor of warm gas in that region. 

Figure \ref{fig:outflowdata_hot} shows that the outflow rates of hot gas at t=30 Myr rise rapidly in the inner region, reaching $50\%$ and $100\%$ of the peak at $r\sim 0.5$ kpc and $1.2\,$ kpc, before gradually decreasing at $r\gtrsim 1.2\,$ kpc. The peak value of $\dot{M}_{hot}$ in simulations with and without gas return are approximately $5-10\,\rm{M_{\odot}/yr}$ and $1-2\,\rm{M_{\odot}/yr}$, respectively. The trend and magnitude in simulations without gas return align closely with the results of in \cite{schneider2020physical} (scaled for a full $4\pi$ solid angle) and with the high hot-phase density scenario in \cite{2022ApJ...924...82F}, while being about twice as large as the results of \cite{schneider2024cgols}. In comparison, observations of the starburst region estimate the hot gas outflow rate to be $1.6-3.6\, \rm{M_{\odot}/yr}$ at $r\sim 0.5\,$ kpc in \cite{2009ApJ...697.2030S}, which is consistent with our simulations that include gas return. The momentum and energy outflow rates of hot gas, $\dot{P}_{hot}$ and $\dot{E}_{hot}$, rise sharply in the inner region, stabilizing at around $10^{33.5}-10^{34.5}$ dynes and $10^{41.5}-10^{42.0}\, \rm{erg/s}$, respectively, in the outer region. The gas return process significantly increases hot gas outflow rates, while higher initial gas mass, $\rm{M_{gas,t}}$, leads to moderate increases. However, $\dot{M}_{hot}$ in the M3R16DGF and M3R16HGS simulations exhibited exhibits a sharp decline beyond $r=1$ kpc.

In our simulations, the outflow velocity of the hot gas ,$V_{out,hot}$, ranges from 800-1600 km/s in simulations without gas return, and 500 - 800 km/s in those with gas return, across $r=0.2$ kpc and $r=4 $ kpc. The enhanced hot gas outflow rate with gas return is primarily driven by the increase in number density (see Section 4.1). $V_{out,hot}$ in simulations without gas return is slightly slower than the results in \cite{schneider2020physical} and \cite{2022ApJ...924...82F}. \cite{2009ApJ...697.2030S} estimated that the hard X-ray emission gas (at $3-8\times 10^7$ K) in the starburst region of M82 would have a terminal velocity of $\sim 1400-2200$ km/s, assuming negligible radiative losses and minimal mass loading. \cite{2024arXiv240815327B} report that the soft X-ray emission gas ($\rm{T}\sim 3\times10^6$ K) has outflow velocities exceeding 2000 km/s. In our simulations, the outflow velocity of gas hotter than $2\times 10^6$ K is comparable to that of the whole hot phase($T>5\times 10^{5} $ K). Thus, while hot gas in simulations without gas return reaches velocities comparable to observed values, its velocity in simulations with gas return is slower than observations suggest.

\begin{figure*}[htbp]
    \centering
    \includegraphics[width=0.90\textwidth]{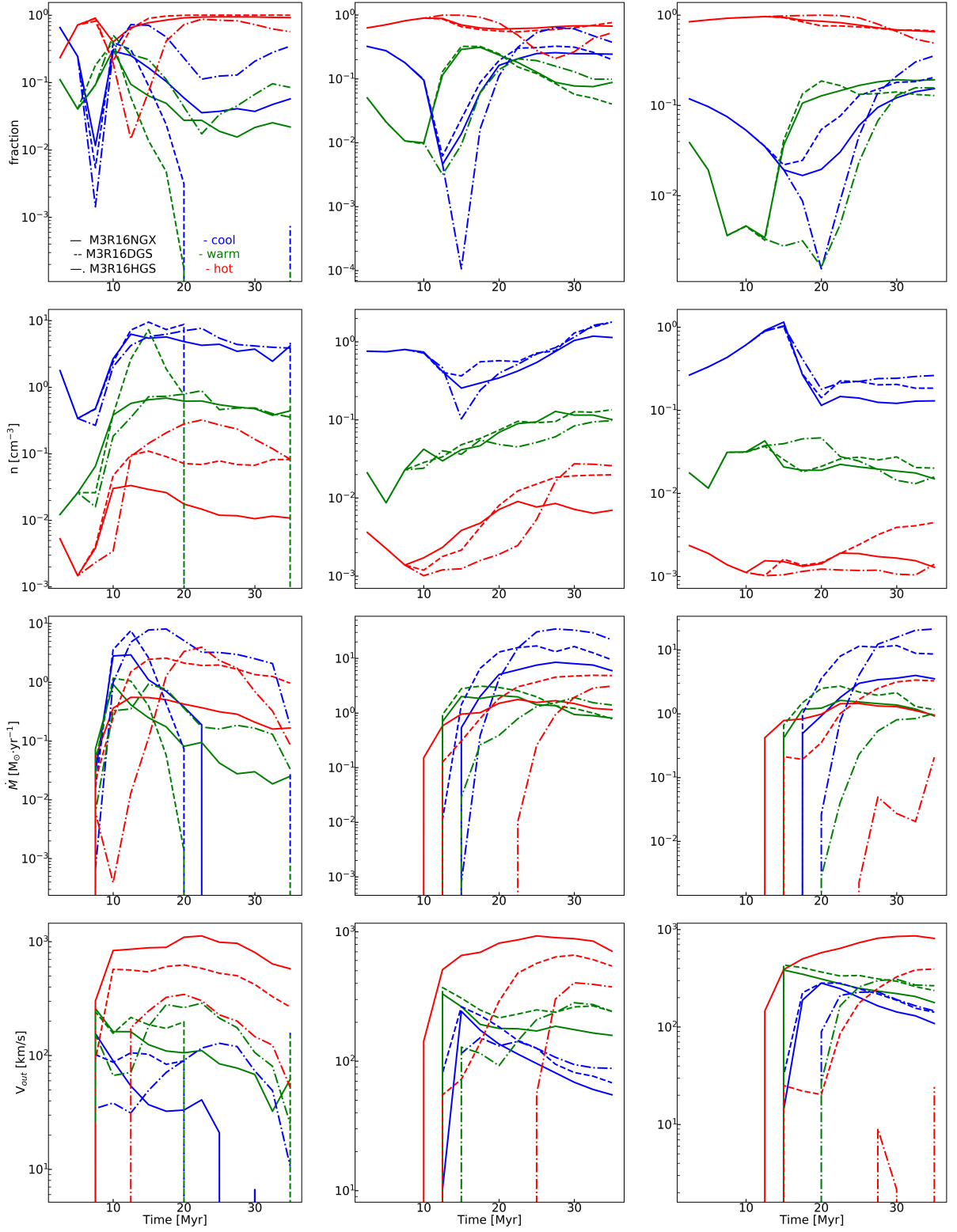}
    \caption{From top to bottom, shows the volume fraction, number density, mass outflow rate and velocity of cool (blue), warm (green) and hot (red) gas as functions of time in simulations M3R16NGX (solid lines), M3R16DGS (dashed lines) and M3R16HGS (dotted-dashed lines). The left, middle and right column correspond to radial distances of $r=$ 0.5, 1.5 and 2.5 kpc, respectively.}
    \label{fig:prof_time}
\end{figure*}

In summary, the total mass outflow rate, $\dot{M}_{tot}$, of the three phases ranges from about 20 to 40 $\rm{M_{\odot}/yr}$ at $r\sim1.5\,$ kpc and from 6 to 12 $\rm{M_{\odot}/yr}$ at $r\sim4.0\,$ kpc at $t=30$ Myr in simulations with gas return. The corresponding mass loading factor, defined as $\dot{M}_{tot}/SFR$, is about 6 to 12 at $r\sim1.5\,$ kpc and about 2 - 4 at $r\sim4.0\,$ kpc, with a trend toward further reduction over time or at greater distances. The mass outflow rates for the cool and hot phases in simulations with gas return are consistent with observations of M82. Without gas return, $\dot{M}_{tot}$ drops by about $40\%-50\%$, but still remains comparable to results from \citealt{schneider2020physical,2022ApJ...924...82F,schneider2024cgols}. In most cases, the cool and warm gas in the outflow is expected to fall back onto the disc, as their velocities remain below the escape velocity (around 500 km/s), meaning only the hot phase can escape. This conclusion is consistent with other studies suggesting a galactic fountain model for cool and cold outflows in M82 (e.g., \citealt{Adam2015};\citealt{2018ApJ...856...61M}
). However, some recent works favor an alternative scenario (e.g. \citealt{2023MNRAS.518.4084Y}). 

\subsection{Evolution of Profiles and Outflow Rates}

Figure \ref{fig:prof_time} shows the time evolution of volume fraction, number density, mass outflow rates and velocities of the three phases in three simulations M3R16NGX (solid lines), M3R16DGS (dashed lines) and M3R16HGS (dotted-dashed lines). The left, middle and right column correspond to radial distances of $r=$ 0.5, 1.5 and 2.5 kpc, respectively. Blue, green and red color indicate cool, warm and hot phases. The cool, warm, and hot phases are represented by blue, green, and red, respectively. Before $t\sim 8-10$ Myr, the net outflow rate and velocity are either zero or negative, as most of the gas is still falling onto the disc. The volume fraction and number density of the three phases evolve in a complex manner due to the interplay of outflow, inflow, and radiative cooling processes. 

At $r\sim 0.5$ kpc, outflow begins to surpass inflow around 
$t\sim8$ Myr, marked by rapid increases in number density, outflow rate, and velocity. After this 'turnaround' point, the volume fraction of cool and warm gas rises. For $r=1.5$ and $2.5$ kpc, the turnaround occurs approximately 2 and 5 Myr later than at $r=0.5$ kpc, if gas return is not considered. The inclusion of gas return delays the turnaround, with a higher gas return fraction causing a stronger delay, especially for the hot phase. After the turnaround, the growth in number density at r=1.5 and 2.5 kpc is somewhat slower than at $r=0.5$ kpc. The mass outflow rates peak around $t\sim 15-20$  Myr at $r=0.5$ kpc and around $t\sim25-30$ Myr at $r=1.5$ and 2.5 kpc, coinciding with the maximum outflow velocity of the hot gas. However, the outflow velocities of the cool and warm phases peak around the turnaround epoch and then decrease over time, primarily due to gravitational drag. In some simulations, inflow may again dominate outflow for the cool and warm phases at $r=0.5$ kpc after $t\sim 20$ Myr. 

\subsection{X-ray Emission of Hot Phase}

\begin{table}[htbp]
        \tablenum{3}
	\centering
	\caption{The total thermal X-ray luminosity of hot gas in different simulations at $t=30$ Myr.}
	\begin{tabular}{lccr}
		\hline
		Simulation & Total luminosity (erg/s) & Ratio to observation \\
		\hline
		M2R16NGX& $3.83\times 10^{39} $ &0.140  \\
        M3R16NGX& $4.37\times 10^{39} $ & 0.160  \\
        M3R16DGS& $1.55\times 10^{40} $ & 0.569  \\
        M3R16DGF& $3.91\times 10^{40} $ & 1.431  \\
        M3R16HGS & $1.77\times 10^{40} $ & 0.650 \\
        M4R16DGS& $2.04\times 10^{40}  $ & 0.746  \\
        M5R16NGX& $7.20\times 10^{39}  $ & 0.263  \\
        M5R16DGS& $2.42\times 10^{40}  $ &0.885  \\
        M3R8NGX& $6.49\times 10^{39} $ & 0.237  \\
        M3R8DGS & $1.23\times 10^{40} $ & 0.452 \\
        M3HZ& $1.64\times 10^{40} $ & 0.602  \\
        M3BOOST3 & $1.61\times 10^{40} $ & 0.590 \\     
		\hline
	\end{tabular}
	\label{tab:Lx} 
\end{table}

\begin{figure}[htbp]
    \centering
    \includegraphics[width=0.95\columnwidth]{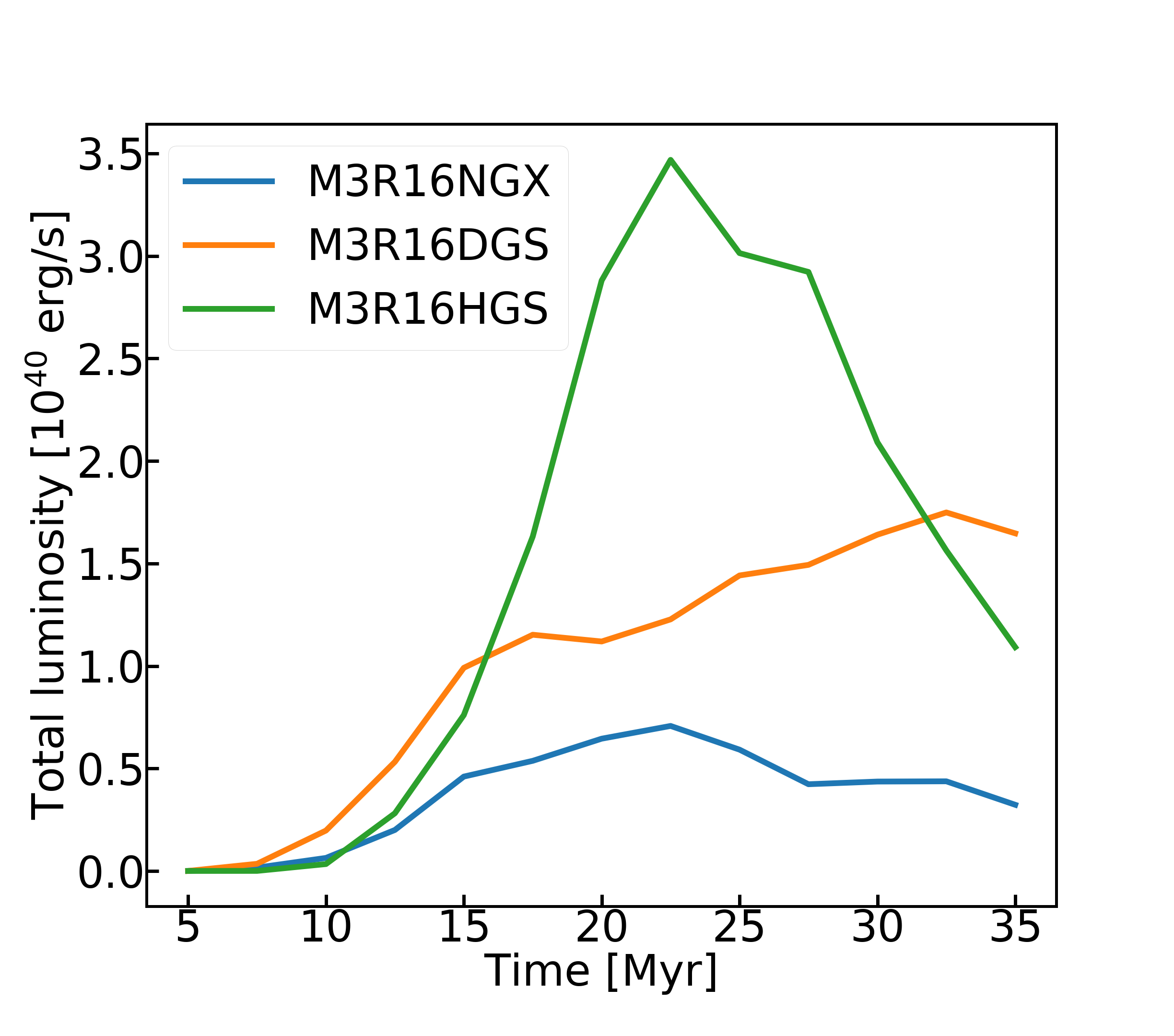}
    \caption{The total thermal X-ray luminosity of hot gas in simulations M3R16NGX, M3R16DGS and M3R16HGS as functions of time.}
    \label{fig:xray_te}
\end{figure}

In addition to the profiles and outflow rates, the X-ray emission of the hot gas offers valuable insights and constraints on the outflow dynamics. We use the pyXSIM\footnote{https://hea-www.cfa.harvard.edu/~jzuhone/pyxsim/index.html} tool to generate broad-band (0.5 - 7.0 keV) X-ray emission across the entire simulation region. For the thermal source, we adopt the apec collisional ionization equilibrium (CIE) model, with the metallicity set to the average value for gas hotter than $5\times 10^{5}$ K in our simulation. Based on Chandra observation, \cite{lopez2020temperature} estimated a total X-ray luminosity of  $3.90\times 10^{40}$ erg/s in the band 0.5 - 7.0 keV. The contributions from point sources, charge exchange and nonthermal component are around $30\%$. Thus, the thermal X-ray emission from the hot gas is around $2.73\times 10^{40}$ erg/s.

\begin{figure*}[htbp]
    \centering
    \includegraphics[width=0.80\textwidth,trim=20 5 5 10, clip]{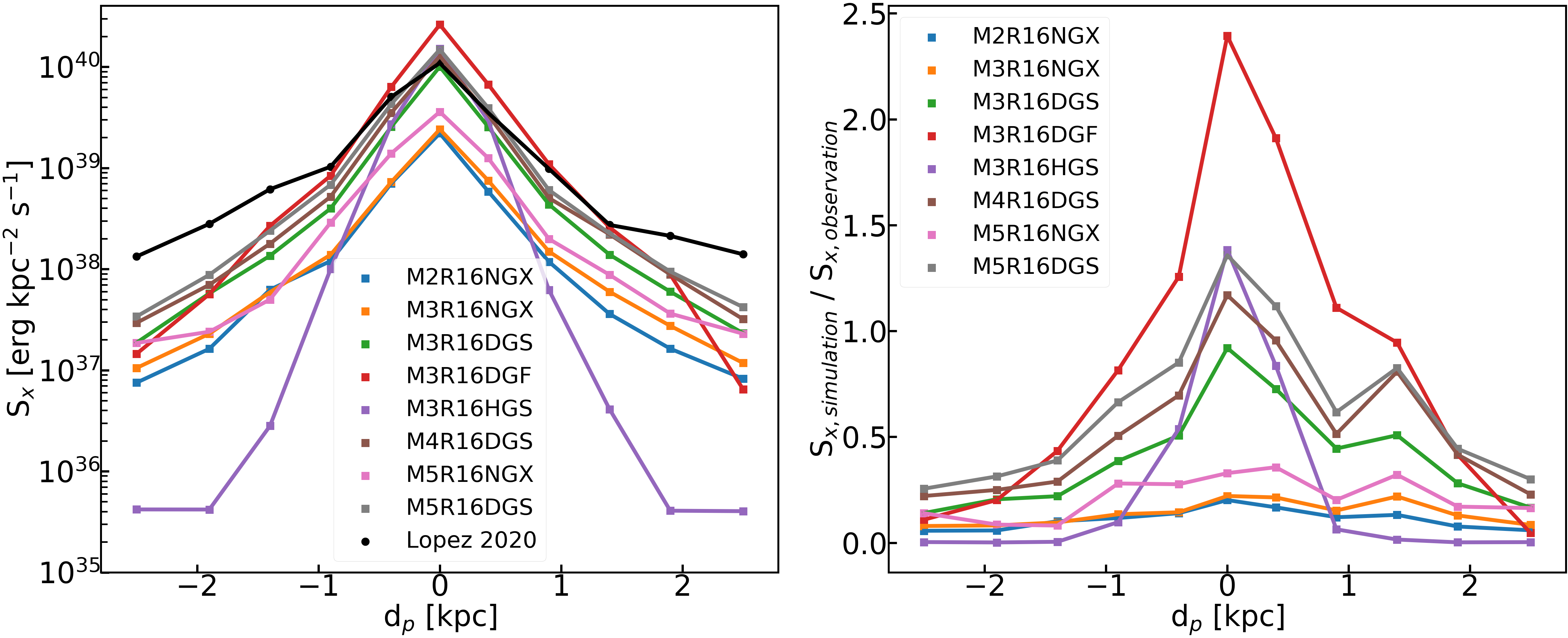}
    \caption{Broad-band (0.5 - 7.0 keV) X-ray surface brightness $S_{X}$ profile along the
M82 minor axis (left) and the ratio of simulation (30 Myr) to observation (right). The black dot indicates observation results in \citealt{lopez2020temperature}.}
    \label{fig:xray}
\end{figure*}

The corresponding X-ray luminosities from our simulations at $t=30$ Myr are listed in Table \ref{tab:Lx}. In simulations with gas return, the total thermal X-ray luminosity is comparable to observational values. In contrast, simulations without gas return produce only about 14 - 26 $\%$ of observations. Figure \ref{fig:xray_te} further illustrates the evolution of total thermal X-ray luminosity, $\rm{L_{X,tot}}$, over time in three representative simulations, M3R16NGX, M3RDGS and M3R16HGs. $\rm{L_{X,tot}}$ rises from near zero at $t=5$ Myr, reaching its peak between $t=25$ Myr and $t=35$ Myr. In the absence of gas return, the peak luminosity is approximately $\sim20\%-40\%$ of the observed value.While increasing the initial gas mass moderately enhances X-ray luminosity, this alone is insufficient to match the observations. However, including gas return from sink particles raises the maximum value to $\sim60\%-150\%$ of the observation. 

We also generated the edge-on surface X-ray brightness profile, $S_{x}$, for the hot wind, assuming an inclination angle of 77 degrees. Following the procedure in \cite{lopez2020temperature},  we select 11 regions with a projected distance $d_{\text{p}}<2.5$ kpc from the central plane along the minor axis. The left panel of Figure \ref{fig:xray} displays the $S_{x}$ profile along the minor axis at $t=30$ Myr from our simulations, compared with the observational results from \cite{lopez2020temperature} (black dot, where contributions from charge exchange and non-thermal power-law components have been subtracted). The right panel of Figure \ref{fig:xray} shows the ratio of $S_{x}$ from our simulations to that reported by \cite{lopez2020temperature}.

In simulations with gas return (except for M3R16HGS), $S_{x}$, can roughly match the observed profile within $r\lesssim 1.0$ kpc, though it remains lower than observed by a factor of 3-5 beyond 1.0 kpc.  In contrast, simulations without gas return show significantly lower $S_{x}$ values compared to observations across the entire region. Among the simulations, M3R16DGF, which features a faster gas return model, exhibits the highest surface brightness within $d_{\text{p}} < $ $\pm$ 1.5 kpc, exceeding $2\times 10^{40} $ $\rm erg$ $\rm kpc^{-2}$ $\rm s^{-1}$ in the central region. M3R16HGS closely matches observations at $d_{\text{p}}\leq 0.5\,$kpc, but its brightness drops sharply beyond this point, even falling below the levels seen in simulations without gas return. Increasing the initial total gas mass moderately enhances $S_{x}$ under the same gas return setup.

The differences in X-ray emission between the simulations arise from several factors. The X-ray emission from hot gas, calculated using the collisional ionization equilibrium model, depends on temperature, metallicity, and is proportional to the square of the gas density. Consequently, including gas return increases the density of gas hotter than $2\times 10^6$ K in the outflow (see Figure \ref{fig:outflowdata_nz}), leading to higher $S_{x}$ values compared to simulations without gas return, particularly in the inner region. However, the timing of the X-ray luminosity peak in simulations with gas return is influenced by multiple factors. Additionally, a higher $\rm{M_{gas,t}}$ generally enhances the metallicity, number density, and average temperature of the hot gas in the wind to a moderate degree, which can further elevate X-ray luminosity. 

\section{Discussion} \label{sec:discus}
\subsection{Impact of Resolution, Gas Metallicity, and Boost Factor}
\begin{figure*}[htbp]
    \centering
    \includegraphics[width=0.90\textwidth]{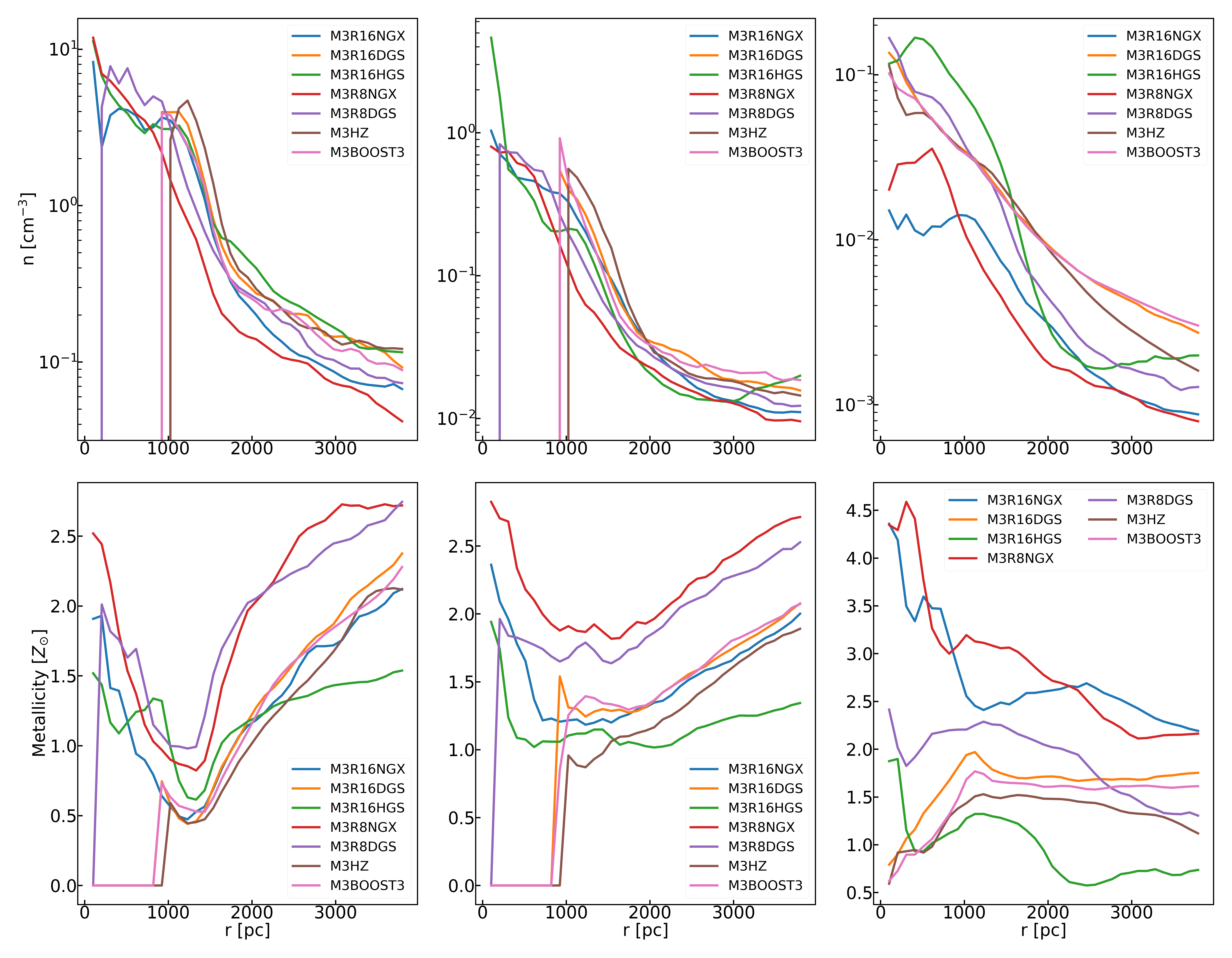}
    \caption{Radial distribution of volume weighted number density and mass-weighted metallicity for gases of three phases at 30 Myr in simulations with a resolution of 8 pc, high initial gas metallicity, and a boost factor of 3. The left, middle and right columns are the cool, warm and hot gas, respectively.}
    \label{fig:outflowdata_nz2}
\end{figure*}

\begin{figure*}[htbp]
    \centering
    \vspace{-0.35cm}
    \hspace{-1.2cm}
    \includegraphics[width=0.35\textwidth,trim=40 40 150 150, clip]{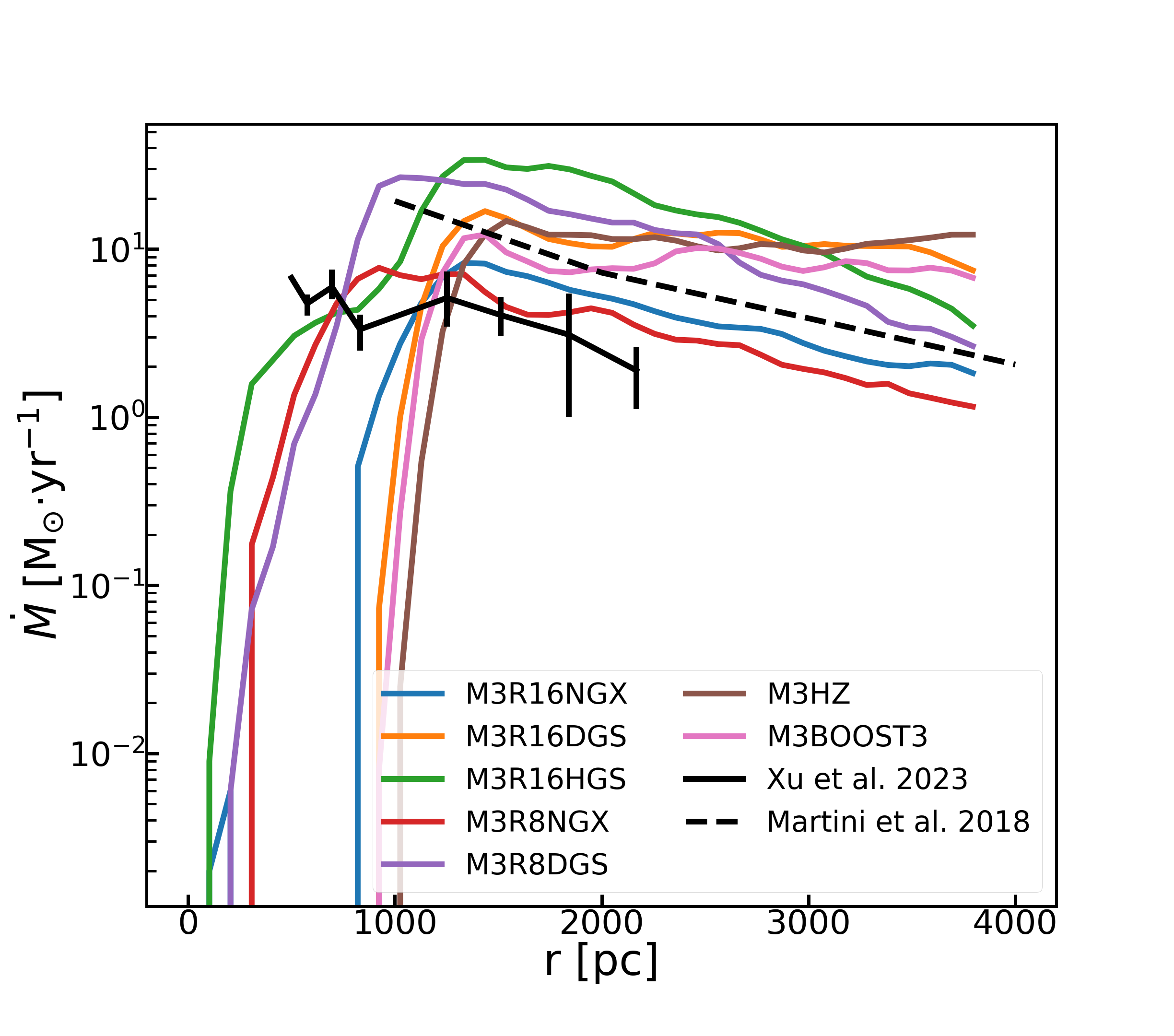}
    \includegraphics[width=0.35\textwidth,trim=40 40 150 150, clip]{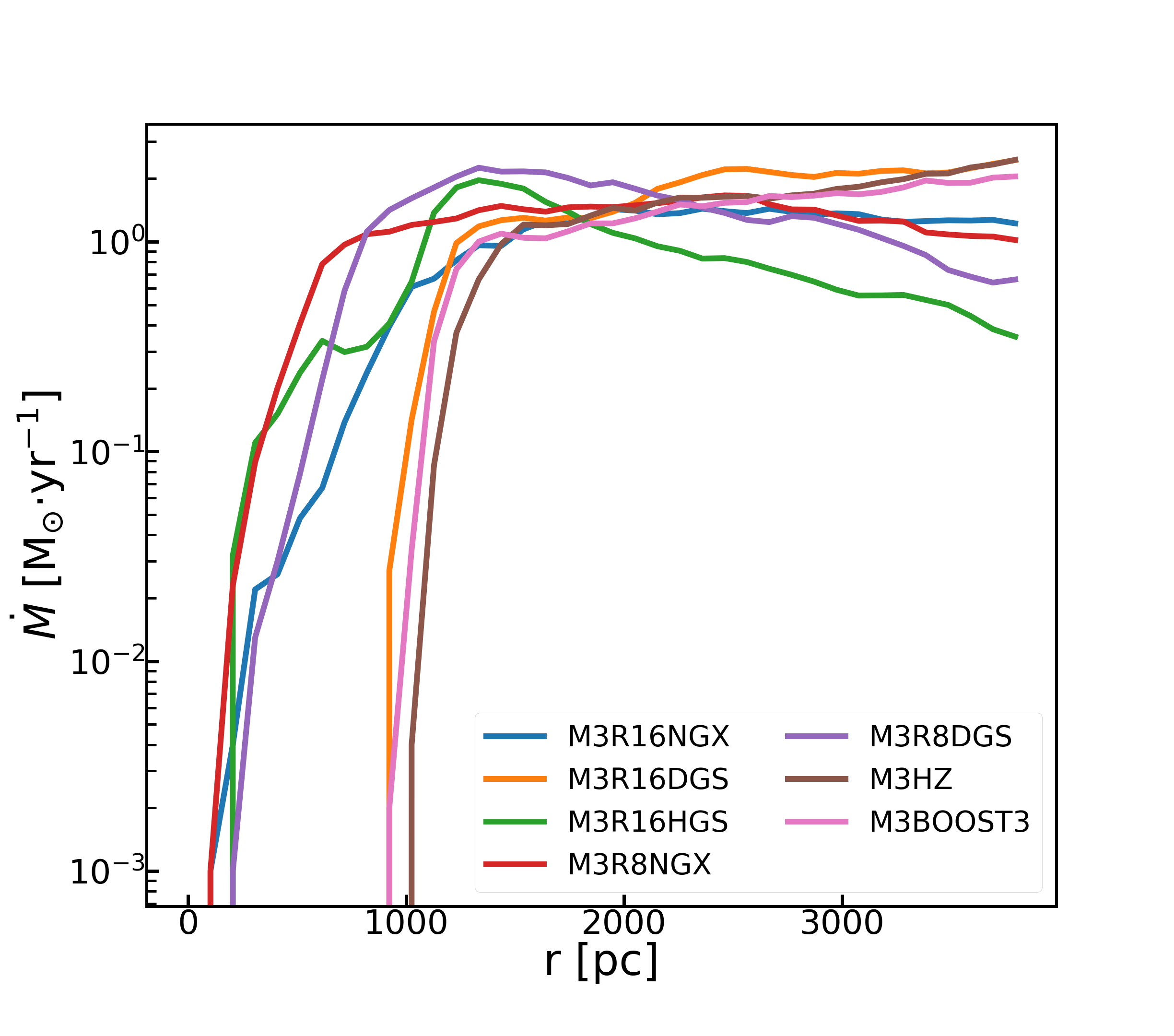}
    \includegraphics[width=0.35\textwidth,trim=40 40 150 150, clip]{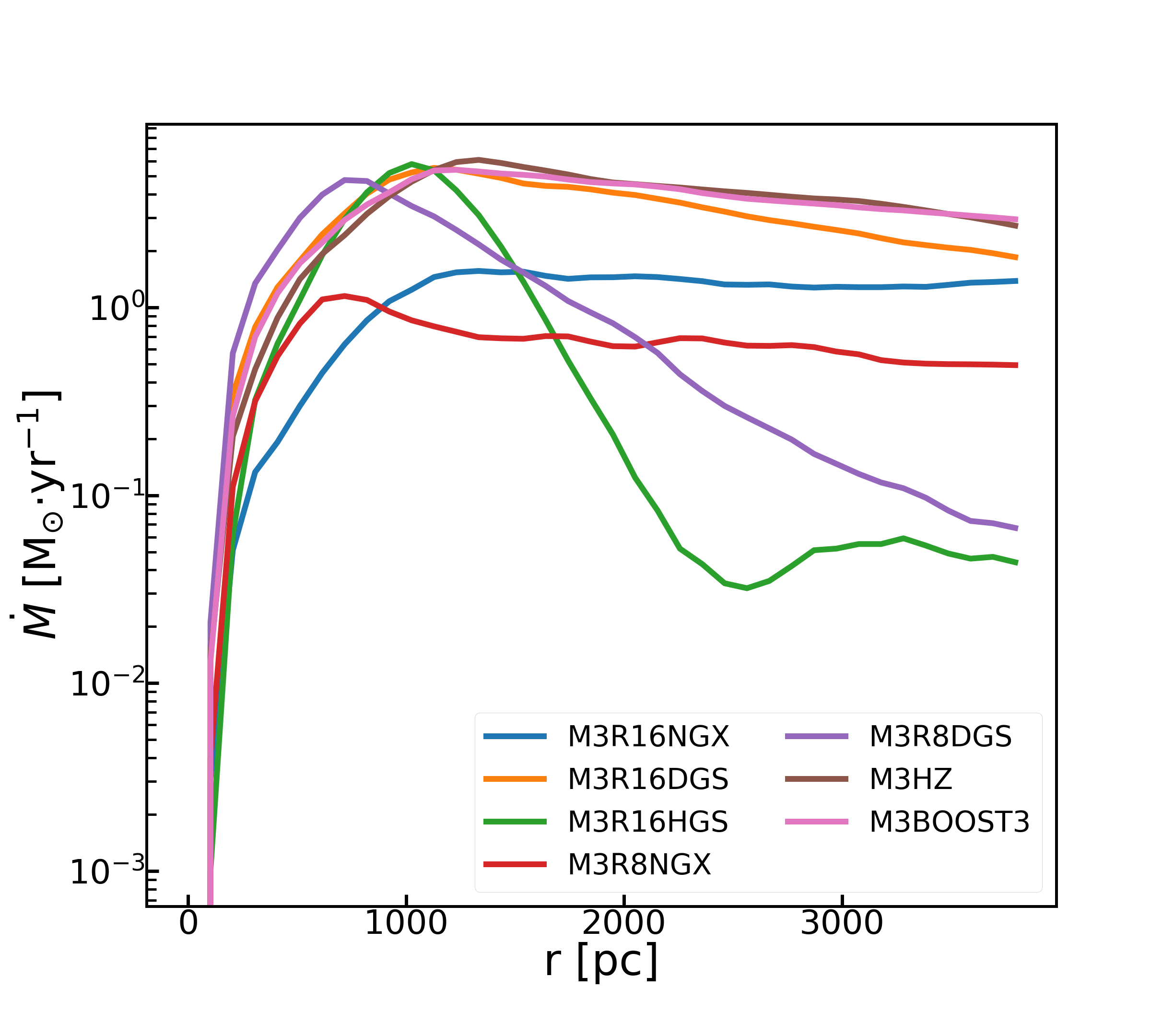}\\
    \vspace{-0.5cm}
    \hspace{-1.2cm}
    \includegraphics[width=0.35\textwidth,trim=40 40 150 150, clip]{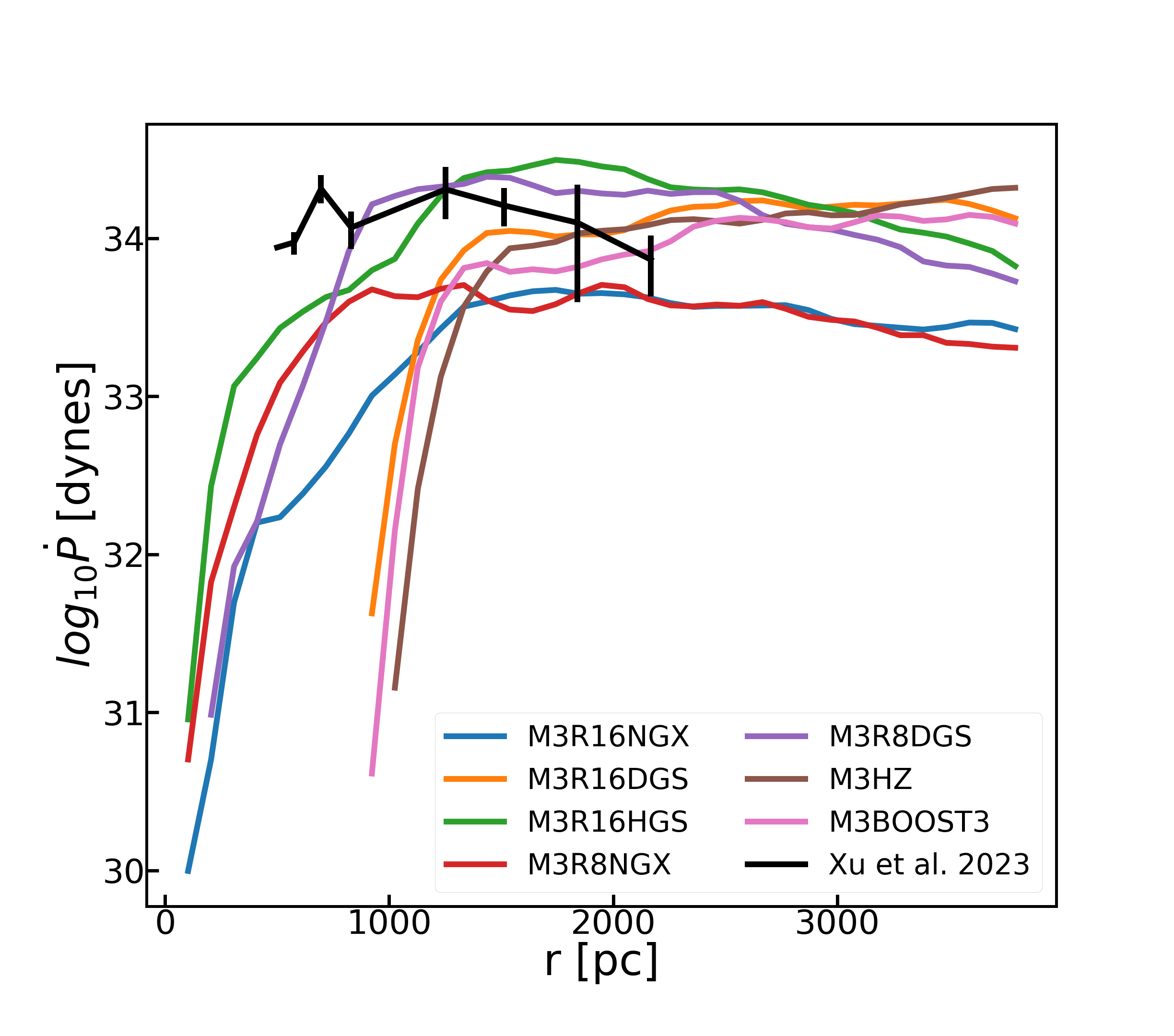}
     \includegraphics[width=0.35\textwidth,trim=40 40 150 150, clip]{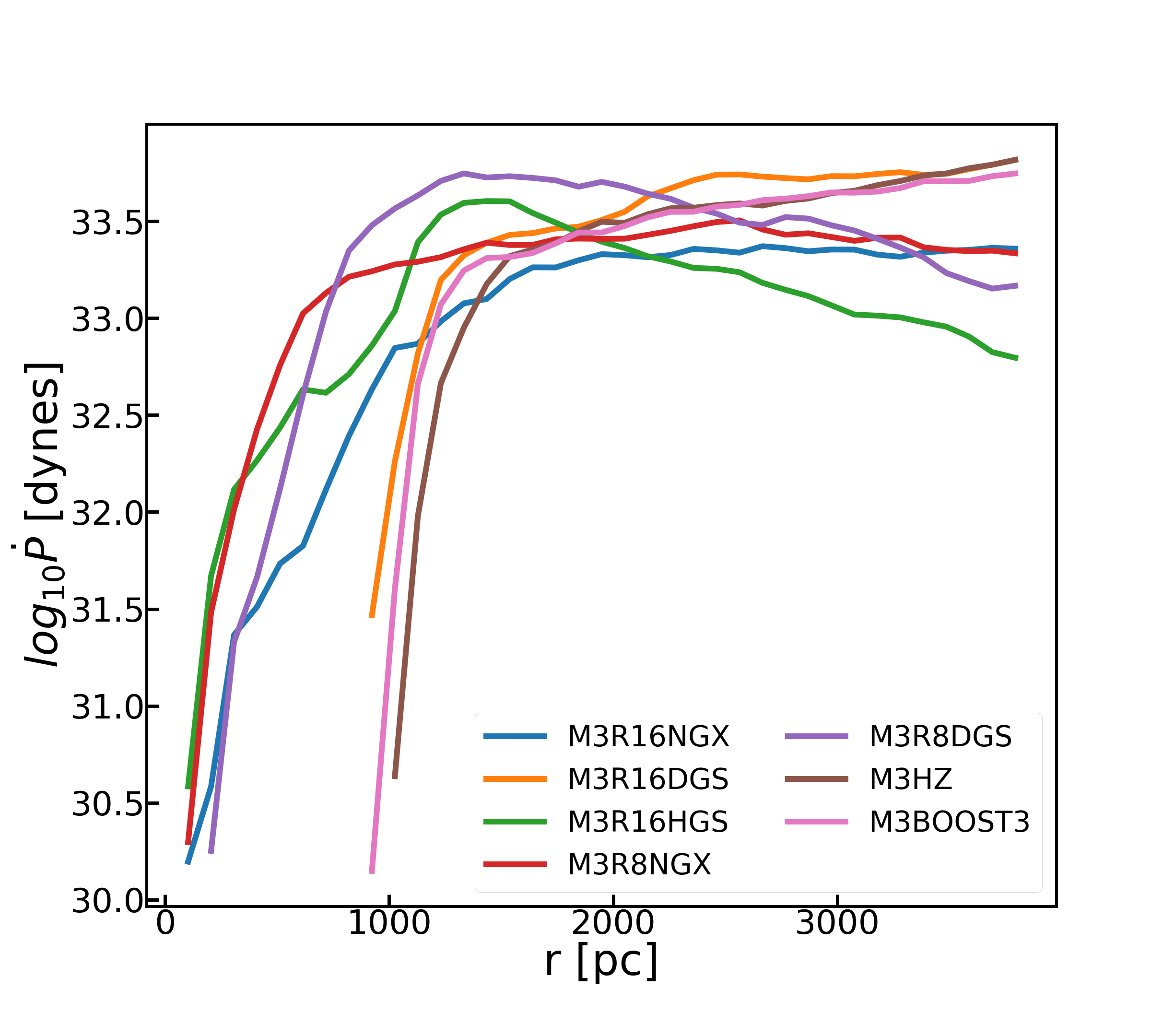}
    \includegraphics[width=0.35\textwidth,trim=40 40 150 150, clip]{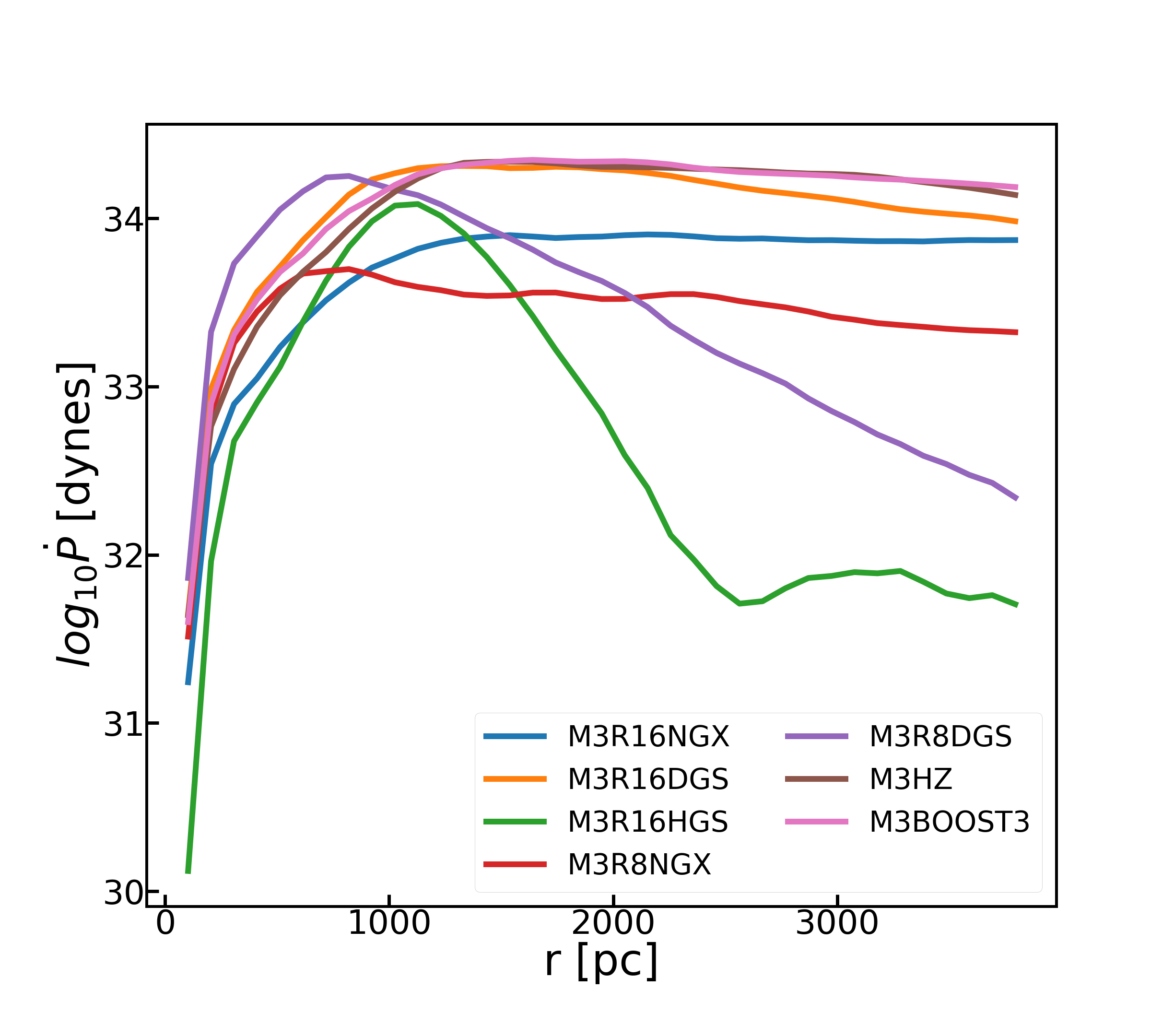}\\
    \vspace{-0.5cm}
    \hspace{-1.2cm}
    \includegraphics[width=0.35\textwidth,trim=40 40 150 150, clip]{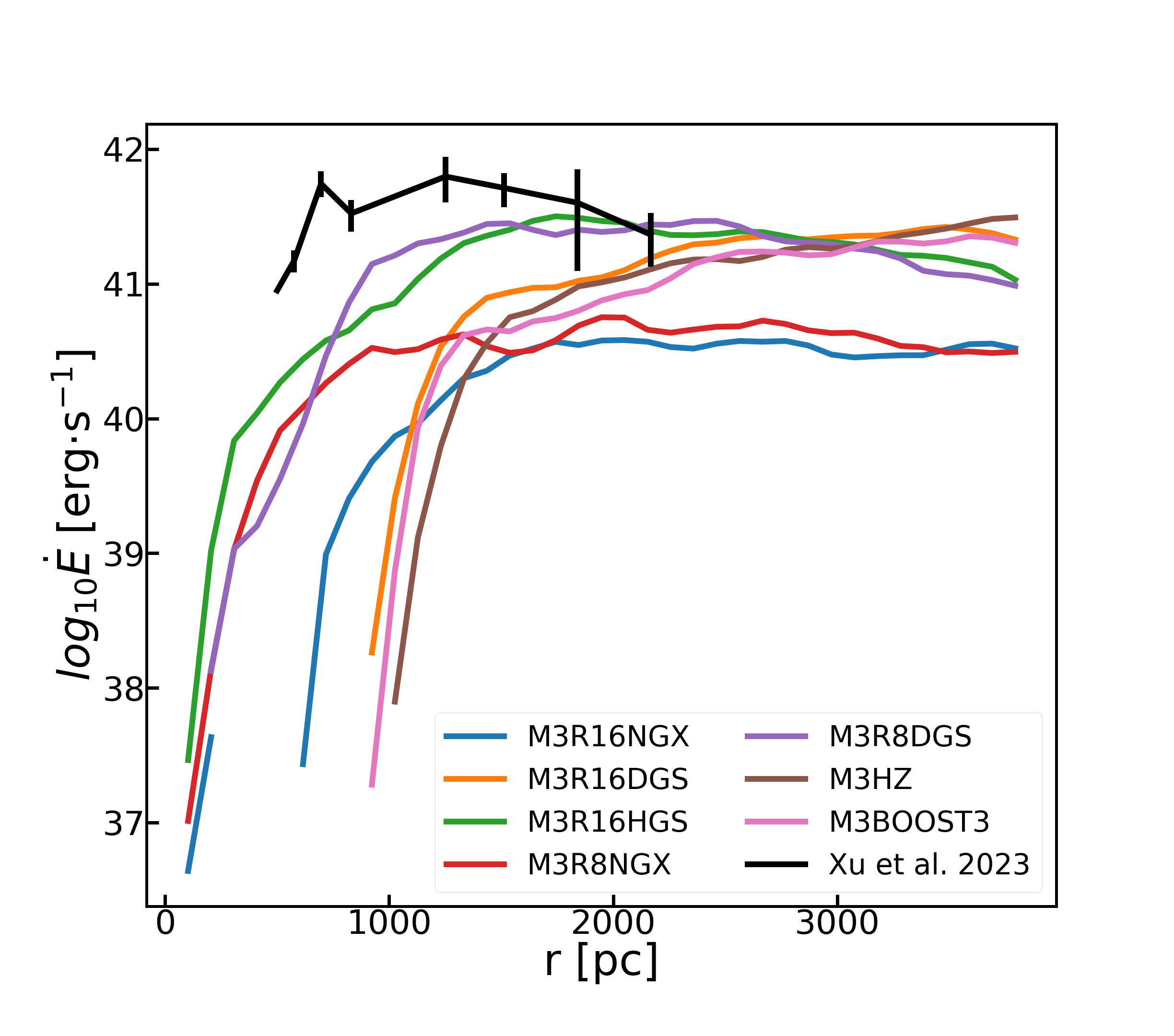}
    \includegraphics[width=0.35\textwidth,trim=40 40 150 150, clip]{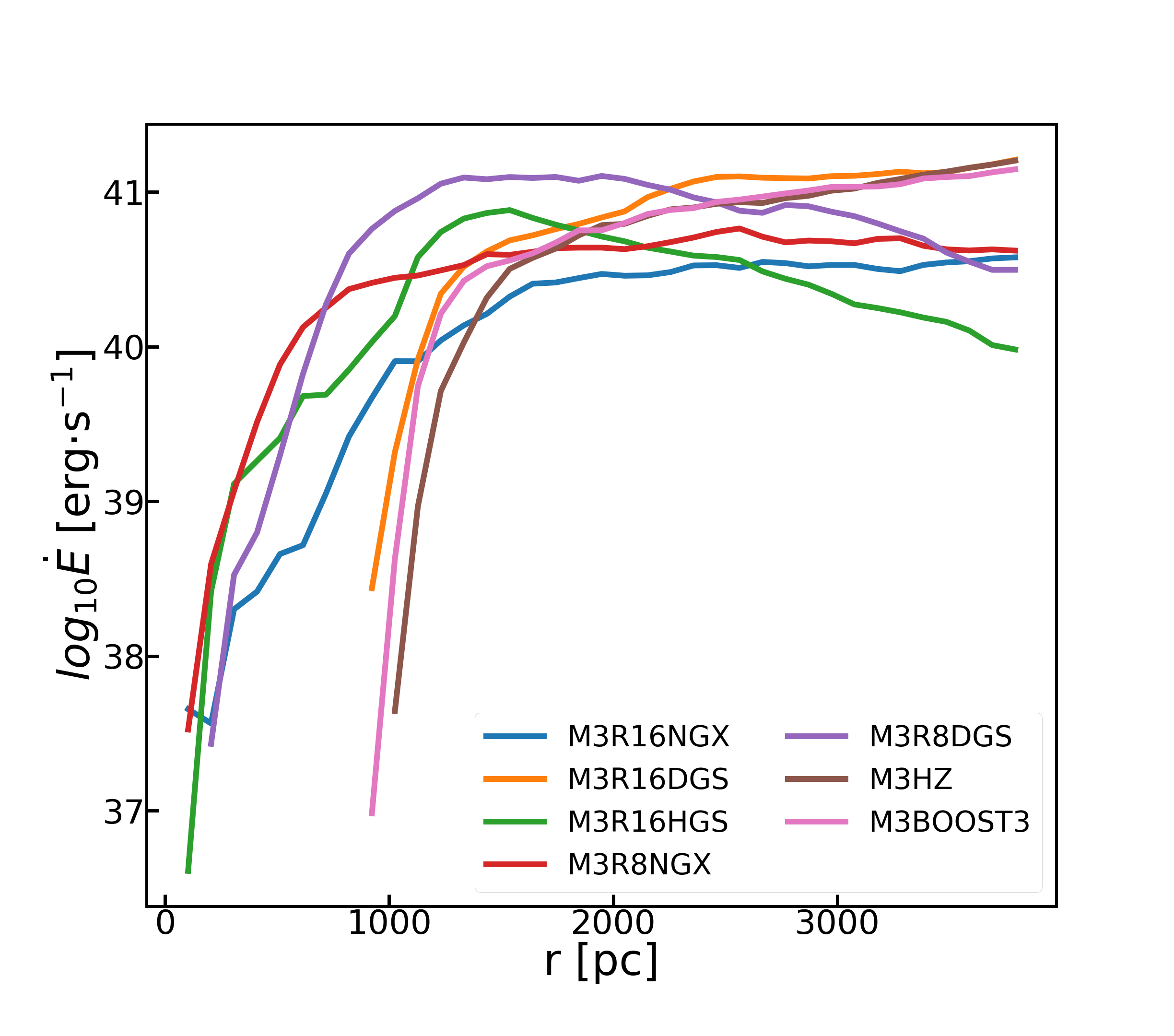}
    \includegraphics[width=0.35\textwidth,trim=40 40 150 150, clip]{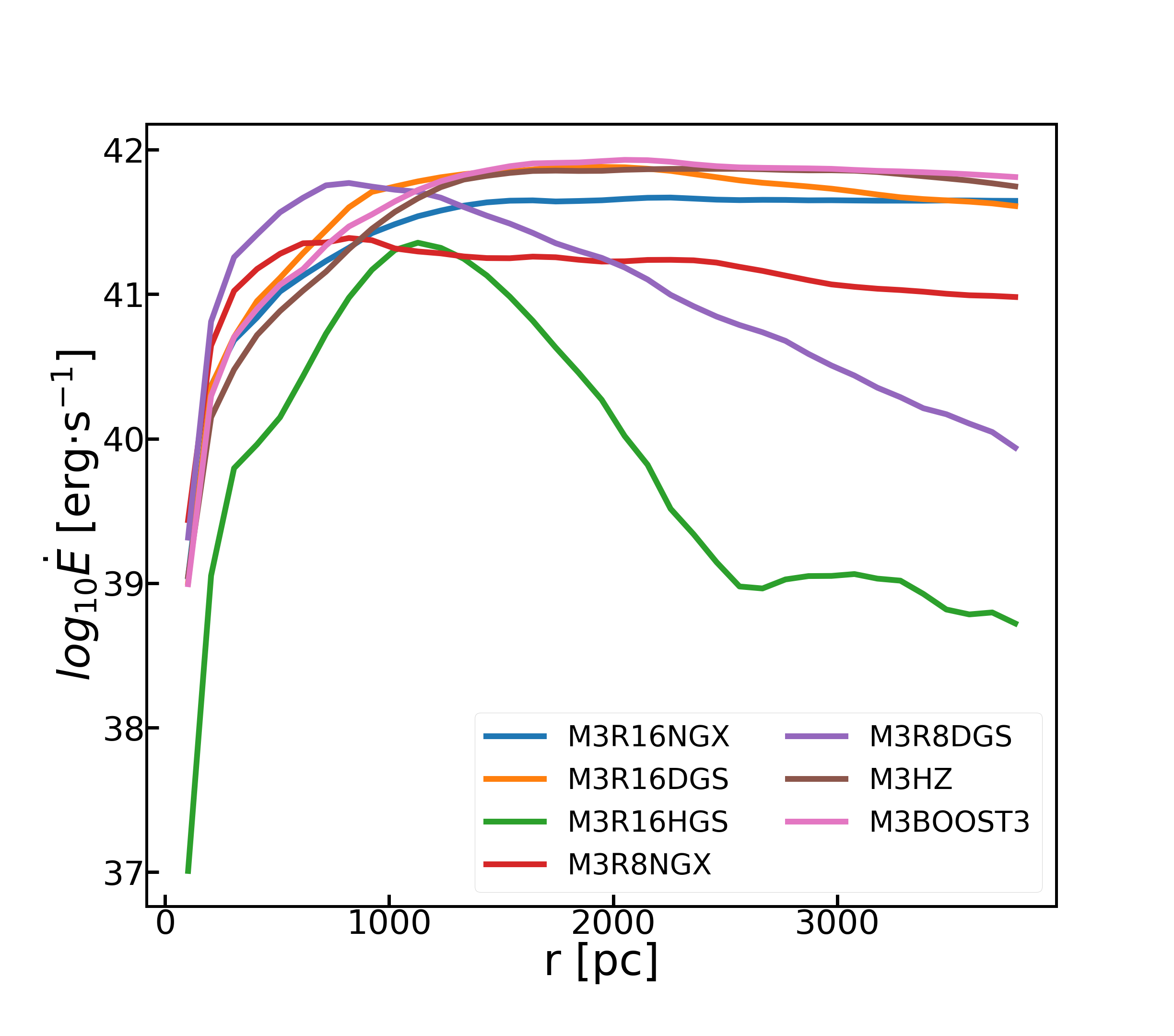}\\
    \vspace{-0.5cm}
    \hspace{-1.2cm}
    \includegraphics[width=0.35\textwidth,trim=40 40 150 150, clip]{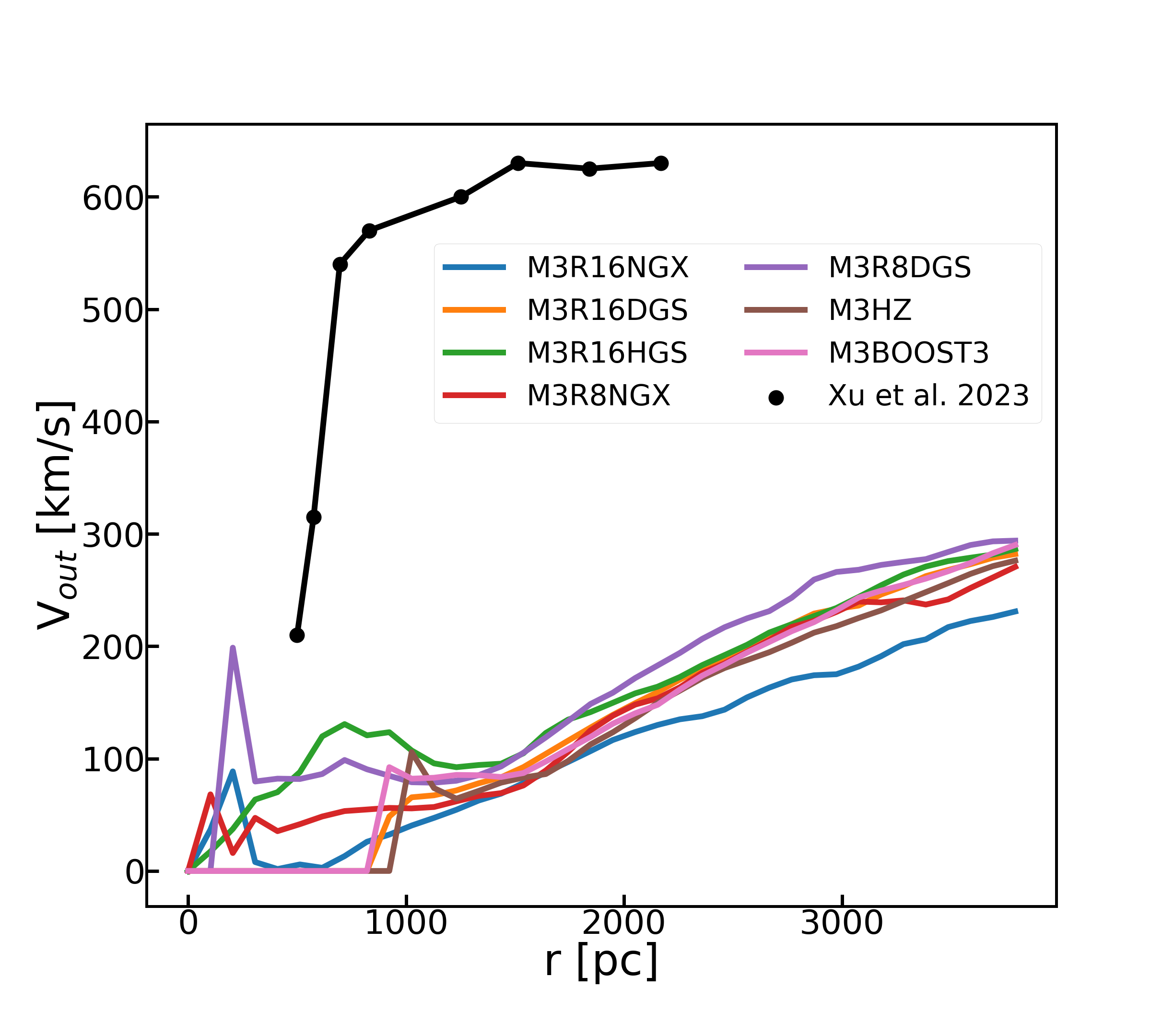}
    \includegraphics[width=0.35\textwidth,trim=40 40 150 150, clip]{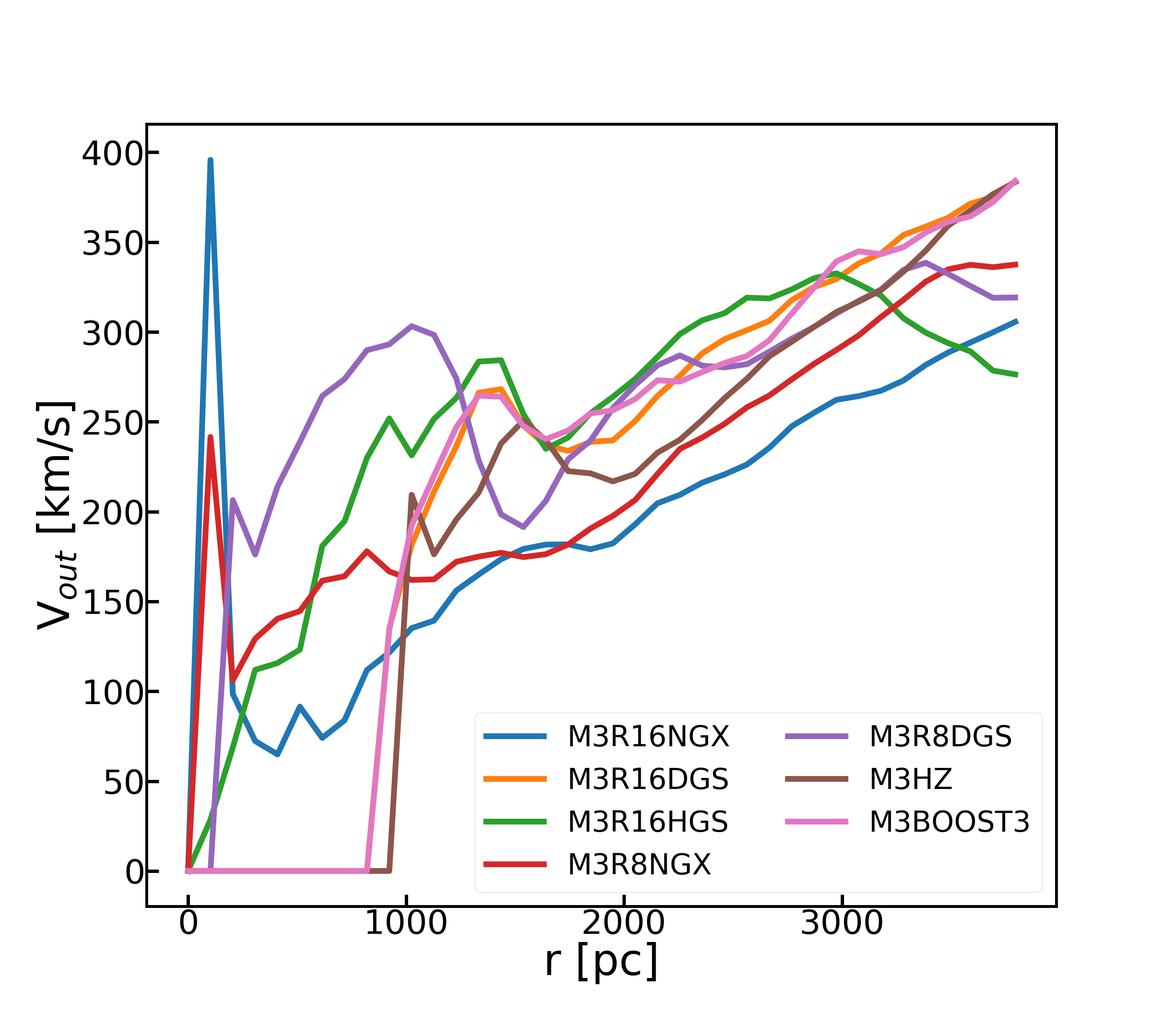}
    \includegraphics[width=0.35\textwidth,trim=40 40 150 150, clip]{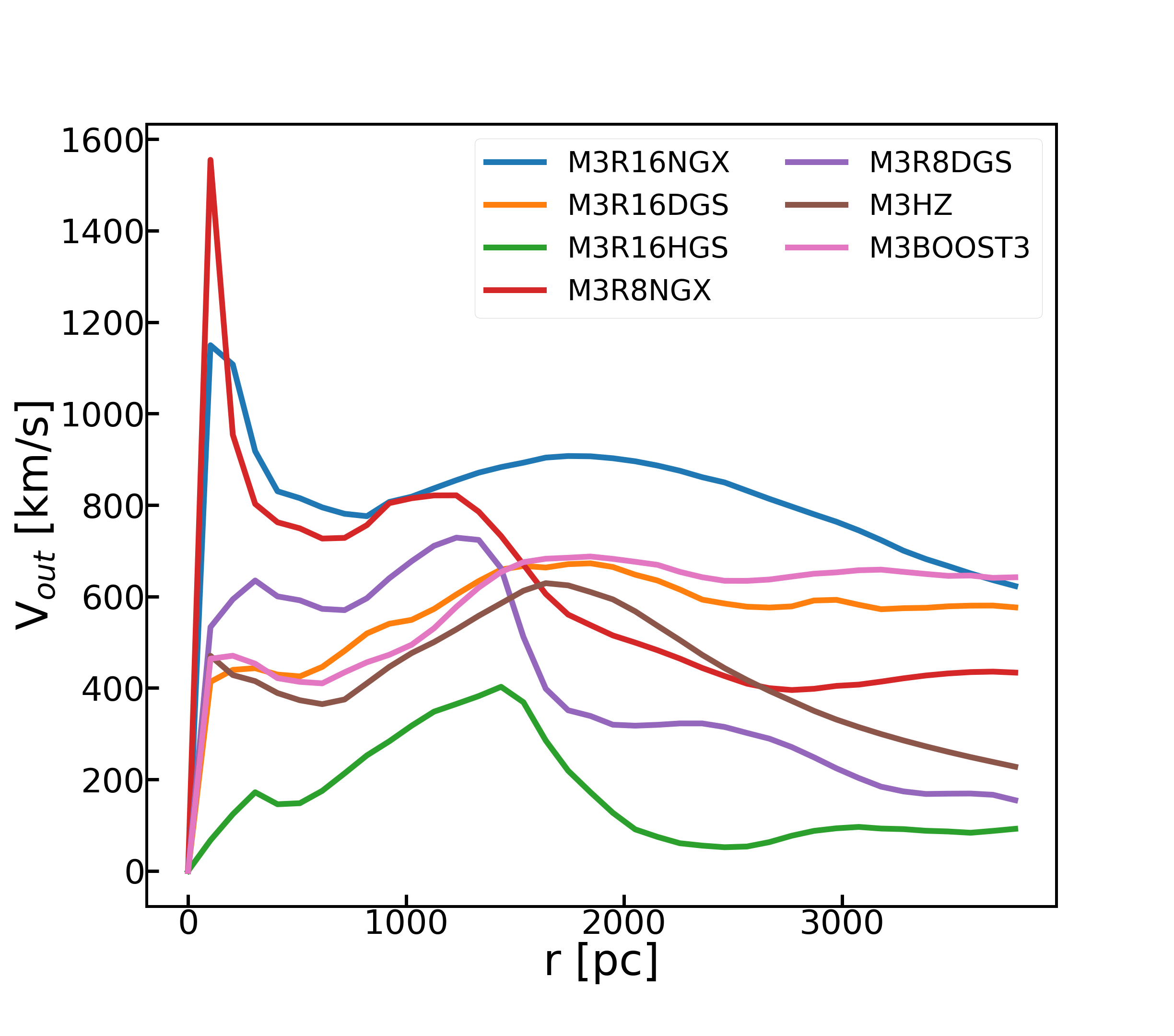}   
    \caption{Radial distribution of mass, momentum, energy outflow rates and outflow velocity for the cool (left), warm (middle) and hot (right) phases at 30 Myr in simulations with a resolution of 8 pc, high initial gas metallicity, and a boost factor of 3. The black line represents the outflow rates of gas with $T\sim 10^4$ K that inferred from observations.}
    \label{fig:outflowdata_vs1_4.1}
\end{figure*}

In Section 4, we investigated how gas return and the initial total gas mass influence outflow development in simulations with a 16 pc resolution. However, other factors may also affect the results. Here, we explore the effects of resolution, initial gas metallicity, and the boost factor in the supernova feedback model.  Figures \ref{fig:outflowdata_nz2} and \ref{fig:outflowdata_vs1_4.1} present the density and metallicity profiles, along with the outflow rates and velocities of cold, warm, and hot gas in simulations covering these additional factors.

When the resolution increases from 16 pc to 8 pc, the gas density across all three phases in the inner region ($r\lesssim 1.0$ kpc) of the outflow is enhanced. This occurs because less gas is accreted onto sink particles, allowing more gas to remain in the disc (see Section 3.3). Beyond $r>1$ kpc, the gas density decreases, as the outflow faces stronger resistance. Additionally, the metallicity of the outflow increases at all radial distances with higher resolution. The mass, momentum, and energy outflow rates for all three gas phases are also elevated at $r\lesssim$ 1 kpc, particularly within $r<0.5$ kpc, in the 8 pc resolution simulations (M3R8NGX and M3R8DGS) compared to their 16 pc resolution counterparts. This is primarily driven by the increased gas density, which also contributes to the higher outflow velocity of cool and warm gas within $r < $ 1 kpc in M3R8NGX and M3R8DGS. 

In most cases, the outflow rates in M3R8NGX and M3R8DGS peak around $r\sim 1$ kpc, then decline with increasing radius at varying slopes, eventually falling below the results from lower resolution simulations between $1-2.5$ kpc. The main reason is that, at $r\gtrsim 1$ kpc, the gas density in the outflow decreases as the resolution increases. However, in some cases, the momentum and energy outflow rates in the 8 pc resolution simulations are comparable to, or even exceed, those of the 16 pc simulations at larger radii, driven by the higher outflow velocity. For cool gas, the mass outflow rates in M3R8NGX and M3R8DGS can also match the observations from \cite{xu2023radial} and \cite{2018ApJ...856...61M} at $r\gtrsim 0.8$ kpc. The momentum outflow rate of cool gas in M3R8DGS approaches the values from \cite{xu2023radial} at $r > $ 0.8 kpc. However, the energy outflow rates in both 8 pc simulations remain lower than those in \cite{xu2023radial}, because the outflow velocity is still lower than the estimated value in \cite{xu2023radial} by a factor of 2. Additionally, increasing the resolution does not improve $\rm{V_{out,hot}}$ at $r>1.0$ kpc, and instead slightly widens the gap between simulation and observation.  

We conclude that simply increasing the resolution, while keeping other aspects of the simulation unchanged, provides only modest improvements to certain outflow properties and fails to resolve key issues. Notably, the starburst timescale remains moderately longer than observed, and the outflow velocities is slower than observations by approximately $\sim 30\%-60\%$. Two primary factors limit the expected benefits of higher resolution. First, as discussed in Section 3, simulations with higher resolution generate more CCSNe, but also retain more gas within the disk, rather than allowing it to be accreted by sink particles. This effect is amplified in simulations with the gas return process, where the return fraction depends on the sink particles' gas mass. In higher resolution simulations, sink particles generally accumulate less gas, leading to a higher gas return fraction. As a result, more gas remains near star particles and within the nuclear region of the disk, preventing the outflow velocity from increasing substantially to align with observations. Additionally, outflow rates in the outer regions are slightly suppressed, further limiting the overall improvement from increased resolution.  

On the other hand, we find that changing the boost factor in the momentum injection of the SN feedback model from 10 to 3 has a very minor impact on the outflow rates and velocities across all three gas phases (as shown by the pink and orange lines in Figure \ref{fig:outflowdata_vs1_4.1}). Similarly, increasing the initial gas metallicity from 0.02 to $0.10$ $\rm{Z_{\odot}}$ has a minor effect on on both the outflow rates and velocities. 

\begin{figure*}[htbp]
    \centering
    \includegraphics[width=0.90\textwidth]{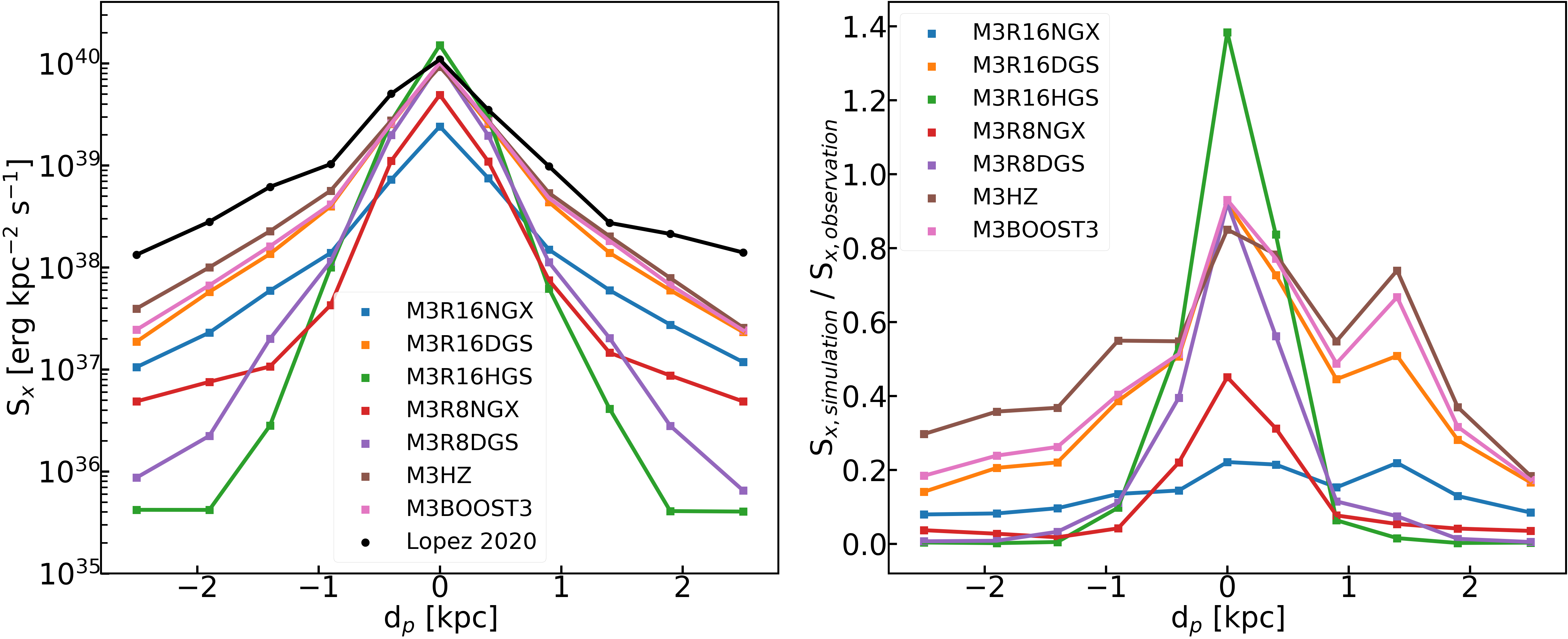}
    \caption{Broad-band (0.5 - 7.0 keV) X-ray surface brightness $S_{X}$ profile along the
M82 minor axis(left) in simulations with a resolution of 8 pc, high initial gas metallicity, and a boost factor of 3, and the ratio of simulation (30 Myr) to observation in \citealt{lopez2020temperature}.}
    \label{fig:xray_4.1}
\end{figure*}

We further explore the effects of these factors on the X-ray emission and surface brightness profile. The profiles $S(x)$ for simulations M3R8NGX, M3RDGS, M3Hz, M3BOOST3 are shown in Figure \ref{fig:xray_4.1}, with their total X-ray luminosities listed in Table \ref{tab:Lx}. When gas return from sink particles is excluded, increasing the resolution results in higher X-ray surface brightness in the central region but lower values beyond $r > $ 1 kpc. In contrast, when gas return is included, $S_{x}$ in M3R8DGS is lower than in M3R16DGS, and it declines steeply from $r=0$ to 1 kpc, primarily due to the rapid drop in hot gas density. At a fixed resolution of 16 pc, increasing the initial gas metallicity moderately enhances X-ray emission at $r>0.5 $ kpc, as the emission increases with metallicity. A boost factor of 3 results in a slight increase in X-ray emission. 

\subsection{Comparison with Simulations and Models in the Literature}

Over the past two decades, several two- and three-dimensional simulations of M82-like starburst galaxies have been conducted  (e.g., \citealt{Strickland2000}; \citealt{2008ApJ...674..157C}; \citealt{melioli2013evolution}; \citealt{schneider2020physical}; \citealt{schneider2024cgols}). These simulations have successfully produced outflows that share some similarities with the observed wind in M82 in certain specific properties. Notably, the simulations by \cite{schneider2020physical}; \cite{schneider2024cgols} were able to drive multiphase outflows on scales of 10 kpc with velocities comparable to the M82 wind, although the mass outflow rates were somewhat lower than observed in M82. Our simulations generate 4 kpc-scale outflows that closely match key properties of the M82 wind, particularly in terms of the mass outflow rates for both cool and hot phases, as well as the X-ray emission. There are several important revisions and improvements in our study that enhance these results.

A key improvement in our simulation is the self-consistent solution of the starburst in the nuclear region and the associated supernova feedback. The collapse of gas clumps and subsequent star formation are explicitly modeled using a built-in sink particle module, incorporating advances from the literature (e.g., \citealt{1995MNRAS.277..362B,2004ApJ...611..399K,Federrath2010,Gong2012,Howard2016}). Star evolution and feedback from massive stars are implemented based on the \cite{Kroupa01} IMF and the stellar evolution model from \cite{Sukhbold2015}, as detailed in Section \ref{sec:sn}. Another key feature of our work is the incorporation of gas return from giant molecular clouds (GMCs) back into the interstellar medium (ISM) in our simulations. Furthermore, our simulation resolves the evolution of gas metallicity, which is integrated into the radiative cooling process, accounting for its dependence on metallicity.

We would like to provide further details on the starburst and associated supernova energy feedback to highlight the effectiveness of the star formation module in our work. Previous studies often employed various assumptions about the intensity, time evolution, and spatial distribution of starburst activity, as well as the properties of the SN feedback from recently formed star clusters. For example, \cite{Strickland2000} considered both an instantaneous starburst and a burst with more complex SFH spread over 10 Myr, with a total injected mechanical energy of $10^{57}\, $ erg and an average injection rate of $3\times 10^{42}\, \rm{erg/s}$. 
\cite{2008ApJ...674..157C} used a fixed energy and mass injection rate of $10^{42}$ $\rm erg/s$ and $1$ $\rm M_{\odot}/yr$, respectively. The distribution of injected energy and mass in \cite{2008ApJ...674..157C} was clumpy based on the initial gas density profile, which roughly approximated the effect of star clusters.


\cite{melioli2013evolution} employed a fixed energy injection rate of $10^{42}$ $\rm{erg/s}$, varying the number of SSCs and the duration of the star formation process. In contrast, \cite{schneider2020physical} adopted a variable SN injection rate aligned with an idealized starburst model, where stars form at a fixed rate of $20$ $\rm M_{\odot}/yr$ between 5 to 35 Myr, and $5$ $\rm M_{\odot}/yr$ from 35 to 70 Myr. In their model, star clusters are primarily concentrated within the central disk region ($r_{as}<1 $ $\rm kpc$) , with each cluster having a total star mass of $10^{7}$ $\rm M_{\odot}$. The total feedback energy is about $\sim 6 \times 10^{57}$ erg in \cite{schneider2020physical}, with an average energy injection rate of around $6.0 \times 10^{42}$ $\rm{erg/s}$ between 5 and 35 Myr. \cite{schneider2024cgols} further revise the assumed SFH into a distributed burst model,considering a broader star cluster distribution and variable cluster masses, with an average energy injection rate of  $\sim 6 \times 10^{42}$ $\rm{erg/s}$ over 40 Myr.

By $t=30 $ Myr, when the outflows are well developed, the total stellar mass formed in the recent starburst in our simulations ranges from $1.62\times 10^8 $ to $3.34 \times 10^8$ $\rm M_{\odot}$, which is in good agreement with observations. The corresponding total injected energy spans from $1.14\times 10^{57}$ to $2.4\times 10^{57}$ erg, yielding an average energy injection rate between $1.21 \times 10^{42}$ and $2.54 \times 10^{42}$ erg/s. While this rate is comparable to values used in most prior studies, it is around $25\%$ to $40\%$ of that in \cite{schneider2020physical} and \cite{schneider2024cgols}. The outflow velocities in these studies are similar to the observational estimates for M82, and exceed our results by about $20\%-50\%$ for the hot phase and $20\%-60\%$ for the cool phase, which probably have benefited from their higher energy injection. However, despite the slower velocities, the outflow rates in our simulations without gas return are comparable to those in \cite{schneider2020physical} and \cite{schneider2024cgols}, while the overall density profiles are similar. The cool outflow in our simulations without gas return likely occupies a larger volume, which compensates for the slower velocity, leading to a comparable mass outflow rate. The inclusion of gas return from sink particles further boosts the density of cool and hot gas in the wind, enhancing the mass outflow rate. 

In contrast to the single phase pure adiabatic expansion model (\citealt{1985Natur.317...44C}), the profiles and outflow rates in our simulations are influenced by the presence of gravity, radiative cooling, and multiphase interactions. Many characteristics of the outflows in our simulations without gas return resemble the high hot phase density scenario of the multiphase outflow model proposed by \cite{2022ApJ...924...82F}. The properties of the outflow in our simulations with gas return from sink particles (giant molecular clouds), including the density profile and mass outflow rate, could be partially explained by the model in \cite{2022ApJ...924...82F}, when the initial mass loading factors of the hot and cold phases are larger than $\sim 0.5$. As \cite{schneider2020physical} and \cite{2022ApJ...924...82F} have demonstrated, the mixing and interactions between the cool and hot phases play a critical role in shaping the multiphase outflows seen in our simulations.

\subsection{Future Improvement}
Our simulations successfully reproduce a starburst event with a total stellar mass closely matching estimates for M82, and drive a galactic-scale outflow that shares several key properties with the observed wind in M82, including morphology, multiphase structure, mass outflow rates of cool and hot phases, and X-ray emission, particularly when incorporating gas return from sink particles (gas clumps harboring newly formed stars). However, several shortcomings and caveats remain. First, the outflow ·velocities of cool and hot gases are slower than the observational estimates by approximately $30\%$-$60\%$ when gas return is implemented. Second, the duration of the starburst in our simulations spans about 20-25 Myr, longer than the estimated age of nuclear star clusters in M82, typically around 10 Myr according to previous studies, although some observational features suggest that the nuclear burst in M82 may have been active for longer (\citealt{2009ApJ...697.2030S}). Third, certain critical modules such as sink particle creation and accretion, star formation thresholds, feedback injection, and gas return processes are resolution-sensitive, leading to non-converging results with increased resolution. Additionally, aspects of the gas return process from sink particles, including return fractions and timescales, are somewhat arbitrarily defined. Furthermore, our results rely on specific models of stellar feedback. Variations in the implementation of feedback mechanisms, such as stellar winds and supernovae, could influence the outcomes to some extent. Additionally, certain processes, such as cosmic rays, are not included in our simulations.

In the near future, we plan to explore strategies to reduce the resolution dependence of the star formation model more effectively. Additionally, we aim to implement a more sophisticated model for gas return from sink particles (destruction of star forming GMCs). Furthermore, since the recent starburst in M82 is thought to have been triggered by its interaction with M81, possibly driven by gas inflow along M82’s bar, we will try to incorporate these factors into our simulations. This may shorten the starburst duration, concentrate star formation more, and accelerate the outflow velocities. At the same time, nonnegligible uncertainties remain in several properties of M82 and its wind that inferred from observations, such as the total gas mass, the total stellar mass formed during the recent starburst, the current star formation rate, and the mass outflow rate of cold gas. More accurate measurements from ongoing and future observations will be crucial in refining our models and improving the accuracy of our results.

\section{Summary} 
\label{sec:summary}

In this study, we performed a suite of three-dimensional hydrodynamical simulations with a resolution of 8–16 parsecs to explore the development of the multiphase galactic wind in M82. The mass model and initial gas disk were tailored to align with the latest observational data. The star burst and the associated feedback processes were resolved self-consistently through a sink particle module, with core-collapse supernovae serving as the primary energy source. These were implemented using a combined kinetic and thermal injection method based on \cite{kim2017three}. We examined the influence of several factors on the evolution and properties of the outflows, including gas return from the destruction of star-forming gas clumps to the ISM, the initial mass of the gas disk (representing varying initial gas distributions in the nuclear region), simulation resolution, and the initial gas metallicity. Our findings are summarized as follows

\begin{enumerate}
\item Our simulations can generate a starburst with a total stellar mass between 1.62 and 3.34 $\times 10^8$ $\rm{M_{\odot}}$ over $\sim25$ Myr, with the majority of stars forming within the central 1000 pc. The total SN energy injected in the simulations ranges from $1.14\times 10^{57}$ to $2.4\times 10^{57}$ $\rm{erg}$, corresponding to an average energy injection rate of $1.21 \times 10^{42}$ to $2.54 \times 10^{42}$ erg/s. These values align well with observations. However, the duration of the starburst in our simulations, approximately 25 Myr, is longer than the $\sim10$ Myr burst period suggested by several earlier studies. 

\item Supernova (SN) feedback heats a portion of the cool gas in the starburst region, transforming it into warm and hot phases. This drives a multiphase galactic outflow that mirrors the winds in M82 in terms of morphology, mass outflow rate, and X-ray luminosity, particularly when gas return from sink particles is included. In simulations with gas return, the peak total mass outflow rate ranges from 20 to 40 $\rm{M_{\odot}/yr}$ at $r\sim1.5\,$ kpc and from 6 to 12 $\rm{M_{\odot}/yr}$ at $r\sim4.0\,$ kpc, with hot gas contributing about $20\%$. The corresponding mass loading factor is 6 to 12 at $r\sim1.5$ kpc and 2 to 4 at $r\sim4.0$ kpc. The maximum X-ray emission luminosity reaches $1.5-3.5 \times 10^{40}\rm{erg/s}$. Without gas return, the mass outflow rate and X-ray emission decrease by approximately $\sim 40\%-60\%$ and $\sim 60\%-70\%$, respectively. However, the outflow velocities in our simulations are $\sim 20\%-60\%$ slower than than those inferred from observations. In simulations with (without) gas return, the typical outflow velocities are 100 - 300 (100 - 200) for cool gas, 200 - 500 (100 - 300) for warm gas, and 600 - 800 (800 - 1400) km/s for hot gas. In our simulations, most of the cool and warm gas in the wind eventually falls back onto the disc. 

\item Gas return from the destruction of star-forming clumps to the interstellar medium (ISM) significantly influences the properties of outflows. It reduces the opening angle and slows down the wind while increasing the density and outflow rates of the cool, warm, and hot gas phases, and enhancing X-ray emission. The initial mass of the gas disk, representing the initial gas concentration in the nuclear region, has a moderate effect on the development of outflows. In contrast, the initial gas metallicity and the boost factor in the momentum feedback model have only minor effects on the wind. 

\item The wind properties in our simulations cannot be adequately described by the pure adiabatic expansion model of a single-phase wind (\citealt{1985Natur.317...44C}). Instead, the multiphase model proposed by \cite{2022ApJ...924...82F} particularly the high hot-phase density scenario, explains many of the wind characteristics in our simulations without gas return. When gas return is included, the model from \cite{2022ApJ...924...82F} with initial mass loading factors for the hot and cold phases greater than $\sim0.5$, may partially account for the wind properties. Our findings support the idea that interactions between the cool and hot phases play a key role in shaping the structure of the multiphase outflow driven by starbursts.

\item Our current star formation and gas return models face challenges with convergence as resolution increases, highlighting the need for further refinement. Additionally, improvements are necessary to reduce the starburst duration and increase outflow velocities in our simulations. On the observational side, several uncertainties regarding M82 and its wind properties require clarification in future studies to better constrain the physical mechanisms of galactic winds by starbursts.
\end{enumerate}

\acknowledgments
We thank the anonymous referee for her/his very useful comments and suggestions that improved the manuscript. We thank the helpful discussions with Antonios, Katsianis, Wen-Shen, Hong, and Cheng, Li. This work is supported by the National SKA Program of China (Grant Nos. 2022SKA0110200 and 2022SKA0110202), by the National Natural Science Foundation of China (NFSC) through grants 11733010 and 12173102, and by the China Manned Space Program through its Space Application System. The hydrodynamic simulation was performed on the Tianhe-II supercomputer and the HPC facility of the School of Physics and Astronomy, Sun Yat-Sen University.
%




\vspace{15mm}


\appendix
\section{Supernovae feedback}
\label{sec:sn_feedback}
Here, we provide a detailed description of the supernova feedback module in our simulations. For each star particle (representing a star cluster), we define a spherical feedback region with a radius of three grid cells, $r_{snr}=3\Delta X$,  where $\Delta X$ is the size of a nearby grid cell. This region serves as the supernova remnant (SNR) deposit zone. Based on the properties of the gas within this zone, we apply three different prescriptions for handling supernova feedback. At each time step, we calculate the total mass $\sum M_{ijk}$ and volume $\sum V_{ijk}$ of the grid cells within the deposit zone, where $ijk$ are the indices of the cells. We then identify any star particle that has triggered supernova events and calculate the ejected mass $M_{ej}$, energy $\Delta \rm{E_{sn}}$ and metal mass $M_{metal,ej}$, as outlined earlier in Section 2.2. The mean density of the SNR region is computed as $\rho_{snr}=[\sum M_{ijk} + M_{ej}] / \sum V_{ijk}$. Following \cite{Kim_2015}, the mass at which shell formation occurs after a supernova event is given by $M_{sf} = 1679M_{\odot}(n_{\text H}$/cm$^{-3})$$^{-0.26}$, where the hydrogen number density $n_{\text H} = \rho_{snr}/(\mu_{\text H}m_{\text H})$, with $\mu_{\text H} = 1.427$. Finally, we compute the ratio of the initial SNR mass to the shell formation mass, $R_{M} = M_{snr}/M_{sf}$, and determine how the supernova feedback is handled based on this ratio.

For a given star particle, if $R_{M}>1$, it indicates that the Sedov-Taylor phase of supernova events cannot be fully resolved. In this case, we inject momentum into the grid cells within the feedback deposit zone, with the total final radial momentum given by $p_{snr} = 2.8\times 10^{5} \rm{M_{\odot}}\,$km s$^{-1}$ ($n_{H}$/cm$^{-3})$$^{-0.17} \cdot f_{boost}$, where $f_{boost}$ is the boost factor accounting for clustered supernovae (\citealt{Gentry17}). The momentum distribution to the individual grid cell is determined by mass-weighting, using ($f_{ijk} = \rho_{ijk} \times V_{ijk} / \sum M_{ijk}$), where 
ijk represents the cell indices. The momentum increment for each grid cell in the x, y, z directions is then calculated as $\Delta \overset{\to }p_{ijk} = f_{ijk} \times p_{snr} \times \overset{\to }{n}$, where $\overset{\to }{n}$ is is the unit vector pointing from the star particle to the grid cell center. The corresponding maximum expected energy increment is $\Delta e_{ijk} = f_{ijk} \times \Delta \rm{E_{sn}}$. To ensure both momentum and energy conservation, we apply the method from \cite{Hopkins18}. After updating the momentum for each grid cell, we calculate the change in kinetic energy $\Delta E_{k,ijk}$. If $\Delta E_{k,ijk}$ exceeds $\Delta e_{ijk}$, a correction factor $\alpha=\sqrt{(E_{k0,ijk} + \Delta e_{ijk}) / (E_{k0,ijk} + \Delta E_{k,ijk})}$ is applied, where $E_{k0,ijk}$ is the initial kinetic energy of the grid cell. The final momentum of the grid cell is then updated as $\overset{\to }p_{ijk} = \alpha \times (\overset{\to }p_{0,ijk} + \Delta \overset{\to }p_{ijk})$, where $\overset{\to }p_{0,ijk}$ is the  initial momentum. In cases where $\Delta E_{k,ijk} < \Delta e_{ijk}$, the remaining energy is added to the internal energy, $\Delta E_{ine,ijk} = \Delta e_{ijk} - \Delta E_{k,ijk}$.

If $0.027<R_{M}<1$, the Sedov-Taylor phase can be resolved in the simulation. In this case, we distribute the total energy $\Delta e$ released from the star particle among the grid cells within its supernova feedback zone, allocating $72\%$ to internal energy and $28\%$ to kinetic energy to ensure energy conservation (\citealt{Kim_2015}). Firstly, First, we calculate the shock front velocity as $v_{sn} = \sqrt{2\times0.28 \Delta \rm{E_{sn}} /M_{ej}}$, which allows us to compute the velocity and the corresponding kinetic energy increment in the x, y and z directions for each grid cell within the initial SNR region as
\begin{equation}
    \Delta v_{(x,y,z), ijk} = v_{(x,y,z),p} + \overset{\to }{n}_{(x,y,z)} \times v_{sn}  
\end{equation}
\begin{equation}
    \Delta E_{k, (x,y,z), ijk} = \frac{1}{2} \Delta m_{ijk} \times \Delta v_{(x,y,z), ijk}^{2}  
\end{equation}
where $\Delta m_{ijk} = f_{ijk}\times M_{ej}$ and $v_{(x,y,z),p}$ is the velocity of star particle, respectively. The subscripts (x,y,z) represent three directions for the grid cells. Subsequently, we calculate the final energy and momentum components in each direction for a given grid cell as follows:
\begin{equation}
    m_{ijk, fin} = m_{ijk, init} + \Delta m_{ijk}  
\end{equation}
\begin{equation}
    E_{k, ijk, fin} = E_{k, ijk, init} + \Delta E_{k, ijk} 
\end{equation}
\begin{equation}
    E_{ine, ijk, fin} = E_{ine, ijk, init} + 0.72\Delta \rm{E_{sn}}  
\end{equation}
\begin{equation}
    \overset{\to }p_{ijk, fin} = \frac{\overset{\to }v_{ijk, init}}{\left | \overset{\to }v_{ijk, init} \right | }  \times \sqrt{2E_{k, ijk, fin} \times m_{ijk, fin}}   
\end{equation}

For star particles located in very low-density regions where $R_{M}<0.027$, we apply the same procedure used for the case of $0.027<R_{M}<1$, but allocate $100\%$ of the supernova energy to kinetic energy. 

In addition, the metal increment of each grid cell within the SNR is given by $\Delta m_{metal,ijk} = f_{ijk} \times M_{metal,ej}$.



\bibliography{main}{}
\bibliographystyle{aasjournal}



\end{document}